\documentclass[aps, prd, nofootinbib, amsmath, preprintnumbers, superscriptaddress,secnumarabic, notitlepage,twocolumn]{revtex4-2} 
\usepackage{amssymb,amsmath,mathrsfs,graphicx,latexsym,amsthm,slashed,eso-pic,hyperref}
\usepackage{bbold}
\usepackage{aas_macros}
               
        \newcommand{\cO}{{\cal O}}

\newcommand{\pL}{\left(} \newcommand{\pR}{\right)} \newcommand{\bL}{\left[} \newcommand{\bR}{\right]}    
\newcommand{\beq}{\begin{equation}} \newcommand{\eeq}{\end{equation}}
\newcommand{\bea}{\begin{eqnarray}} \newcommand{\eea}{\end{eqnarray}}

\newcommand{\Eq}[1]{Eq.~(\ref{#1})}

 \DeclareMathOperator{\tr}{Tr}

\newcommand{\be}{\begin{equation}}
\newcommand{\ee}{\end{equation}}
\newcommand{\nn}{\nonumber}

\def\sfrac#1#2{{\textstyle{#1\over #2}}}

\definecolor{orange}{rgb}{1,0.5,0}

\usepackage{slashed}
\usepackage{appendix}
\usepackage{siunitx}

\begin{document}

\title{Late-Time Dark Matter Oscillations and the Core-Cusp Problem}
\author{James M.\ Cline}
\affiliation{McGill University, Department of Physics, 3600 University Street, Montr\'eal, QC H3A 2T8, Canada}
\author{Guillermo Gambini}
\affiliation{McGill University, Department of Physics, 3600 University Street, Montr\'eal, QC H3A 2T8, Canada}
\affiliation{Instituto de F\'isica Gleb Wataghin, UNICAMP, Rua S\'ergio Buarque de Holanda 777, 13083-859, Campinas-SP, Brasil}
\author{Samuel D.\ McDermott}
\affiliation{Theory Division, Fermi National Accelerator Laboratory, Kirk Road, Batavia, IL 60510, U.S.A}
\author{Matteo Puel}
\affiliation{McGill University, Department of Physics, 3600 University Street, Montr\'eal, QC H3A 2T8, Canada}

\date{\today}

\begin{abstract}
The core-cusp problem persists as an unresolved tension between the predictions of $\Lambda$CDM cosmology and observations of dark matter (DM) profiles in dwarf spheroidal and other galaxies.  We present a novel scenario for converting cusps into cores through reactivation of DM annihilation in galaxies at late times.  This can happen in asymmetric DM models when there is a very small DM-number violating mass term that causes oscillations between DM and its antiparticle. Using analytic methods as well as gravitational N-body simulations, we show that this mechanism can robustly eliminate cusps from galactic DM profiles for light fermionic DM of mass $m_\chi\sim (0.1-1)$ GeV and a lighter  mediator into which the DM can annihilate. We identify regions of parameter space where annihilation of DM particles is more efficient than elastic scattering at reducing the inner density of the DM profile. Dark matter annihilation is therefore a qualitatively distinct alternative to the mechanism of elastic self-interacting dark matter for addressing the cusp-core problem.
\end{abstract}

\preprint{FNAL-PUB-20-556-T}

\maketitle

\section{Introduction}
In many respects the standard $\Lambda$CDM paradigm of cosmology gives an extremely good description of the observed universe.
But it has long been recognized that simulations of structure formation that neglect the presence of baryons predict singular (cuspy) density profiles of the dark matter (DM) toward the centers of galaxies, whereas observations suggest flatter (cored) distributions~\cite{Salucci:2018hqu}.  More recent simulations include the effects of baryonic feedback, which can expel material from denser regions and help to ameliorate this discrepancy, but there is not yet any consensus that this provides a complete solution.  Moreover in systems like dwarf spheroidals, where baryons are relatively scarce, one does not expect baryons to have a significant impact on the small scale structure.  These issues have been reviewed in ref.\ \cite{Weinberg:2013aya}. 

Another proposed solution is self-interacting dark matter (SIDM)  \citep{Spergel:1999mh,Tulin:2017ara}, with a scattering cross section at the level of 
\be
    {\frac{\sigma}{m_\chi}}\sim (0.1-1)\,{\frac{{\rm cm}^2}{\rm g}}
    \label{sigmam}
\ee
that is close to upper bounds from colliding galaxy clusters, such as the Bullet Cluster \cite{Markevitch:2003at}, even though it may be not so robust~\cite{Kim:2016ujt}.
N-body simulations incorporating such interactions have shown that cross sections consistent with eq.\ (\ref{sigmam}) can produce cored DM profiles in a wide range of systems 
\citep{Rocha:2012jg,Vogelsberger:2012ku}.  However, more recent studies indicate that a constant cross section is not the ideal solution, since then $\langle\sigma v\rangle$ increases with the size of the system, contrary to the observation that cores are less pronounced on the scales of galactic clusters.
A weak velocity dependence of the form $\sigma\sim 1/v$ is found to give a better fit to the full range of structures \cite{Kaplinghat:2015aga}. 

The most common assumption is that the DM self-interaction is in the form of elastic scattering, but a more exotic possibility was proposed in ref.\ \cite{McDermott:2017vyk}, in which fusion of DM particles into bound objects is the interaction leading to cored profiles.  Like other exothermic processes, this has the advantage of predicting a cross section with $\sigma v$ remaining constant at low velocities, as desired for fitting the DM profiles of both large and small galactic structures. 
Other interesting possibilities to achieve the correct velocity-dependence of SIDM have been studied, for instance in the context of resonant SIDM~\citep{Chu:2018fzy,Chu:2019awd,Tsai:2020vpi}, puffy DM~\cite{Chu:2018faw}, self-heating DM~\citep{Chu:2018nki,Kamada:2019wjo,Kamada:2017gfc,Kamada:2018hte}, maximally SIDM~\cite{Kamada:2020buc} and DM bound states produced in the early universe by three-body recombination~\citep{Braaten:2018xuw,Braaten:2019ohj}.

Here we explore a different alternative, motivated by the
fact that DM annihilation is also an exothermic process with $\sigma v$ becoming constant as $v\to 0$.  The challenge for such a scenario is to explain how annihilations could go out of equilibrium in the early universe, but then come back at late times \cite{Kaplinghat:2000vt}.  In fact, a mechanism to do this is well known in the context of asymmetric dark matter, where there is an asymmetry between the DM $\chi$ and its antiparticle $\bar\chi$.  By allowing for a small mass term that violates the conservation of DM number, oscillations between $\chi$ and $\bar\chi$ can reactivate the annihilations at late times \cite{Cohen:2009fz, buckley2012, Cirelli:2011ac, Tulin:2012re}.

The reactivation of DM annihilation at late times is usually seen as a danger to be avoided, since it is known that the DM density should not change appreciably between the era of the CMB (redshift $z\sim 1100$) and structure formation \cite{Poulin:2016nat,Bringmann:2018jpr}, but in the present work we demonstrate that this mechanism can efficiently produce cored profiles in galaxies without changing the total DM density significantly.  The reason is that the efficiency of oscillations leading to regeneration of the anti-DM component can depend strongly on density, so that annihilations are effective in the centers of {galaxies} but not in the outer regions.

For a more quantitative investigation, one should integrate quantum Boltzmann equations for the density matrix, that account for the coherence of states undergoing oscillations, analogous to those used for the study of neutrino oscillations in a medium.  This formalism was initially worked out for DM in ref.\ \cite{Cirelli:2011ac}, and some important corrections were realized in ref.\ \cite{Tulin:2012re}, which we follow closely in the present work.  

We consider two models of quasi-Dirac fermionic DM $\chi$ of mass $m_\chi$. In the first, the dark matter couples to a lighter vector boson $V^\mu$ (Model 1), with effective Lagrangian
\be
   {\cal L}_1 \supset -\sfrac12 m_V^2 V_\mu^2 - g'\bar\chi\slashed{V}\chi\,.
 \label{eqLag1}
\ee
In Model 1, the dark matter freeze-out and the late-time depletion are both allowed by the annihilation process $\chi\bar\chi\to VV$.
In the second model, dark matter couples to a complex scalar $\Phi = \phi + ia$ (Model 2),
\be
   {\cal L}_2 \supset -\sfrac12 m_\phi^2\phi^2 -\sfrac12 m_a^2 a^2
  -g'\bar\chi(\phi + ia\gamma_5)\chi\,.
   \label{eqLag2}
\ee
Model 2 allows freeze-out and late-time depletion from $\chi\bar\chi\to\phi a$ (which unlike $\chi\bar\chi\to\phi\phi$ or $\chi\bar\chi\to a a$ is $s$-wave, hence not suppressed at low velocities). The coupling between $\chi$ and either kind of
boson is denoted as $g'$, and its associated fine-structure constant is $\alpha' = g'^2/4\pi$.
The DM-violating mass term is 
\be
    {\cal L}_m = \sfrac12\delta m \left(\bar\chi\chi^c + {\rm H.c.}\right)\,.
    \label{eqdm}
\ee
 The parameter $\delta m$ violates not only dark matter number, but also the gauge symmetry of Model 1, which
 is additionally broken by the Stueckelberg 
 mass term for the vector.  It would be possible to replace both of these explicit breakings by a Higgs mechanism,
 but for simplicity we adopt the simpler effective theory.

We begin by making preliminary estimates to identify viable regions of the parameter space, in section \ref{sec:analyt}.
The essential details of the density matrix Boltzmann equation formalism are reviewed in section\ \ref{sec:form}.
In section\ \ref{sec:early} we will show that, for appropriate choices of the model parameters, integration of the Boltzmann equations in the early universe leads to the conventional freeze-out of DM annihilations, leaving only the asymmetric component of the DM.  This justifies the initial conditions for the second step,
described in section\ \ref{sec:struc}, where we re-solve the
analogous Boltzmann equations in the background of an
already-formed galaxy and show how an initial cusp gets erased by reactivated annihilation following $\chi$-$\bar\chi$ oscillations.  This is a somewhat crude approach since it considers formation of the galaxy to happen suddenly and neglects the role of gravity in shaping the DM halo. In section\ \ref{sec:sim}, we improve on this by carrying out a gravitational N-body simulation of galaxy evolution, in a code adapted to properly account for the new physics effects.
Conclusions are given in section\ \ref{sec:conc}, and details of the
quantum Boltzmann and N-body simulation methods are presented in the appendices.

\section{Analytic estimates}
\label{sec:analyt}
Before embarking on a detailed analysis, we analytically estimate the regions of parameter space that are of interest for our mechanism.  First, the annihilation cross sections at threshold for the two models {are}
\be
\langle\sigma v\rangle_a = {\pi\,\alpha'^2\over m_\chi^2}
\times\left\{\begin{array}{cc} (1-r_m^2)^{3/2}/(1-r_m^2/2)^{2},& \hbox{Model 1}\\ (1-r_m^2/4),& \hbox{Model 2}\end{array}\right.
\label{eqxsects}
\ee
where $r_m$ is the ratio of the mediator to the DM mass, 
$r_m = m_V/m_\chi$ for $\chi\bar\chi\to VV$ (Model 1) or $r_m = m_\phi/m_\chi$ for $\chi\bar\chi\to\phi a$ (Model 2).  In the latter, we have assumed for simplicity that 
$m_a\ll m_\phi$, and neglected 
the $p$-wave suppressed channels $\chi\bar\chi\to \phi\phi,\, a a$. 

To compare eq. (\ref{eqxsects}) to the desired cross section (\ref{sigmam}), consider a reference
velocity $v_0=100$\,km/s characteristic of DM in a Milky-Way-like galaxy, 
and the upper value in the range (\ref{sigmam}), giving $\sigma v/m_\chi\sim 100$\,cm$^2$\,km/s/g $\sim 0.2 \times(100{\,\rm MeV}/m_\chi)$\,GeV$^{-2}$.  Equating this to $\langle\sigma v\rangle_a$ suggests the
constraint
\be
   \alpha' \cong 0.7\,\left(m_\chi\over{\rm GeV}\right)^{3/2}
\label{eqalphap}
\ee
For example with $m_\chi = 100\,$MeV, $\alpha' \cong 0.02$; we will adopt these as approximate benchmark values.
However nothing prevents us from taking somewhat heavier DM, up to
$m_\chi\sim 1\,$GeV; above this, the theory starts to be
strongly coupled.

It is impossible to avoid $\chi\chi$ elastic scattering
mediated by the annihilation products, and we choose to constrain these cross sections so that they are below the
level that would change the DM density profile independently
of the annihilation effect, which is the focus of this work.
The elastic scattering cross sections at low velocities are
\be
    \sigma_s \cong 4\pi\alpha'^2\,{m_\chi^2}\left\{\begin{array}{cc} m_V^{-4},& \hbox{Model 1}\\
    m_\phi^{-4} + (5/4)v^4 m_a^{-4},& \hbox{Model 2}
    \end{array}\right.
    \label{eqsigmas}
\ee
where $v = v_{\rm rel}/2$ is the center-of-mass velocity. 
Here, all the relevant channels contributing to the $\chi \chi$ and $\chi \bar{\chi}$ scatterings are included and the cross sections for these two processes turn out to be the same in the low-velocity limit.
To avoid that the scattering self-interactions play a leading role in the galactic dynamics, we require that $\sigma_s v_{\rm rel} \ll \langle\sigma_a v\rangle$.  This implies $(m_{V,\phi}/m_\chi)^4 \gg 4 v_{\rm rel}$, which is most stringent for large systems,
{galaxy} clusters, that have the highest DM velocities.
For example, the cluster A2537 has 
velocity dispersion $\sigma_v\sim 1000\,$km/s \cite{Newman:2012nv},
with $v_{\rm rel} = (4/\sqrt{\pi})\sigma_v$ (assuming the velocity is Maxwell-Boltzmann distributed), and demanding that $\sigma_s v_{\rm rel} < 0.3\,\langle\sigma_a v\rangle$ gives the
constraints
\bea
  0.6 &<&  r_m < 0.94,\hbox{\ Model 1}\nn\\
  0.6 &<&  r_m  < 1.99,\hbox{\ Model 2}
 \label{eq:consistent}
 \eea
where the upper limits come about because of phase space suppression of the annihilation.\footnote{We assumed $m_a\ll m_\phi$ in the last limit, for the process $\chi\chi\to\phi a$.}\ \  

The pseudoscalar mass $m_a$ should not be arbitrarily small, since its virtual contributions can become Sommerfeld enhanced if $m_a\ll\alpha'm_\chi$,
$v m_\chi$
\cite{Bellazzini:2013foa,Kahlhoefer:2017umn,Agrawal:2020lea}.  In the present work we avoid these complications by considering
$m_a\sim m_\chi/10$, which is small enough to ignore it in phase space integrals and
 its $d$-wave suppressed contribution to scattering in eq. (\ref{eqsigmas}),
  but large enough to avoid nonperturbative effects, as well as cosmological problems
  in the era of Big Bang Nucleosynthesis.

The $\chi$-number violating mass $\delta m$ must be small enough so that $\chi$-$\bar\chi$ oscillations have not yet
started at the time of DM freeze-out, $T_{\chi{\rm,fo}}\sim m_\chi/20$, where we allow for a lower temperature
\be
    T_\chi \equiv \xi T < T
\ee
in the dark sector, as discussed in more detail in Sec.~\ref{sec:const}. For annihilations to recouple during structure formation, the oscillations should
start before the epoch of structure formation,  $t_s  {\sim0.1}\,$Gyr.  On the other hand, we will show in Sect.\ \ref{sec:early} that too-early onset of recoupled annihilations tend to change the DM relic density more than is allowed by CMB constraints
\cite{Poulin:2016nat,Bringmann:2018jpr}.  This leads to a window of allowed values, whose upper limit depends upon details of the scenarios we will discuss,
\bea \label{maj-mass-scale}
&&{1\over t_s} \lesssim \delta m \lesssim \left\{\begin{array}{ll} {16.3\,\delta_\eta^{1/2}\,m_\chi^{1/2}\over
    g_*^{1/4}\overline{\langle\sigma v\rangle}_s\langle\sigma v\rangle_a^{1/2}
    \eta_{\rm DM}^{3/2} M_p^{5/2} },& \hbox{Model\  }1 \\
{342\,\delta_\eta^{1/2}\over g_*^{1/2}\langle\sigma v\rangle_a^2\,\eta_{\rm DM}^2\,M_p^3}, & \hbox{Model\ 2}\end{array}    \right. \\
&&\implies 10^{-31}\,{\rm eV}\lesssim \delta m \lesssim  \left\{\begin{array}{ll} 5\times 10^{-28}\,{\rm eV},&
\hbox{Model\  1}\\ 3\times 10^{-30}\,{\rm eV},&
\hbox{Model\ 2}\end{array}\right.,\nn
\eea
where $\eta_{\rm DM}$ is the DM asymmetry and
$\delta_\eta$ is the fractional change in $\eta_{\rm DM}$ allowed by the CMB constraints.  The numerical values are indicative, based on the limited parameter choices we have investigated here.
It is possible that the upper limits could be relaxed 
in a wider search of parameter space.  The analytic expressions in eq.\ (\ref{maj-mass-scale}) are derived in Appendix \ref{app:dm}.

\section{Oscillation formalism}
\label{sec:form}
In the presence of DM oscillations, the distinction between particle
and antiparticle becomes time-dependent.  If we define a basis
\be
    |\chi\rangle = \left(1\atop 0\right), \quad 
        |\bar\chi\rangle = \left(0\atop 1\right)\,,
\label{basis}
\ee
then it is straightforward to show that the time dependence of a state that is initially pure $|\chi\rangle$ is
\be
   \left |\chi(t)\right\rangle = e^{-i m_\chi t}\,\left(c_\varphi\atop -is_\varphi\right)
   \label{chi_osc}
\ee
with $c_\varphi = \cos\varphi$, $s_\varphi = \sin\varphi$, $\varphi = \delta m\, t$.
To this state we can associate the density matrix for a single-particle state,
\be
    n_1 = \left|\chi(t)\right\rangle\left\langle\chi(t)\right| = \left({c_\varphi^2\atop -ic_\varphi s_\varphi}\,{i c_\varphi s_\varphi\atop s^2_\varphi}\right)\,.
\label{freeneq}
\ee
Naively it might seem like appreciable amounts of $\bar\chi$ appear as soon as $\varphi\sim 1$ and $\chi\bar\chi$ annihilations could recommence, but this need not be true if all the particles in the plasma are oscillating with the same phase.  Ref.\ \cite{Tulin:2012re} showed that recoupling of annihilation depends on the nature of the interactions.  Interactions of fermionic DM with vectors $V$ are called ``flavor sensitive,'' while interactions of $\chi$ with scalars or pseudoscalars are ``flavor blind,'' leading to very different behaviors of the annihilation probabilities.  In the collision of
two particles with respective phases $\varphi$ and $\varphi'$, the annihilation rates are modulated by the factors
\bea
 \label{sin_factorV}
 \chi\bar\chi\to VV: && \ \sin^2(\varphi-\varphi') \hbox{\ (flavor sensitive)}\,, \\
 \chi\bar\chi\to \phi a:  && \ \sin^2(\varphi+\varphi')
 \hbox{\ (flavor blind)}\,.
 \label{sin_factorS}
\eea
In the first case, a bath starting as pure $|\chi\rangle$ and maintaining phase coherence never undergoes annihilations since
$\varphi - \varphi'$ remains zero, despite the oscillations.  In the second, the modulation factor
averages to $1/2$ for fast oscillations, and is therefore effective even when the particles stay in phase with each other.

For a thermal bath, the matrix $n_1$ in eq.~\eqref{freeneq} is replaced by an integral over the corresponding matrix distribution function ${\cal F}(k)$ for the states of momentum $k$,
\be
    n = (2s+1)\int{d^{\,3}k\over(2\pi)^3}\,
    {\cal F}(k)\,,
\label{kint}
\ee
where $s=1/2$ for fermions as we consider.
Then $n_{11}$ ($n_{22}$) represents the number density of particles (antiparticles) defined with respect to the basis $\{ |\chi\rangle, |\bar\chi\rangle \}$ as in (\ref{basis}); the off-diagonal elements keep track of the coherence between these two states.

The Boltzmann equation for $n$ reduces to the usual form when we consider only the diagonal elements, but it has additional terms due to the off-diagonal elements, which depend on whether the interactions are flavor-sensitive or flavor-blind. 

\subsection{Model 1: vector mediator}
We first consider the flavor-sensitive case, applicable to Model 1, for which the Boltzmann equation is
\bea
    \dot n + 3 H n &=& -i[{\cal H}_0,n]- \sfrac32\langle\sigma v\rangle_s \left(\tr n\right)\left({0\atop n_{21}}\,{n_{12}\atop 0}\right) \nn\\
    &-& \langle\sigma v\rangle_a\left({\rm det}\, n 
    - n_{\rm eq}^2\right)\,\mathbb{1}\,,
\label{boltz1}    
\eea
where $H$ is the Hubble parameter, the thermally averaged free Hamiltonian is
\bea
    {\cal H}_0 &=& \langle E\rangle\,\mathbb{1} + \left\langle {m_\chi\,\delta m\over E}\right\rangle \left({0\atop 1}\,{1\atop0}\right)\nn\\
    &\cong& m_\chi \,\mathbb{1} + \delta m\left({0\atop 1}\,{1\atop0}\right)\,,
\label{Hamiltonian}
\eea
$\langle\sigma v\rangle_s$ is the $\chi\chi$ or $\chi\bar\chi$ scattering cross section (that coincides at low energies), $\langle\sigma v\rangle_a$
is the $\chi\bar\chi\to VV$ annihilation cross section,
and $n_{\rm eq}$ is the equilibrium number density.
The scattering term in eq.\ (\ref{boltz1}) is derived in appendix \ref{appA}, while the other terms can be found in ref.\ \cite{Tulin:2012re}.\footnote{Ref.\ \cite{Tulin:2012re} derived the scattering term for
$\chi f\to \chi f$ with $f$ being a different particle in the plasma.}
Eq.\ (\ref{boltz1}) is the appropriate form for cosmology; in section \ref{sec:struc} we will discuss how it can be applied in a galactic environment.

The scattering term in (\ref{boltz1}) has the effect of 
damping the off-diagonal elements of $n$, which destroys the
coherence of the quantum superpositions and effectively measures the state of an oscillating system.  The loss of
coherence results in $\det n\neq 0$, which activates the 
annihilations.  Since $\langle\sigma v\rangle_s$ is proportional to the DM velocity, this makes the effect stronger in systems with large velocity dispersions.  We will see in section \ref{sec:struc} that this is contrary to observations, disfavoring Model 1 taken 
by itself.

The origin of the factor (\ref{sin_factorV})
can be heuristically understood from (\ref{boltz1}) by interpreting the annihilation 
term $\det n\,\mathbb{1}$ as the matrix \cite{Tulin:2012re}
\bea
\det n\, \mathbb{1} &\to& \sfrac12(n_1 \sigma_2  n_2^T \sigma_2 + n_2 \sigma_2 n_1^T \sigma_2)\nn\\
&=& \sfrac12 \sin^2(\varphi_1-\varphi_2)\,\mathbb{1}\,,
\label{deteq}
\eea
where $n_1$, $n_2$ represent the density matrices (\ref{freeneq}) of two particles, having respective phases $\varphi_1$, $\varphi_2$, and $\sigma_2$ is the Pauli matrix.\footnote{In the notation of ref.\ \cite{Tulin:2012re}, $\sigma_1 n^T\sigma_1 = \bar n$ and $\sigma_3 = O_-$. Appendix \ref{thdec} implies that the actual matrix structure is more complicated than (\ref{deteq}), but this form is adequate for our application of it in section \ref{sec:struc}, which can only account for coherence effects in an approximate way. In particular, the off-diagonal elements are not exactly zero, but they average to zero over the ensemble.}  

Similarly, the effect of the scattering term can be understood by replacing $n\to n_1$ everywhere except in 
the trace, where $\tr n\to \tr n_2$, which does not depend on $\varphi_2$, and just represents the total density $n$ of DM scattering on particle 1.  Then the off-diagonal parts of the Boltzmann equation 
determine the damping of $\varphi_1$ as 
\be
    {d\over dt}(c_{\varphi_1}s_{\varphi_1}) \sim - \sfrac32 n\langle\sigma v\rangle_s     c_{\varphi_1} s_{\varphi_1}\,,
\label{vpdamp}
\ee
which has the solution $c_\varphi s_\varphi \sim \exp(-\sfrac32\Gamma_s t)
(c_\varphi s_\varphi)_0$, where $\Gamma_s = n\langle\sigma v\rangle_s$ is the elastic
scattering rate.

\subsection{Model 2: scalar mediators}
For scalar interactions, the Boltzmann equation simplifies, because elastic scattering no longer has any effect on 
the density matrix.  The form of the annihilation term is 
also changed, in a way that makes it lead to decoherence 
by itself
\bea \label{boltz2}   
    \dot n + 3 H n &=& -i[{\cal H}_0,n]\\
   &-& \langle\sigma v\rangle_a\left[
    \left({\det'n\atop (\tr n)n_{21}}\,{(\tr n)n_{12}\atop \det'n}\right)- n_{\rm eq}^2\,\mathbb{1}\right]\nn\,.
\eea
Here we define $\det'n \equiv n_{11} n_{22} + n_{21}n_{21}$.
In contrast to eq.\ (\ref{boltz1}) for the vector model,
there is no dependence on the DM velocity in (\ref{boltz2}), leading one to expect a more similar level of cusp erasure in both
large and small galactic systems, independently of the 
differences in their velocity dispersions.

The analogous reasoning that led to eq.\ (\ref{deteq}) can
be applied in the simpler case where $\varphi_1=\varphi_2=\varphi$ since 
the annihilation term no longer vanishes in that limit, giving
\bea
{\det}'n  &\to& \sfrac12\{n, \sigma_1  n^T \sigma_1\}\nn\\
&=& \sfrac12 \sin^22\varphi\,\mathbb{1} + \sfrac12\sin2\varphi\,\sigma_2\,.
\label{detpneq}
\eea
The diagonal term goes to (\ref{sin_factorS}) when the two phases are
different from each other.  The off-diagonal term leads to phase damping similarly to (\ref{vpdamp}), but now with 
$c_\varphi s_\varphi \sim \exp(-\Gamma_a t)
(c_\varphi s_\varphi)_0$, where $\Gamma_a$ is the conventional annihilation
rate, without the $\sin^22\varphi$ modulation factor.

\begin{figure*}[ht]
  \centerline{\includegraphics[height=0.37\textwidth]{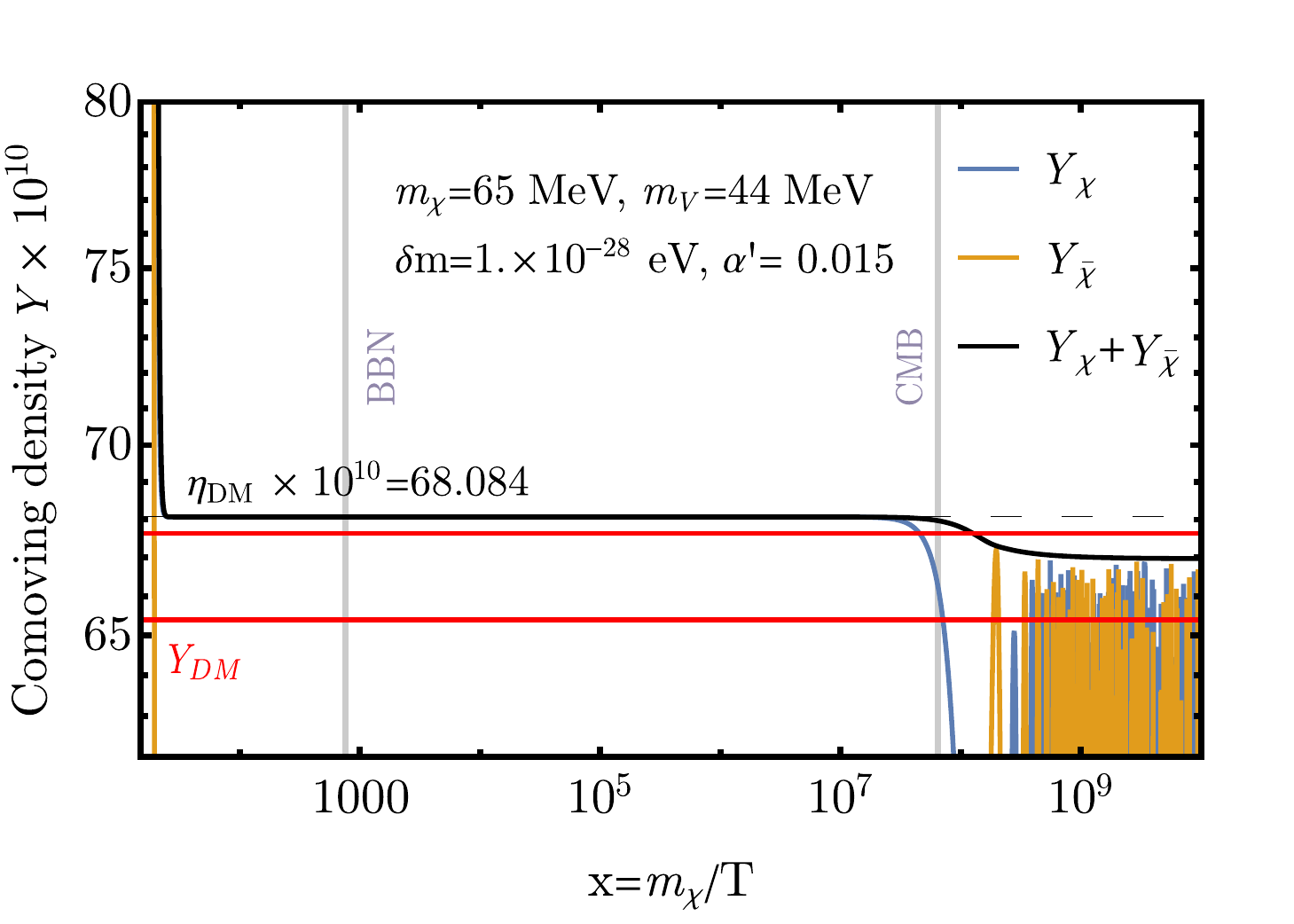}
  \includegraphics[height=0.365\textwidth]{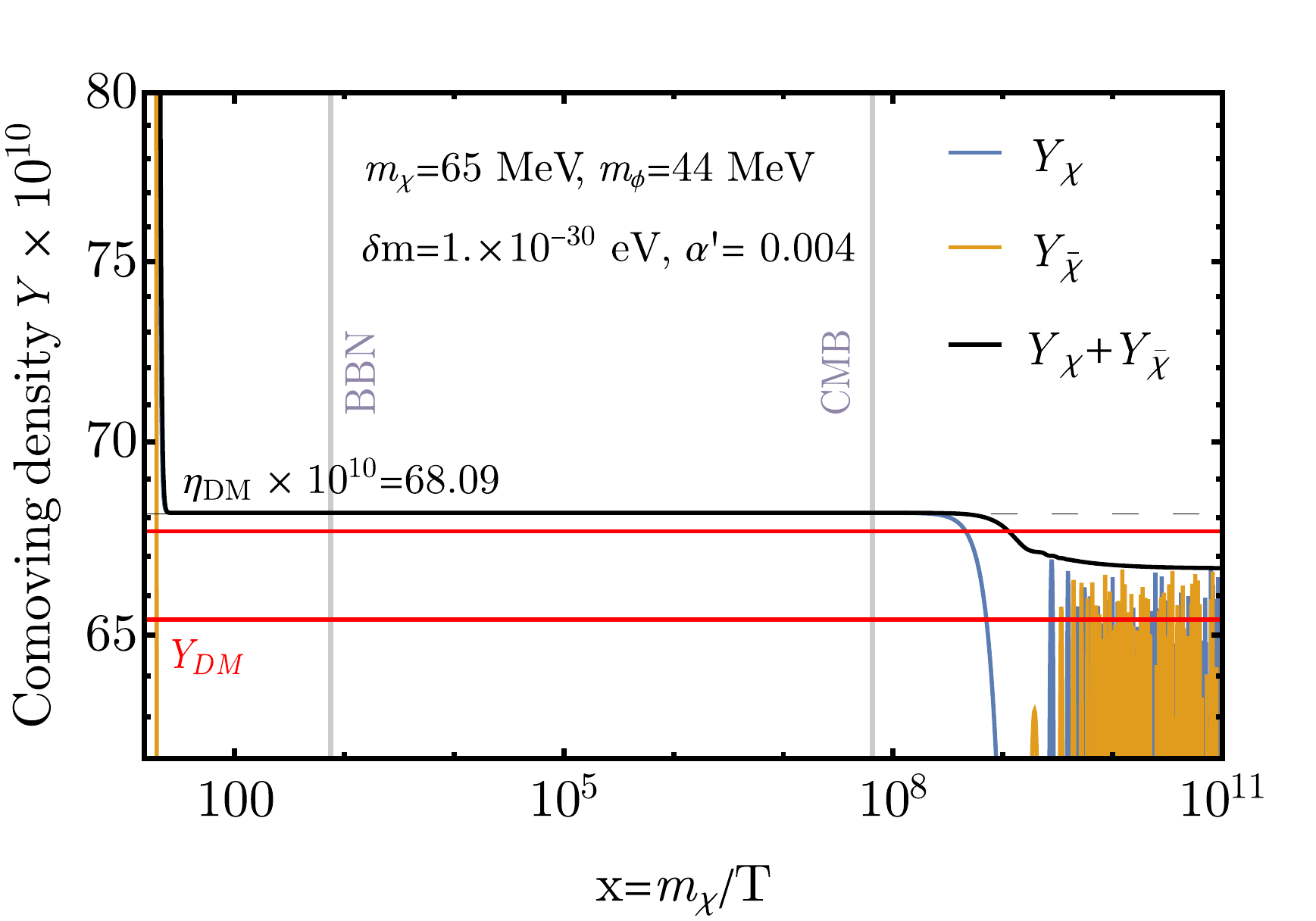}}
  \caption{Cosmological evolution of $\chi$, $\bar\chi$ and total abundances for Model 1 (left) and Model 2 (right).  The model parameter values are indicated in the plots.  We indicate the approximate time of BBN and CMB with faint gray vertical lines. The ratio of dark to visible sector temperatures is taken to be $\xi=1$. }
\label{TvsNTD}
\end{figure*}

\section{Early cosmology}
\label{sec:early}
For  small values of $\delta m \lesssim 10^{-30}\,$eV, 
oscillations are unimportant until the epoch when structure formation begins.  For larger values of $\delta m$ they can cause annihilations to temporarily recouple, further reducing the
density of the asymmetric component, before structure formation begins and annihilations are reactivated once
again.  In this section we illustrate these possibilities
by solving the Boltzmann equation at early times.  This is meant to provide the initial conditions before the effects of oscillations
on structure formation begin, that we will investigate in the following sections.

\subsection{Model 1} 
Like for conventional freeze-out, it is convenient to use
$x \equiv m_\chi/T$ as the independent variable, and the abundance $Y \equiv n/s$ as the dependent variable,
where $s=2\pi^2 g_{*s} m_\chi^3/(45 x^3)$ is the entropy density and $Y$ is now a matrix. The Boltzmann equation becomes
\bea
Y'&=&-\frac{i}{xH}\left[ \mathcal{H}_0, Y \right] -\xi^3{3\langle\sigma v\rangle_s s\over 2 x H}\left({0\atop Y_{21}}\,{Y_{12}\atop 0}\right)\tr Y \nn\\
&-&\xi^3\frac{\left< \sigma v \right>_a s}{xH} \left( \det Y  -Y_{\rm eq}^2 \right)\,\mathbb{1}\,. 
\label{boltz1c}
\eea
{Here} $H\cong 1.66 \sqrt{g_*}m_\chi^2/( M_p x^2)$ is the Hubble parameter, and we have allowed for the DM temperature to differ from that of the standard model by putting the appropriate factors of $\xi = T_\chi/T$. The
averaged scattering cross section is
\bea
\langle\sigma v\rangle_s &=& \sigma_0 \sqrt{\xi\over x},\qquad
\sigma_0 \cong  8\pi\alpha'^2 {m_\chi^2\over m_V^4}\,,
\eea
and the equilibrium abundance is
\be
    Y_{\rm eq} \cong {45\, x^3\over 2\pi^4\,g_{*s}}\,{ K_2(\xi x)\over \xi x}
\ee
in the Maxwell-Boltzmann approximation.  It is the abundance of just the DM particle $\chi$, not including the antiparticle.

There is an additional possible source of decoherence that is not captured by eq.\ (\ref{boltz1c}).  The full Hamiltonian (\ref{Hamiltonian}), before taking the non-relativistic limit, depends
on the momentum $k$ of the state, which is neglected in  (\ref{boltz1c}).  This causes states of different momenta to oscillate at slightly different frequencies $\delta\omega \sim \delta m\, (k^2/2 m_\chi^2)$, giving rise to thermal decoherence even in the absence of scattering.  To fully investigate {this} 
effect would require solving for the full distribution function ${\cal F}(k)$, which is numerically prohibitive.  Instead we model it in an approximate way, by splitting the integral over $k$ in (\ref{kint}) into two bins of small and large
momenta, $Y = Y_s + Y_l$.  The averaged Hamiltonians ${\cal H}_{s,l}$
for the respective bins are shown in eq.\ (\ref{Hsleq}).  The resulting coupled Boltzmann equations are given in eqs.\ (\ref{boltz-2bin}).  They have a more complicated matrix structure
than (\ref{boltz1c}), but the sum of the two agrees with (\ref{boltz1c}).  We found this additional source of decoherence to have a negligible effect, compared to that due to the scatterings.

The left panel of Fig.\ \ref{TvsNTD} shows the effect of thermal decoherence in the  evolution of the oscillating dark matter in the vector model at early, intermediate, and late times. 
We see that after oscillations commence at late times (values of $x \sim 10^7-10^8$ in the two models), annihilations recouple briefly before freezing out again. The dark matter density, $Y = Y_{\bar{\chi}} + Y_\chi$, is reduced by $\lesssim \cO(5\%),$ which is roughly compatible with observations of density perturbations in the CMB \citep{Poulin:2016nat}.  These constraints were refined in Ref.\ 
\cite{Bringmann:2018jpr}, which limits the change in the DM abundance $Y$ as a function of the redshift of the transition as well as its duration,
with respect to CMB data.  We have checked in detail that the examples shown in Fig.\ \ref{TvsNTD} are compatible with the limits found there.

On the other hand, we observe that 
increasing $\delta m$ has the effect of shifting the
recoupling of annihilation to earlier times, and also
inducing much larger changes in $\Delta Y$, that would
be in conflict with the CMB.  By varying $\delta m$ and
comparing to the excluded regions from Ref.\ \cite{Bringmann:2018jpr},
we arrive at the approximate upper bounds 
shown in eq.~\eqref{maj-mass-scale}.
It is possible that models saturating these limits
could ameliorate current tensions in the different measurements of  $H_0$ \cite{Blinov:2020uvz} and $\sigma_8$ \cite{Abellan:2020pmw}. We leave further exploration of these implications for future work.

\subsection{Model 2}  

The cosmological version of the Boltzmann equation (\ref{boltz2})
is
\bea
Y'&=&-\frac{i}{xH}\left[ \mathcal{H}_0, Y \right] \\
&-&\xi^3\frac{\left< \sigma v \right>_a s}{xH} \left[
\left({\det' Y\atop Y_{21}\tr Y}\,{Y_{12}\tr Y\atop \det' Y}\right)  -Y_{\rm eq}^2\,\mathbb{1} \right]\nn\,,
\label{boltz2c}
\eea
where scatterings no longer play any role. {Its solution is shown in the right panel of Fig.~\ref{TvsNTD}.}  The implications of the brief recoupling in dark matter annihilation are similar as in Model 1.

\subsection{Constraints on $N_{\rm eff}$}
\label{sec:const}

As usual for dark matter models coupled to light mediators, indirect detection constraints from X-ray and gamma-ray telescopes require the hidden sector to be largely secluded from the Standard Model. Moreover, the force mediators like the vector $V$ of Model 1 or the scalars $\phi$ and $a$ of Model 2, which are the products of $\chi \bar \chi$ annihilation, must decay into radiation of some sort to avoid dominating the energy density of the universe at low temperatures, for example at the time of BBN. 
The simplest solution to these possible issues is to introduce a dark radiation species, such as massless sterile neutrinos $\nu'$, that couple to the mediators and allow for the decays $V,\phi,a\to\nu'\bar\nu'$.   
As long as these new species have a mass smaller than $\sim {\rm eV}$, they will not come to matter-dominate the universe before the formation of the CMB. 

Even this single light degree of freedom might be detected by precise probes of the energy content of the early universe at BBN \cite{Cyburt:2015mya, Fields:2019pfx} and CMB \cite{Aghanim:2018eyx}, which constrain the number of new relativistic degrees of freedom. For single-parameter extensions of $\Lambda$CDM, the constraints are of order $\Delta N_{\rm eff} \lesssim 0.2$, but known parameter degeneracy with the helium abundance $Y_p$ and possible hints of beyond-$\Lambda$CDM physics such as neutrino masses or the $H_0$ tension can partially relax these constraints to the level of $\Delta N_{\rm eff} \lesssim 0.5$ \cite{Cyburt:2015mya, Fields:2019pfx, Aghanim:2018eyx}.\footnote{see Eqs.\ 68-69 and 81 of
ref.\ \cite{Aghanim:2018eyx} or Fig.\ 11 and Table 5 of ref.\  \cite{Fields:2019pfx}.}   
In any case, these  would robustly exclude the $1.75$ $(3.5)$ degrees of freedom contributed by a fully thermalized Majorana (Dirac) fermion.  This does not occur in our setup because the two sectors are assumed to remain secluded.

If we allow for an initial discrepancy between the Standard Model and dark sector temperatures, $T_{d,0} = \xi_0 T_{\gamma,0}$, the predicted contribution to the effective number of relativistic species at any temperature within our framework is given by
\bea \nn
\Delta N_{\rm eff}(T_\gamma) &=& \frac47\, \xi_0^4\,  g_*^d \bL \pL\frac{11}4 \pR^{\Theta(m_e-T_\gamma)} \frac{g_{*S0}^d}{g_{*S}^d}  \frac{g_{*S}}{g_{*S0}} \bR^{4/3} \\
& =& 0.43\,\xi_0^4 \pL \frac{g_*^d}{7/2} \pR^{-1/3} \pL \frac{g_{*S0}^d}{11} \frac{106.75}{g_{*S0}} \pR^{4/3}
\eea
where $g_*^d$, $g_{*S}^d$ and $g_{*S},$ are the number of degrees of freedom in energy and entropy in the dark sector and in entropy in the visible sector, respectively, and all of them are evaluated at $T_\gamma$. In going from the first to the second line, the Standard Model degrees of freedom in entropy before and after $e^+e^-$ freezeout cancel the change in neutrino temperature, as expected. We normalize to the values appropriate for a dark sector containing one light and one heavy Dirac fermion, a vector, and a complex scalar (required to give the vector a mass), reflecting the field content of Model 1. In the case of a dark sector with two heavy and one light Majorana fermion plus a complex scalar, as in the minimal case to generate the phenomenology of Model 2, one obtains the smaller result $\Delta N_{\rm eff} \simeq 0.31\, \xi_0^4$\,. 

In either case, the CMB and BBN limits in single-parameter extensions of $\Lambda$CDM are in tension with the models if $\xi_0=1$, but  the most stringent BBN and CMB limits are satisfied for $\xi_0 \simeq 0.9$, which requires only a moderate difference in inflationary reheating efficiencies for the two sectors \cite{Adshead:2016xxj,Adshead:2019uwj}. Even if $\xi_0=1$, either model is compatible with CMB and BBN limits once uncertainties in $Y_p$ or other quantities are more conservatively taken into account \citep{Cyburt:2015mya, Fields:2019pfx, Aghanim:2018eyx}. 
In the near future, high-resolution studies of the CMB damping tail will improve these bounds by an order of magnitude \cite{Green:2019glg}.

\section{Structure formation}
\label{sec:struc}

\begin{figure*}[ht]
  \centerline{
  \includegraphics[width=0.5\textwidth]{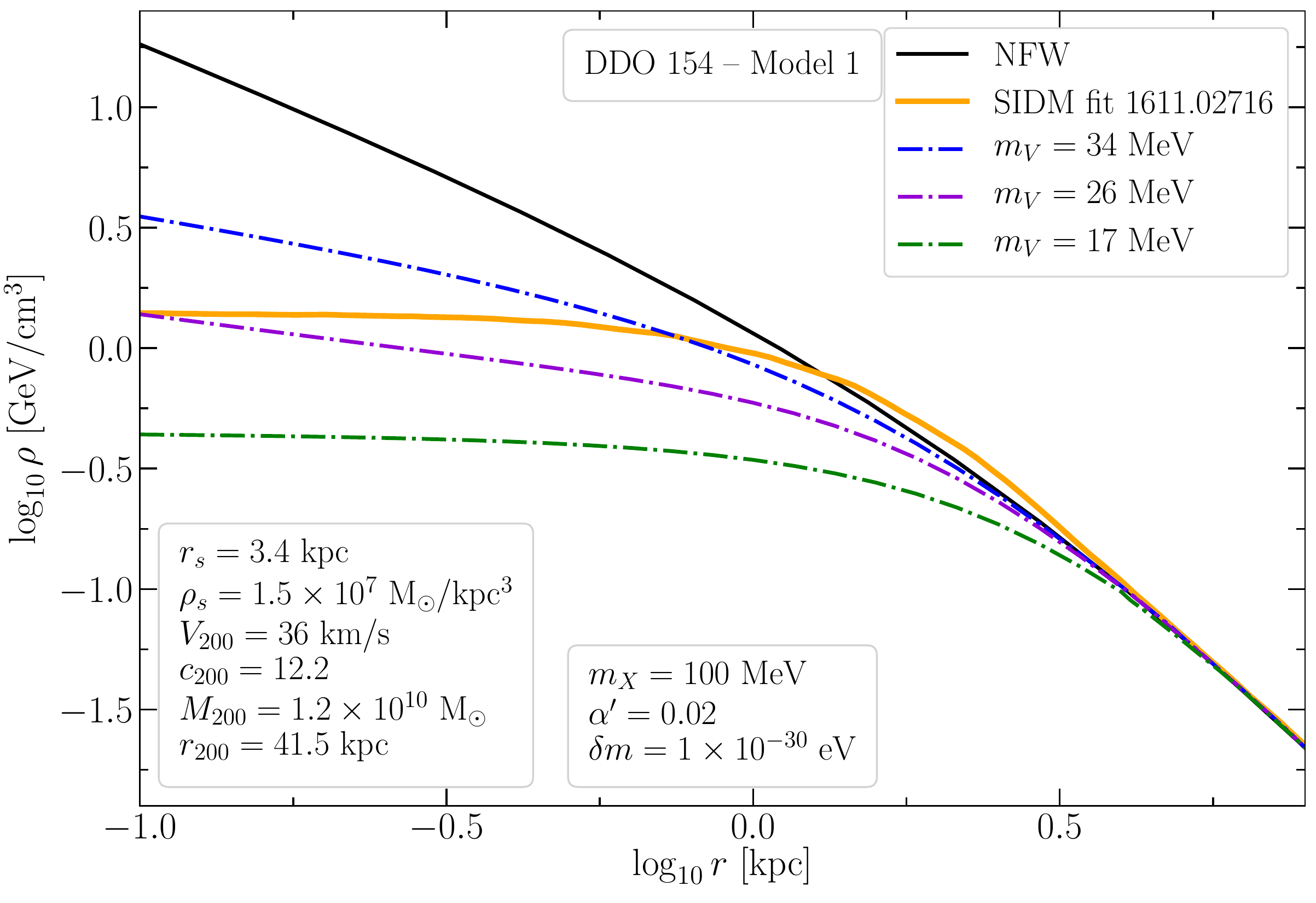}
   \includegraphics[width=0.5\textwidth]{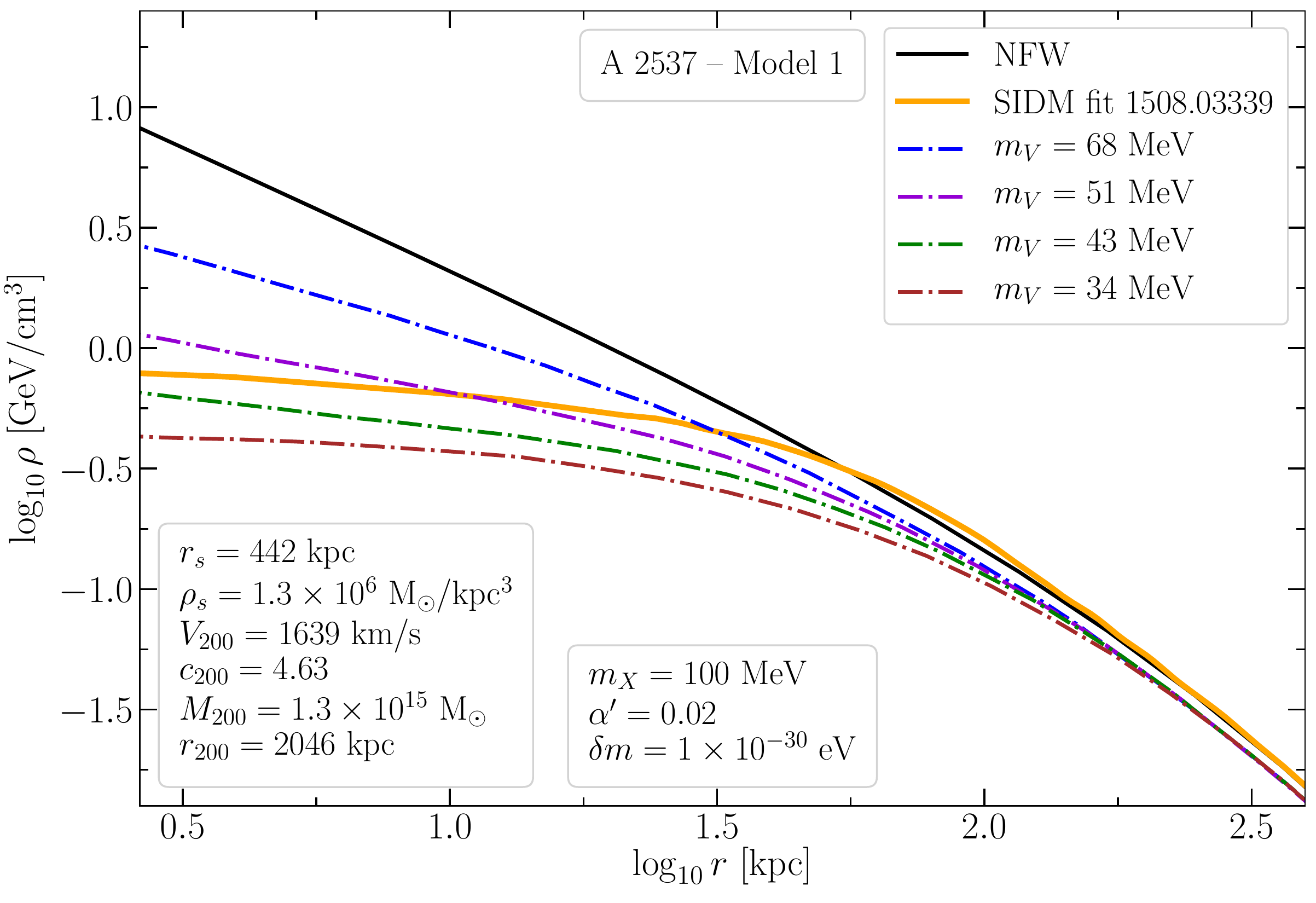}
   }
    \centerline{
  \includegraphics[width=0.5\textwidth]{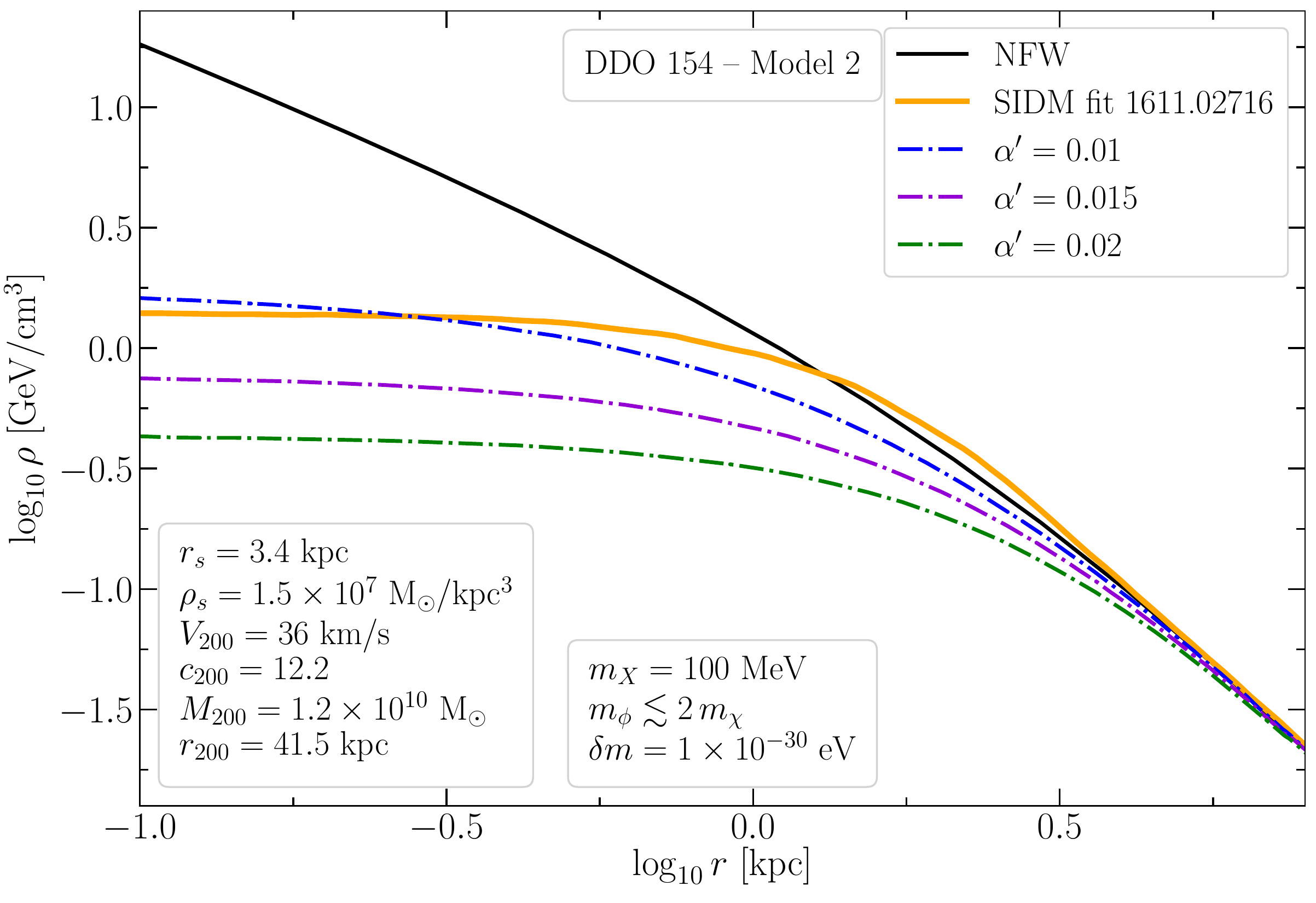}
   \includegraphics[width=0.5\textwidth]{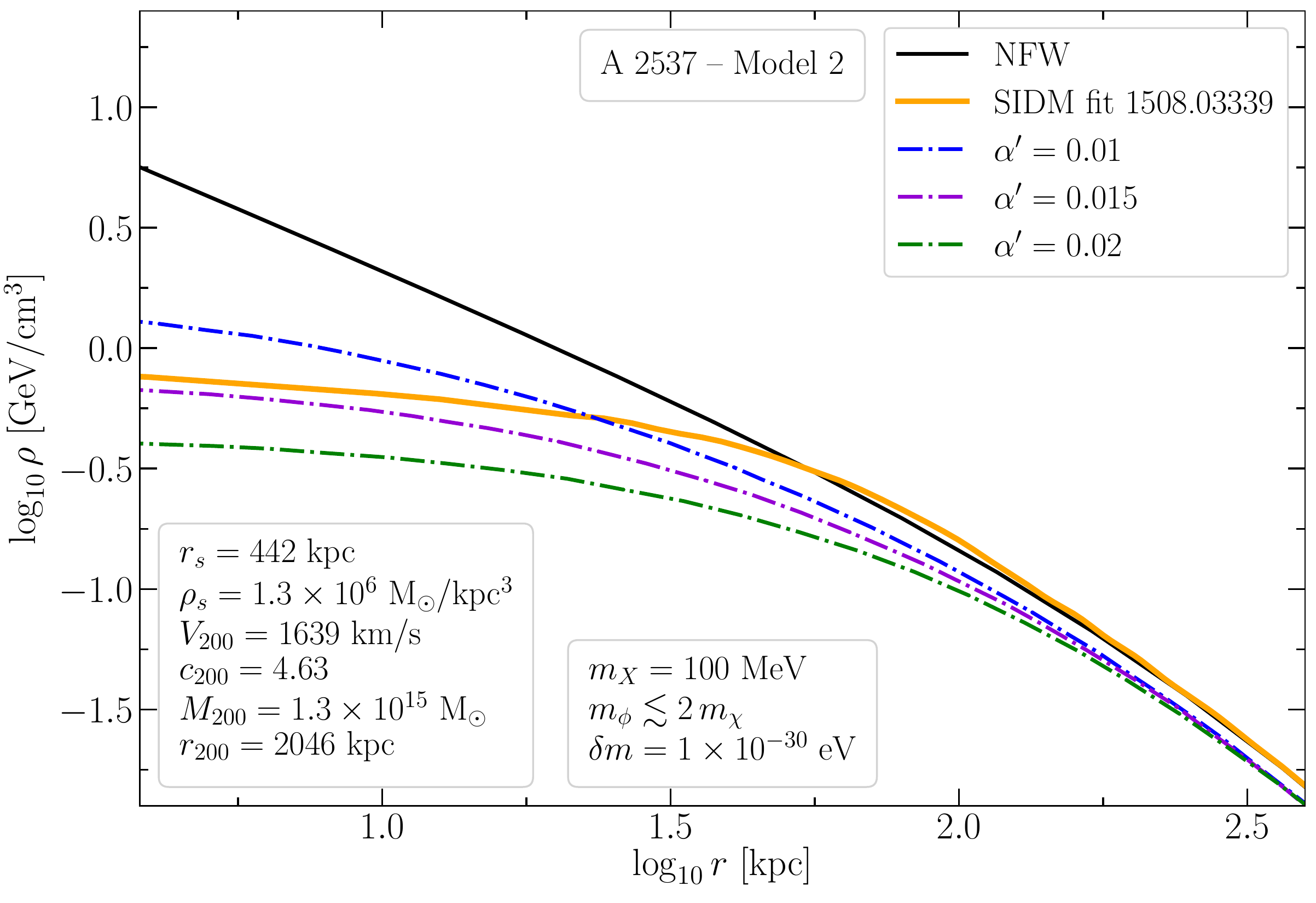}
   }
  \caption{Left: density profiles for dwarf galaxy DDO 154.  NFW and modified profiles from SIDM are from ref.\ \cite{Kamada:2016euw} (solid curves), while dot-dashed curves are the predictions of Model 1 (Model 2) for different indicated values of the vector mediator mass $m_V$ (dark fine-structure constant $\alpha'$). Right: corresponding results for galaxy cluster A2537,
  where SIDM result is from ref.\ \cite{Kaplinghat:2015aga}.  Top row is for Model 1 (vector), bottom for Model 2 (scalar).}
\label{model-res}
\end{figure*}

We start with an approximate treatment of the effect of $\chi$-$\bar\chi$ oscillations on galactic
dynamics, by imagining that
an NFW-shaped halo with
\be
    \rho_{\chi,0} = {\rho_s\over (r/r_s)(1 + r/r_s)^2}
\ee
has already formed at some time
$t_0$, with the initial condition on the matrix density that 
\be
n_{ij}(r;t_0) = {\rho_{\chi,0}(r)\over m_\chi} \,\delta_{i1}\delta_{j1}
\label{ninit}
\ee
at each position $r$ in the collapsed system: this corresponds to a pure $\chi$ state, in which oscillations have not yet had any effect.  

To apply the Boltzmann equation (\ref{boltz1}) in a galactic environment that
has separated from the Hubble expansion, we drop the $3Hn$
term, and set $n_{\rm eq}=0$, since the annihilation products escape without further interactions.    
 For example with fiducial parameters
$\alpha'=0.02$, $m_\chi = 100\,$MeV, and central
densities $\rho_\chi\sim 1\,$GeV/cm$^3$,
the mean free path for $\phi\chi\to\phi\chi$ or $V\chi\to V\chi$ scattering is of order $(\alpha'^2 \rho_\chi/m_\chi^3)^{-1}\sim 10^{21}$\, kpc.  In principle,
the Boltzmann equation in an inhomogeneous environment could contain extra terms, coming from the Liouville operator
\be
  \hat L[{\cal F}] \equiv  \left({\partial\over\partial t} + {\vec k\over m_\chi}\cdot\vec\nabla + \vec F(r) \cdot \vec\nabla_k\right){\cal F}(t,\vec x,\vec k)
    \label{liouv}
\ee
acting on the density matrix
${\cal F}$, where $\vec F(r)$ is the gravitational force  at a given radius in the halo. However in the approximation used in this section, we are assuming as an initial condition
an already-formed NFW halo in 
which the velocity distribution
is isotropic.  Therefore in the
integral of  eq.\ (\ref{liouv})
over $d^{\,3}k$ to convert ${\cal F}\to n$, all terms average to zero except for $\dot n$.  Hence the diffusion of dark matter particles that is modeled in $N$-body simulations is not captured in the Boltzmann
equation (\ref{boltz1}), although the quantum coherence effects are.  We supplement this analysis by a
complementary $N$-body approach
in section \ref{sec:sim}, which will 
corroborate the qualitative features found here.\footnote{To model effects of anisotropic velocity distribution, one could for example
assume that ${\cal F}$  factorizes into spatial and $\vec k$-dependent functions, ${\cal F} =
n(r) f(k_r, k_t)$, where $k_r$ and $k_t$ are the radial and tangential momentum components,
and take an additional moment $\int d^{\,3}k\, k_r$
of the Boltzmann equation to obtain coupled equations for $n$ and $f$.   We have checked that
$f$ is in fact isotropic in the $N$-body simulations described below; hence we do not pursue such a more detailed investigation in the present work.}

We evolve the initial density (\ref{ninit}) at each radial position $r$, up to a final time $t_f$ of order 10\,Gyr. This leads to a modified
density profile $\rho_\chi = (n_{11}+n_{22})\,m_\chi$, that is to be compared to present-day observations.
In addition to knowing the initial density profile, it is also necessary (for Model 1 only) to specify the DM velocity profile, since the
relative velocity enters into the scattering rate through $\langle\sigma v\rangle_s$ (whereas the annihilation rate is
insensitive to $v_{\rm rel}$).  We have adopted the analytic
solution for the radial velocity dispersion $\sigma_r(r)$
from ref.\ \cite{Lokas:2000mu} (see eq.\ (14) of that reference), derived by solving the Jeans equation for an NFW profile.  This determines $\sigma_r(r)$ for given NFW parameters $r_s$, $\rho_s$.  The latter can be related to the virial radius
$r_{200}$, concentration $c_{200}$, mass
$M_{200}$ and
velocity $V_{200}$ through
\bea
\label{eq:NFWparams}
{\rho_s\over \rho_c} &=& {200\over 3}\,c_{200}^3\, g(c_{200})\,,\nn\\
r_{200} &=& c_{200}\, r_s\,,\nn\\
M_{200} &=& {4\pi\over 3}\, 200\, r_{200}^3\, \rho_c\,,\nn\\
V_{200}^2 &=& {G M_{200}\over r_{200}}\,,
\eea
where $\rho_c$ is the present critical density,
and $g(c) = [\ln(1+c) -c/(1+c)]^{-1}$.

\begin{figure*}[t]
  \centerline{
  \includegraphics[width=0.4925\textwidth]{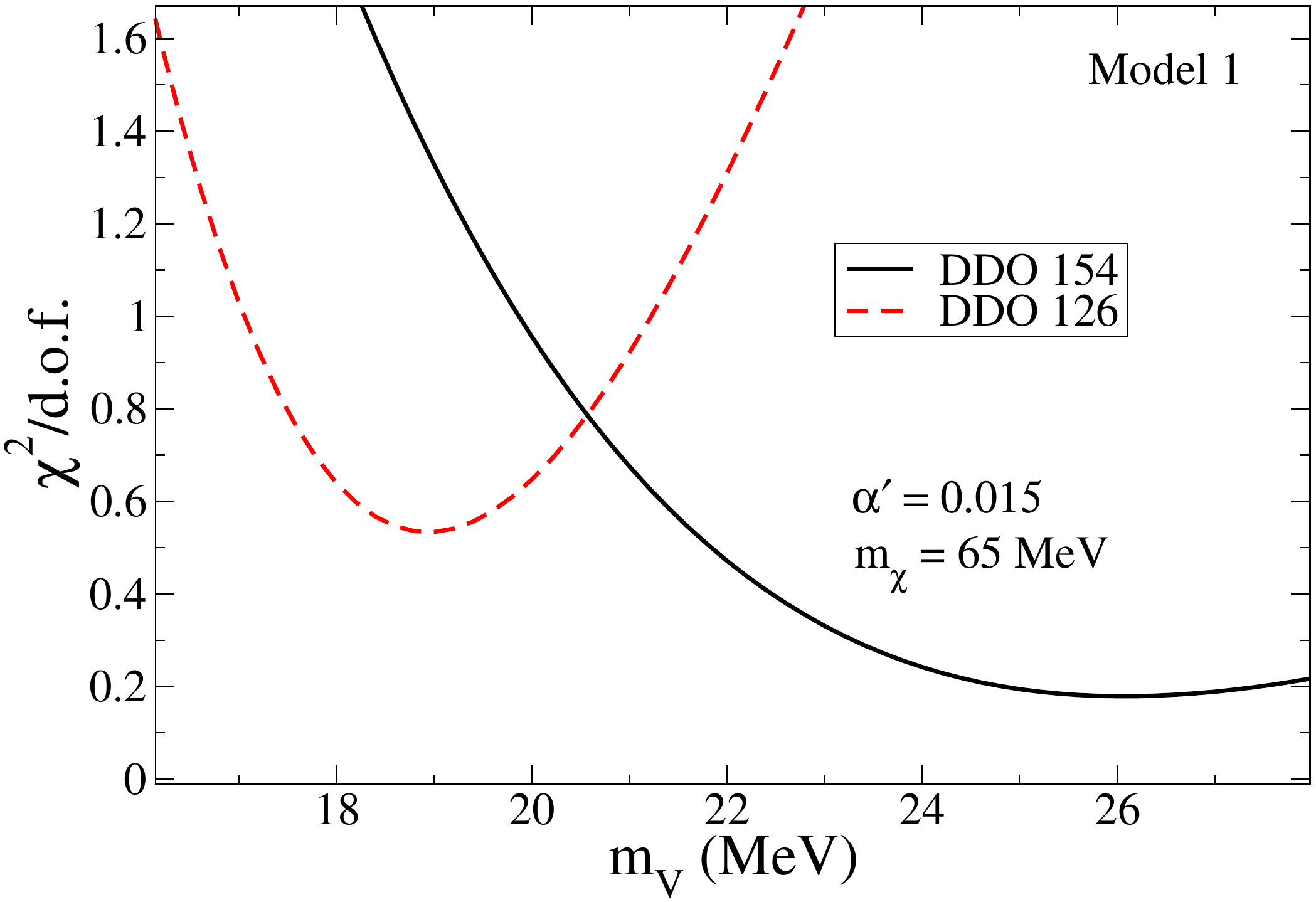}
   \includegraphics[width=0.5\textwidth]{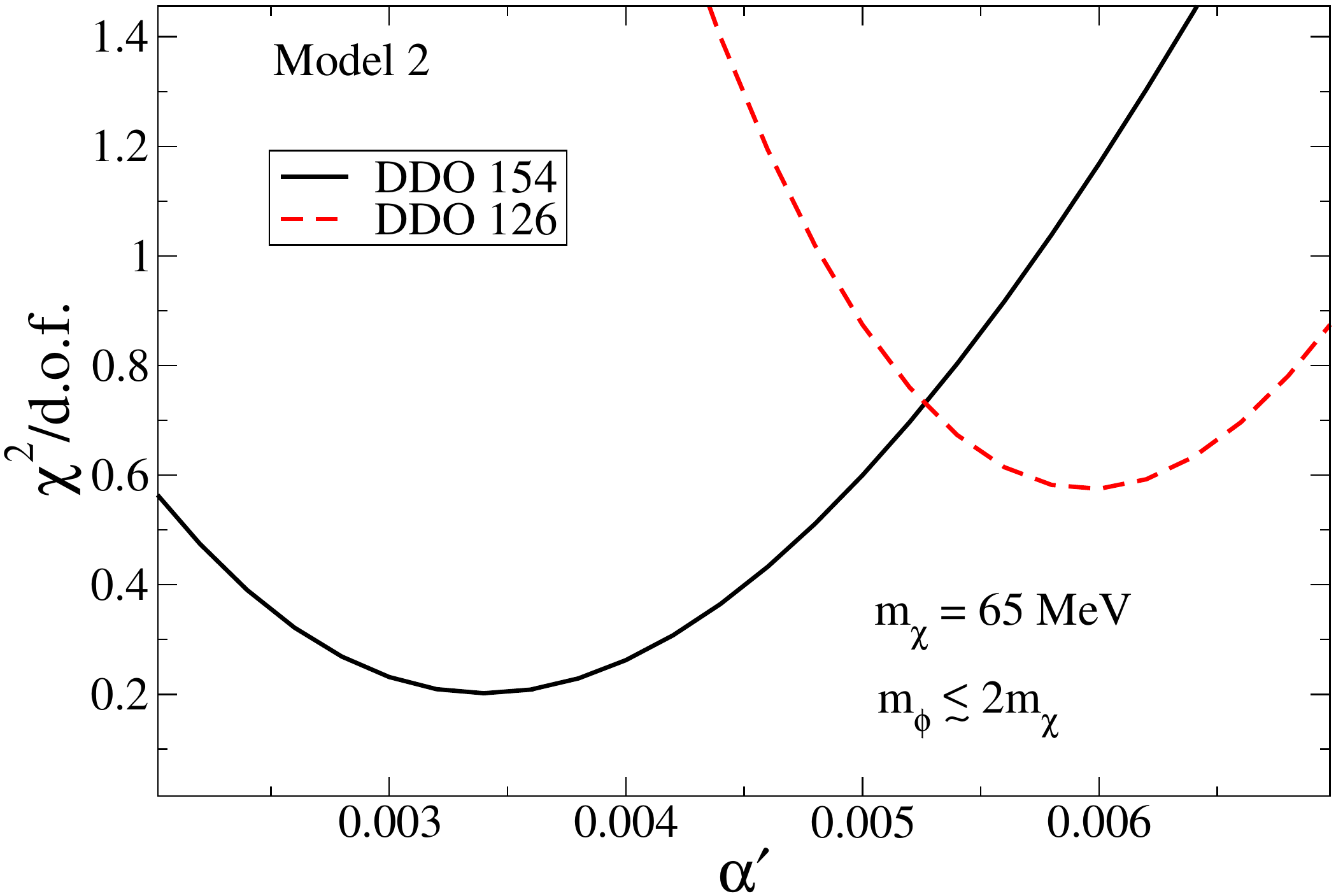}
   }
\caption{Left: $\chi^2$ per degree of freedom versus the vector mediator mass $m_V$ in Model 1, for fits to the circular velocities of dwarf spheroidals DDO 154 and 126,
with DM mass $m_\chi = 65$\,MeV.
Right: similar to left, for Model 2 with varying $\alpha'$.  In either model, acceptable joint fits can be found by taking intermediate values of $m_V$ or $\alpha'$, respectively.}
\label{chi2dwarf}
\end{figure*}

\subsection{Model 1}

We applied this procedure first within Model 1 for a particular
dwarf spheroidal galaxy, DDO 154 \cite{1988ApJ...332L..33C}, that has been
discussed from the point of view of self-interacting dark matter  in ref.\ \cite{Kamada:2016euw}.  There the
NFW parameters were determined using data from 
ref.\ \cite{Oh:2015xoa} and the mass-concentration
relation from ref.\ \cite{Dutton:2014xda}.
The resulting NFW profile is shown in Fig.\ \ref{model-res} (top left).  This profile disagrees with the observed rotation curve in the inner part of the galaxy, whereas the
solid ``SIDM'' curve, which arises from elastic DM self-interactions adjusted to the appropriate cross section, gives a good fit.

The dot-dashed curves show the results for our
model, with $m_\chi = 100\,$MeV, $\alpha'=0.02$, and several
values of $m_V$.  The profile is significantly cored, depending on the value of $m_V$, and has a different shape from that predicted by SIDM.
The closest match between the SIDM profile and ours is produced for $m_V = 34\,$MeV, which however is inconsistent with the constraint (\ref{eq:consistent}).
It means that we should not neglect the effects of elastic scattering by itself, which go into the usual SIDM treatment.
This problem can be overcome by simultaneously increasing
$m_V$ and $\alpha'$; for example $m_V = 60\,$MeV and $\alpha' = 0.1$ gives a reasonable fit.  However neither of these models are consistent with data from galaxy clusters, as
we discuss next.

Ref.\ \cite{Kaplinghat:2015aga} presents evidence for the DM profiles of galaxy clusters also being cored, to a somewhat lesser extent than dwarf spheroidal galaxies.  These larger
systems have much higher velocity dispersions, which leads to a stronger reduction of the central density by our mechanism,
using Model 1.  This is shown for the cluster A2537 in Fig.\ \ref{model-res} (top right), where for the same values $m_\chi = 100\,$MeV and $\alpha' = 0.02$ as before, the best match to the SIDM curve is for $m_V \cong 51\,$MeV; the lower value of $m_V=34\,$MeV, favored by dwarf spheroidals, leads to unacceptably large suppression
of the central density to be compatible with measured stellar velocity profiles.
More detailed quantitative comparisons between the theory and data will be presented in section \ref{sec:sim}, in terms of the predicted versus observed velocity profiles.

One may also wonder to what extent a given model can match the observed properties of different spheroidal dwarf galaxies, whose density profiles can be diverse.  Although an exhaustive comparison is beyond the scope of the present work, we have studied a contrasting example, DDO 126, whose DM density profile (like that of DDO 154) was estimated by ref.\ \cite{Oh:2015xoa}.  The best fits to the circular
velocity measurements for the two galaxies occur at different values of the model parameters, as shown in the left panel of Fig.~\ref{chi2dwarf}, where we fixed $m_\chi = 65\,$MeV, $\alpha'=0.015$ and allowed $m_V$ to vary.  (Notice we have chosen a lower value of $m_\chi$ in this example; it is motivated by the discussion in section \ref{hybrid}.)  However, an acceptable fit to both systems can be found at an intermediate value $m_V \cong 20.6$\,MeV, resulting in $\chi^2$/d.o.f. $\cong 0.8$ for
either system.  We have allowed for systematic uncertainty in the magnitude of the DM density profiles, reflecting an estimated $\sim 25\%$ uncertainty in the baryonic content of the galaxies \cite{Kaplinghat:2015aga}.  Since
the baryons comprise $\sim 10\%$ of these systems, this
translates to a 2.5\% uncertainty in the overall DM densities, that we have marginalized over to slightly improve the fits.

\begin{figure*}[t]
  \centerline{
  \includegraphics[width=0.5\textwidth]{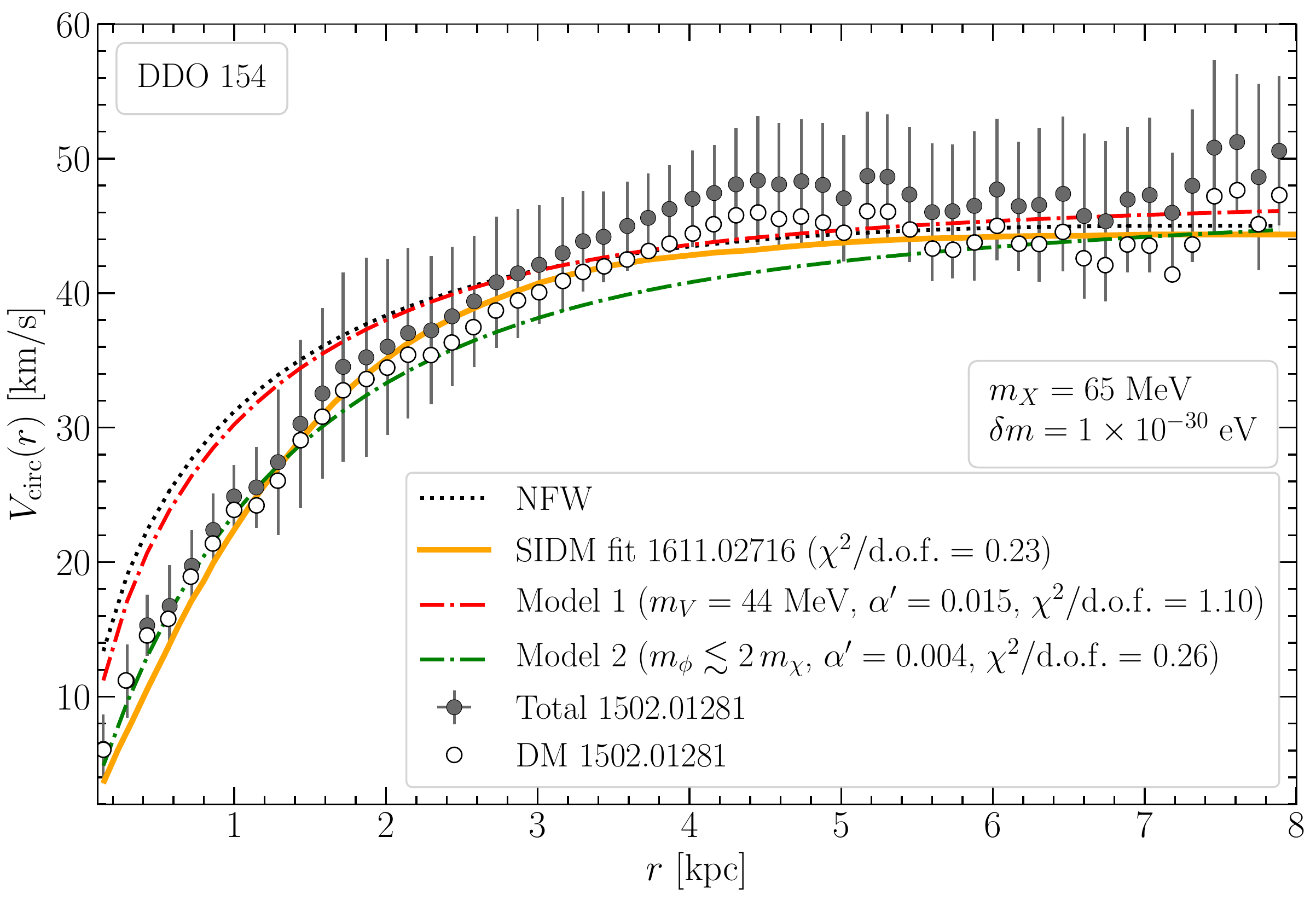}
   \includegraphics[width=0.4925\textwidth]{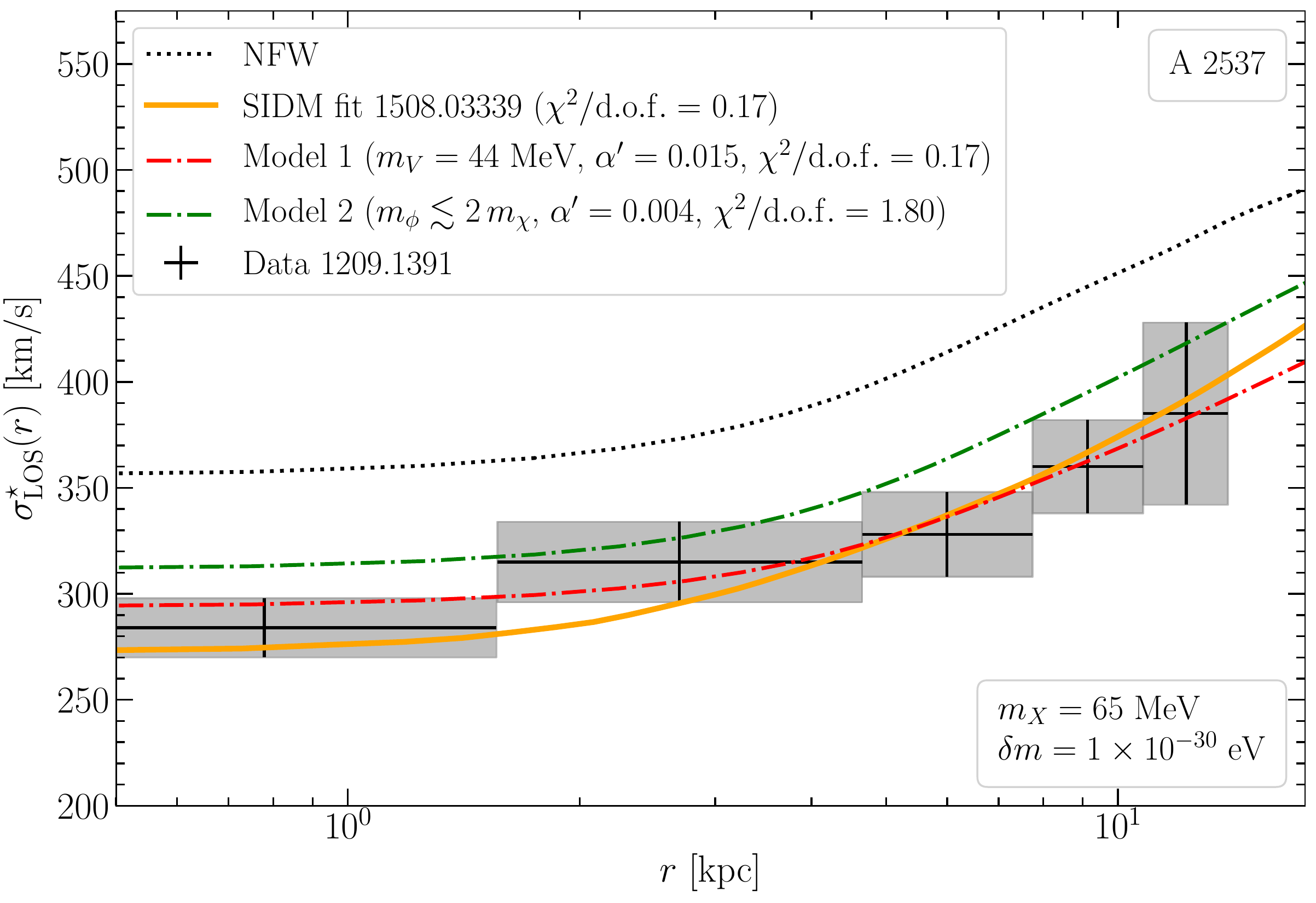}
   }
  \caption{Illustration of how combining vector and scalar mediators could give a good simultaneous fit for both dwarf spheroidals (left) and clusters (right). Left: predicted circular velocities due to the DM component alone from the same two models and from SIDM (ref. \cite{Kamada:2016euw}), and data from ref.\ \cite{Oh:2015xoa}.
  In each case, one mediator dominates the coring effect of the central profile in one system, while having little effect in the other system.  Right: stellar velocity dispersion along the line-of-sight for cluster A2537, with predictions based on the DM density profile from two of our models, from SIDM (ref.\ \cite{Kaplinghat:2015aga}) and data from ref. \cite{Newman:2012nv}.}
\label{fig:vcomp}
\end{figure*}

\subsection{Model 2}

In Model 2, the situation is the opposite, though with a 
smaller discrepancy.  In this case nothing depends on the scalar mass $m_\phi$, as long as it satisfies the consistency
condition (\ref{eq:consistent}).  For a fixed value of $m_\chi$ (here still at 100\,MeV), only $\alpha'$ matters.
The SIDM profile can be approximately matched by taking $\alpha'\cong 0.01$ in the DDO 154 dwarf galaxy, while for the same parameter choices, the predicted inner profile of cluster A2537 lies
somewhat above the SIDM fit, a factor of 1.7 higher as illustrated in the density profiles shown in Fig.\ \ref{model-res}.

We have found that this qualitative difference between scalar and vector mediators is generic: the velocity dependence of the decoherence mechanism in Model 1 makes it more effective for cusp suppression in high-velocity systems (clusters), whereas the 
lack of such dependence in Model 2 leads it to be more efficient in higher density systems (dwarfs).

The mild tension in simultaneously explaining the density profiles of different spheroidal dwarf
galaxies, described for Model 1, is also present in model 2, as illustrated in the right panel of Fig.~\ref{chi2dwarf} for the case of $m_\chi = 65$\,MeV: the best fits occur at different values of $\alpha'$ for the DDO 154 and DDO 126 galaxies.  Like for Model 1, it is not a serious difficulty since an intermediate choice $\alpha'\cong 0.0053$ results in an acceptable $\chi^2$/d.o.f. $= 0.72$ for both systems.
We leave a more exhaustive study, both of the allowed parameter space and including more galaxies, for future work.

\subsection{Hybrid models}
\label{hybrid}
The previous results suggest that the challenges for Models 1 or 2 to simultaneously
fit the rotation curves of both dwarf galaxies and clusters could be overcome in a model with both mediators present.  Here we present an example that supports this hypothesis, leaving for future work a more rigorous or detailed analysis.  

Since it is technically difficult to implement both kinds of mediators simultaneously, we will be content here to give an example in which a vector mediator
gives a good fit to a cluster, while leaving a dwarf galaxy relatively unaffected, and at the same time
a scalar mediator that achieves the opposite.  Since
each model has a relatively small effect on one of the systems, it seems likely that by combining them, one can add the coring effects to both systems in a roughly linear fashion.

For example we find that for lighter DM with $m_\chi = 65\,$MeV,
and Model 1 parameters $\alpha'=0.015$, $m_V = 44\,$MeV, we fit the observed stellar line-of-sight velocity dispersion profile (described in more detail in the next section) for
A2537 extremely well, while leaving the predicted circular velocity $V_{\rm circ} (r)$ in DDO 154 too high 
from not sufficiently reducing the central density in the dwarf system.    On the other hand, choosing a lower coupling $\alpha' = 0.004$ in Model 2 gives an excellent fit to the DDO 154 rotation curve, 
while having a small impact on the inner profile of A2537.
These outcomes are shown in Fig.\ \ref{fig:vcomp}, indicating that by combining the two mediators, it is possible to get as good a fit as an elastic SIDM model with a velocity-dependent cross section that is tuned to fit both systems.  In elastic SIDM, a cross section of 
$\sigma/m \cong 3$\,cm$^2$/g \cite{Kamada:2016euw} is needed to agree with dwarf spheroidals, whereas a smaller value $\sim 0.1$\,cm$^2$/g is used to explain clusters \cite{Kaplinghat:2015aga}.

\begin{figure*}[ht]
  \centerline{
  \includegraphics[width=0.5\textwidth]{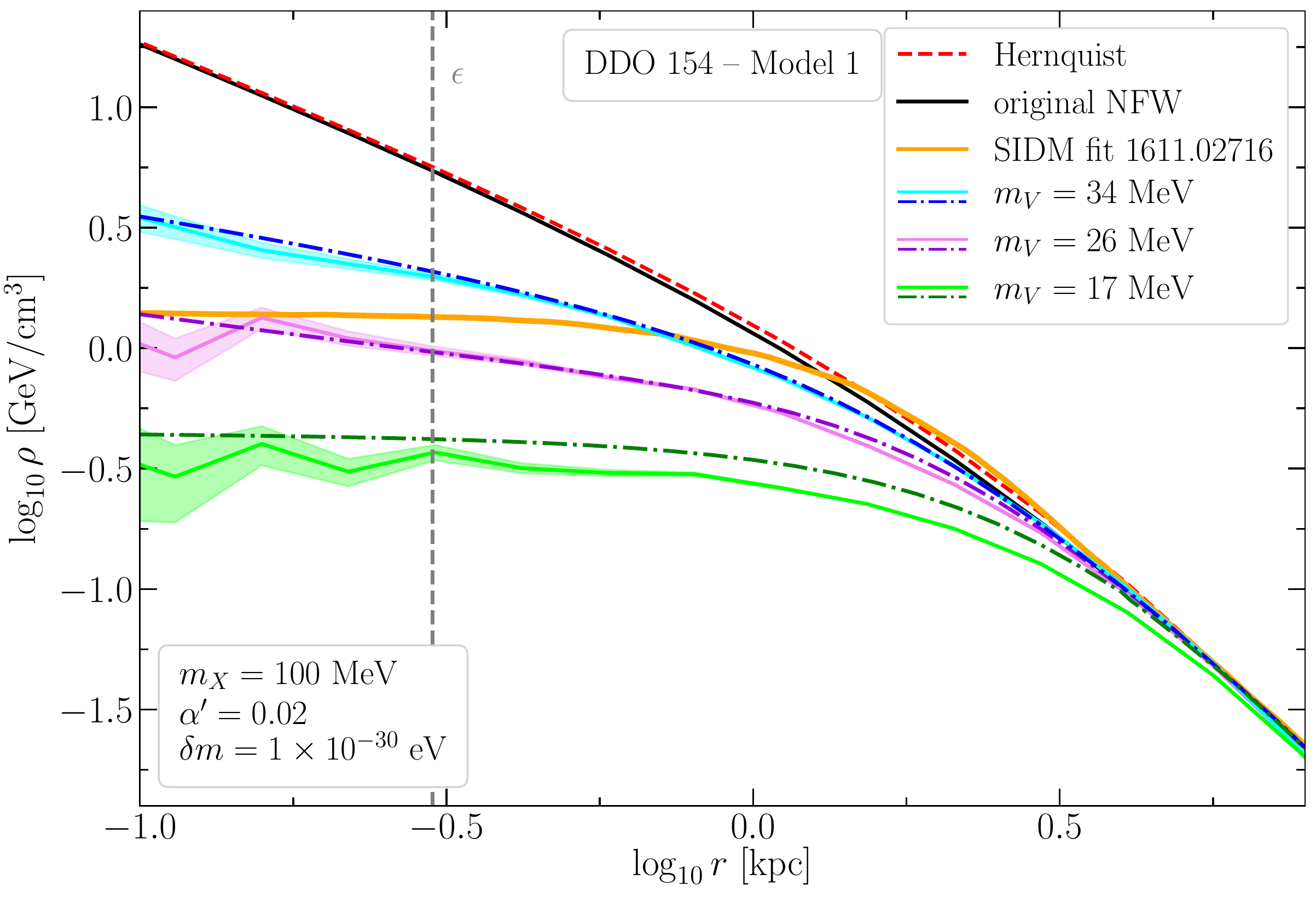}
   \includegraphics[width=0.5\textwidth]{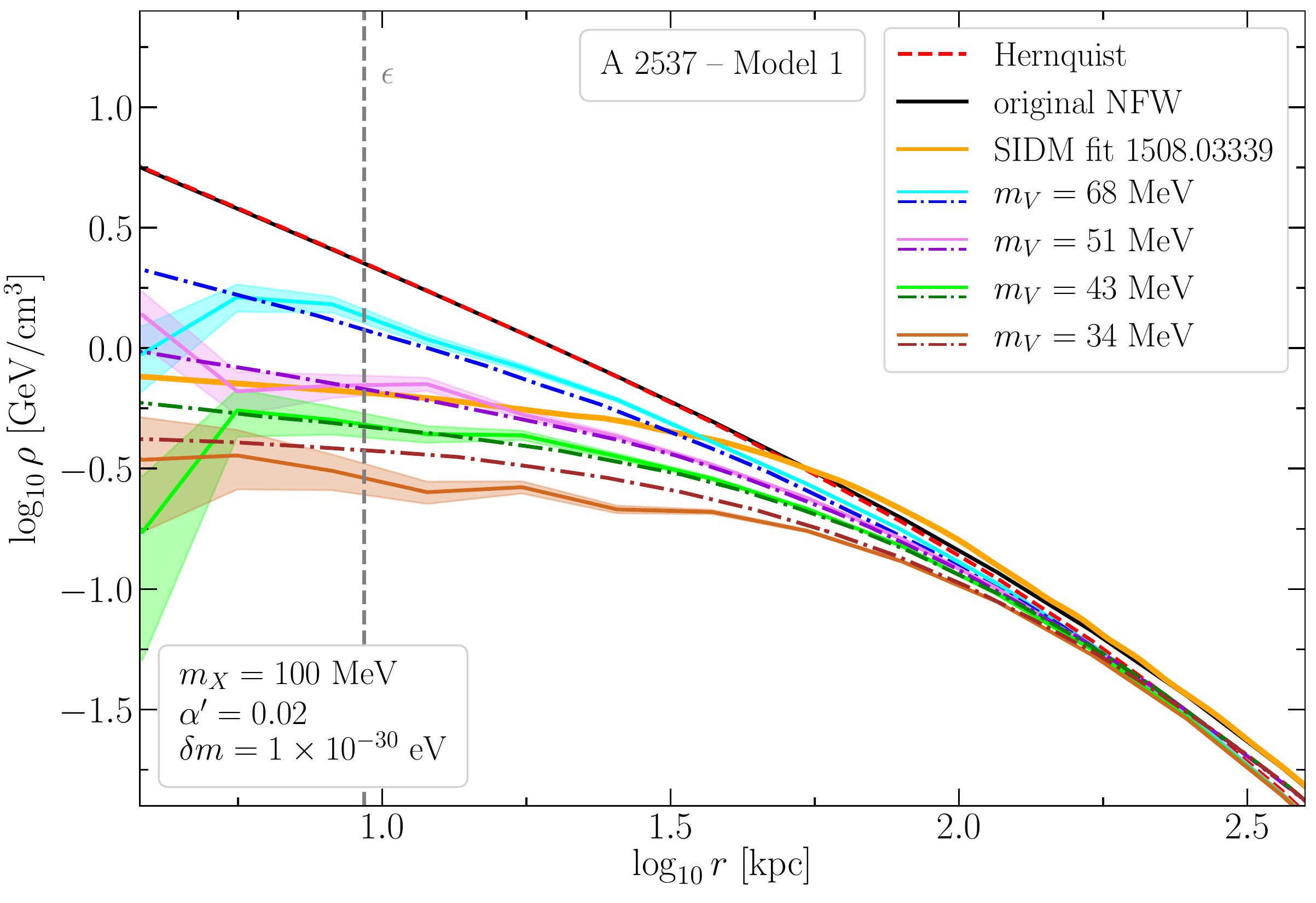}}
    \centerline{
  \includegraphics[width=0.5\textwidth]{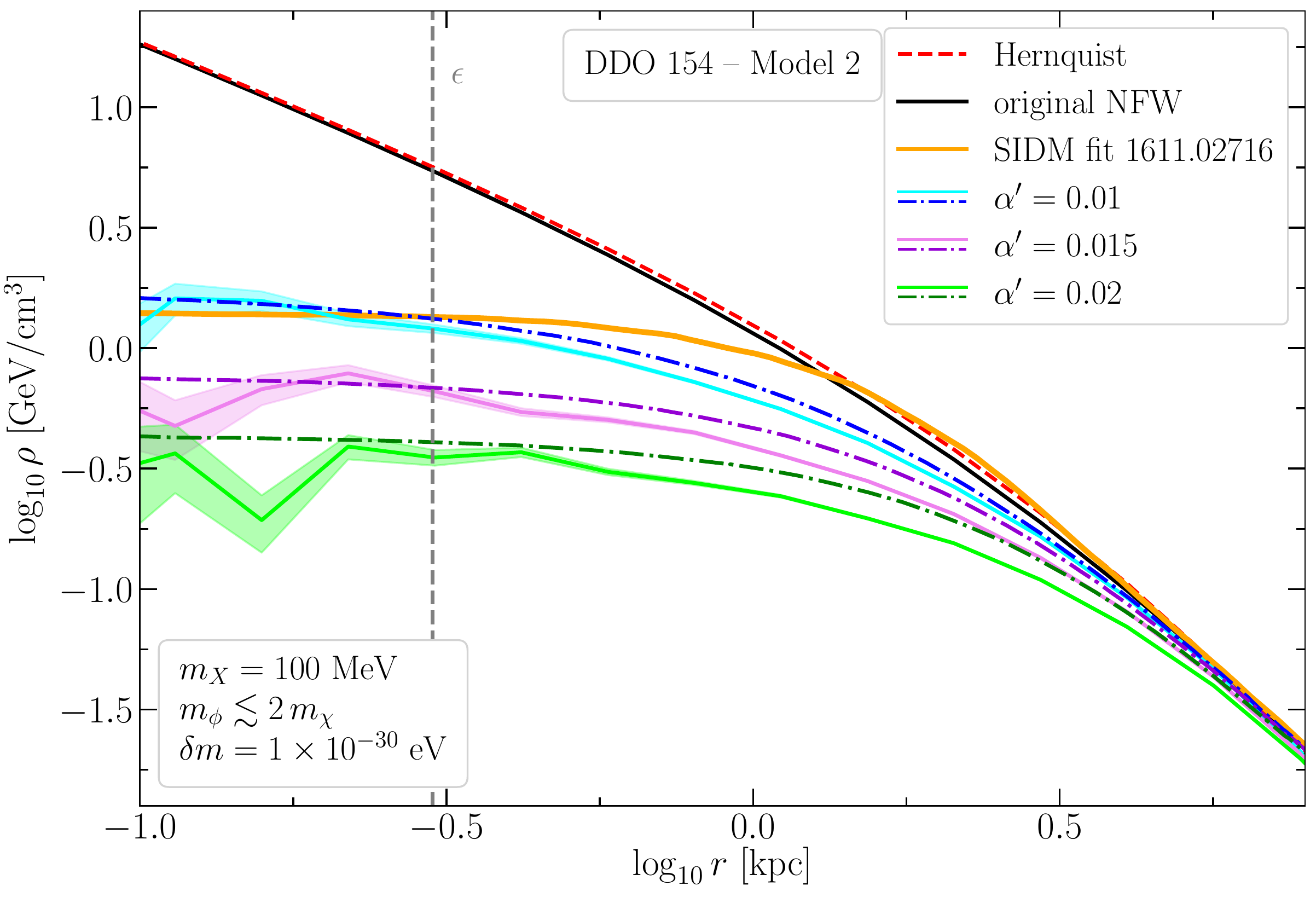}
   \includegraphics[width=0.5\textwidth]{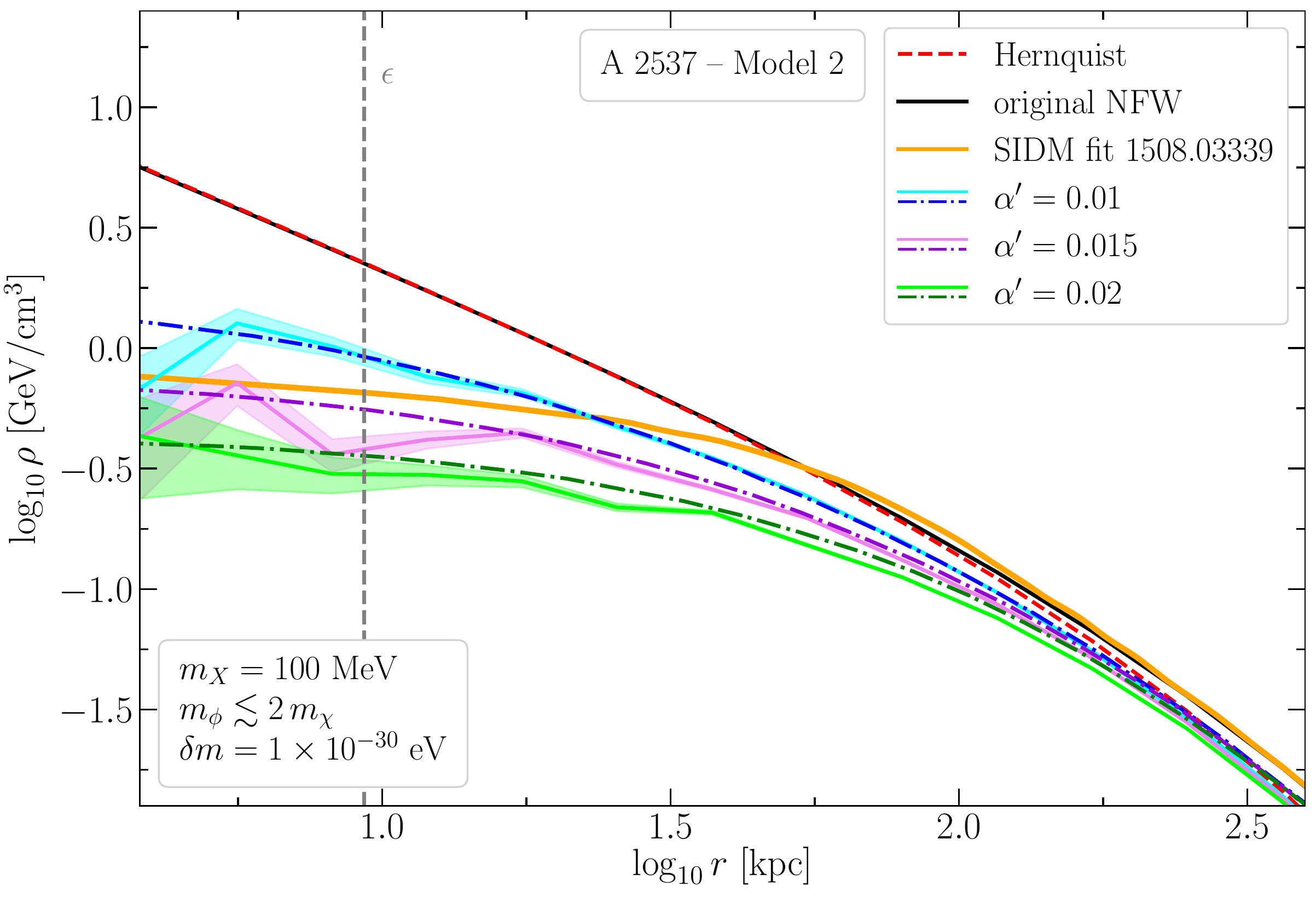}}
  \caption{Like Fig.\ \ref{model-res}, but including comparison with the N-body simulation results. The latter are shown as solid lines surrounded by the $1\sigma$ uncertainty band, obtained by assuming that the number of particles in each bin is Poisson-distributed. The black solid curve corresponds to the original NFW profile, whereas the matched Hernquist profile is shown with the red dashed line. The other dot-dashed curves are the results of Fig.\ \ref{model-res}. The orange solid line is the SIDM prediction from ref.~\cite{Kamada:2016euw} for DDO 154 and from ref.~\cite{Kaplinghat:2015aga} for A2537.
  The dashed vertical line shows the position of the gravitational softening length $\epsilon$ used in the simulations.}
\label{fig:sim_results}
\end{figure*}

\section{N-body simulations}
\label{sec:sim}
To more quantitatively predict the evolution of galactic structures in our scenario, we have performed N-body simulations that take into account the peculiar interactions described by the Boltzmann approach
of the previous sections.  The two approaches should be viewed as complementary since each has its own limitations.  The challenge for N-body simulations, even if modified to account for self-interactions, is that they treat test particles classically, with scatterings occurring probabilistically rather than quantum mechanically.  For conventional self-interactions this is not a serious limitation, but in the present context, the overall coherence of the DM ensemble is of primary importance.

To address this, we have modified the public version of the
\texttt{GADGET-2} code~\citep{Springel:2005mi,Springel:2000yr}, which
is widely used to generate N-body cosmological
simulations.\footnote{\url{https://wwwmpa.mpa-garching.mpg.de/gadget/}}\ \ The novel feature, apart from including DM scattering and annihilation (see appendix~\ref{app:Nbody} for implementation details and code tests), is to keep track of the phase $\varphi$ of each test 
particle, that describes the oscillations as in eq.\ (\ref{chi_osc}).
We assume that all particles in the halo are initially in phase with each other.  Depending on whether the model is flavor-sensitive (Model 1) or flavor-blind (Model 2), this phase plays different roles, and evolves differently.  In the absence of interactions, the phase
of each particle would evolve trivially as $\varphi = \delta m\, t$.  
To mock up the behavior predicted by the
quantum Boltzmann equations while still treating the particles
classically, we implement scattering as follows.

\begin{trivlist}
\item 
{\bf Model 1.}  Elastic scatterings damp the quantum coherence, as described by the off-diagonal elements of the collision term in
(\ref{boltz1}).  Integrating the off-diagonal elements over a collision time $\Delta t = 1/\Gamma_s = (n\langle\sigma v\rangle_s)^{-1}$ leads to a phase change $\Delta\ln(c_\varphi s_\varphi)
= -3/2$, as shown in eq.\ (\ref{vpdamp}).  For strongly damped systems such that $\Gamma_s > \delta m$,
this can be modeled by replacing the phase of each particle undergoing elastic scattering by
\be
\varphi \to (\varphi\,\, \text{mod}\,\, 2\pi)\, e^{-3/2}\,,
\ee
leading to decoherence of the ensemble, that allows annihilations to occur. 
The annihilation probability of two particles with respective phases $\varphi_1$ and $\varphi_2$ is reduced relative to its usual value by the factor $\sin^2(\varphi_1-\varphi_2)$, as derived in eq.\ (\ref{deteq}).

\item
{\bf Model 2.}  In this case, the scattering self-interactions have no effect on the phases, and they play exactly the same role as in conventional SIDM.  Instead, decoherence is caused by the annihilation interactions themselves.  The phase reduction described above now becomes a factor of $e^{-1}$ each time an annihilation {\it would have} occurred, for a fully decoherent mixture of $\chi$ and $\bar\chi$.  The annihilation probability is modulated by the different factor $\sin^2(\varphi_1+\varphi_2)$ as was explained below 
eq.\ (\ref{detpneq}).

\end{trivlist}

\begin{figure*}[ht]
  \centerline{
  \includegraphics[width=0.5\textwidth]{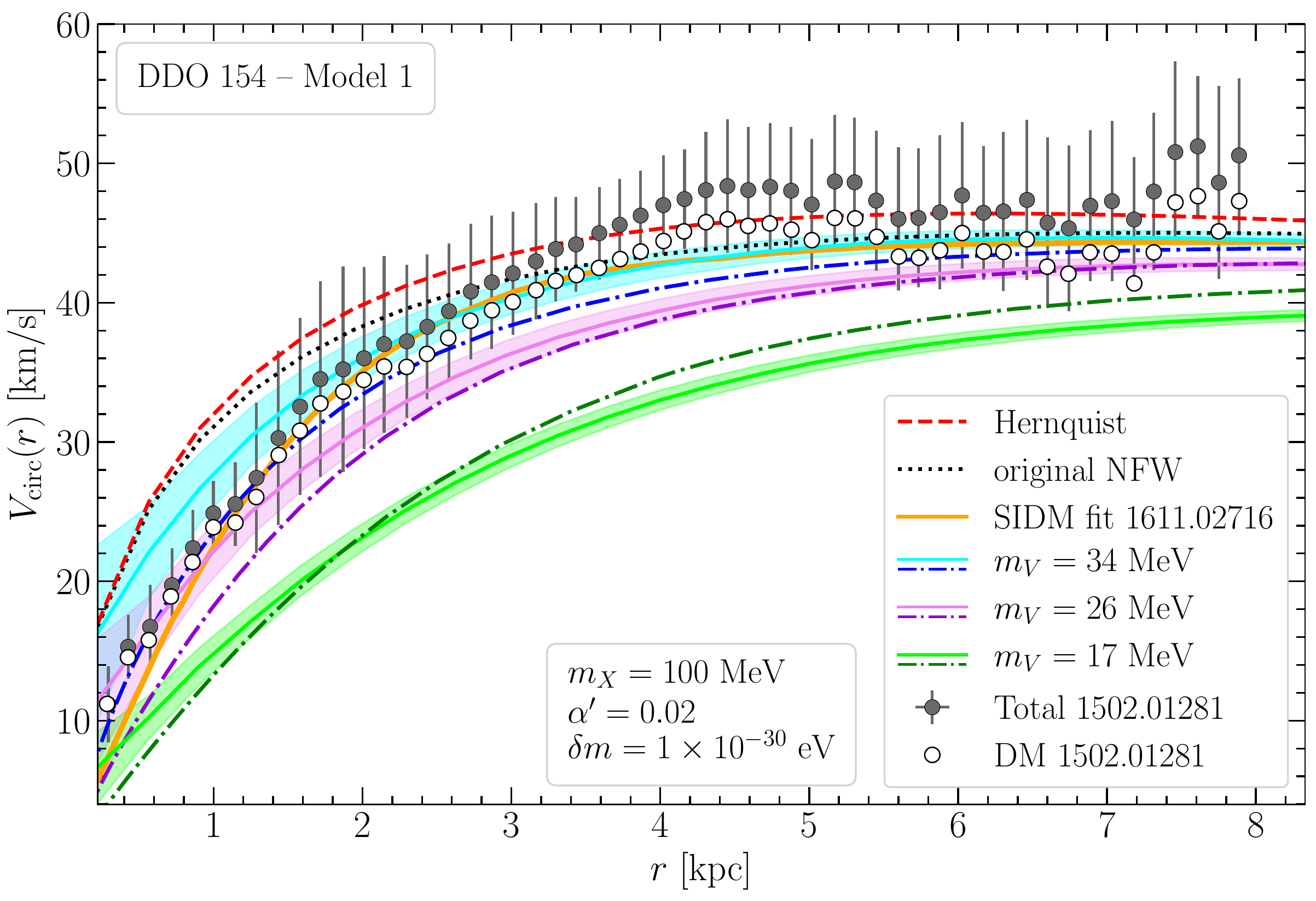}
   \includegraphics[width=0.4925\textwidth]{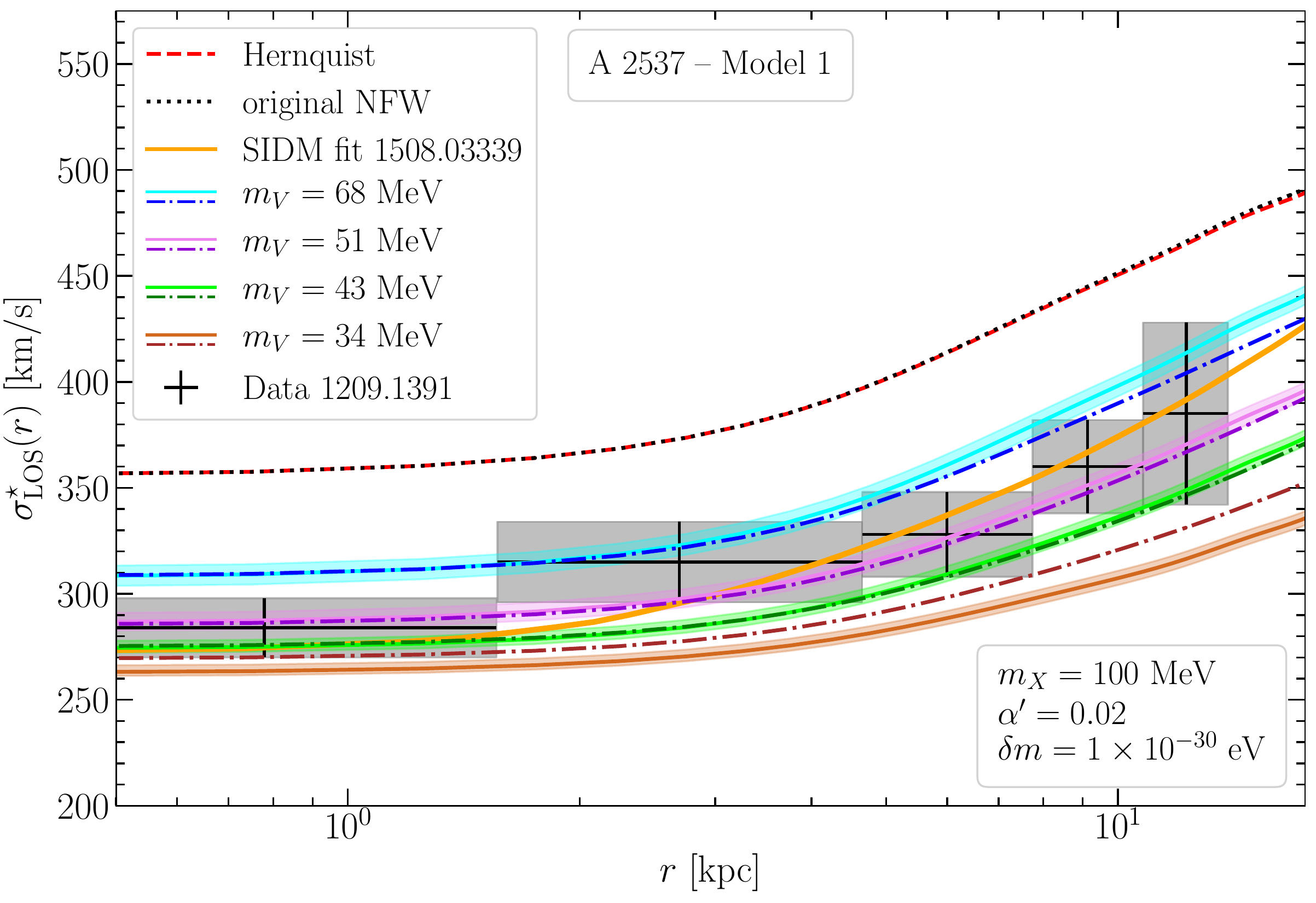}}
    \centerline{
  \includegraphics[width=0.5\textwidth]{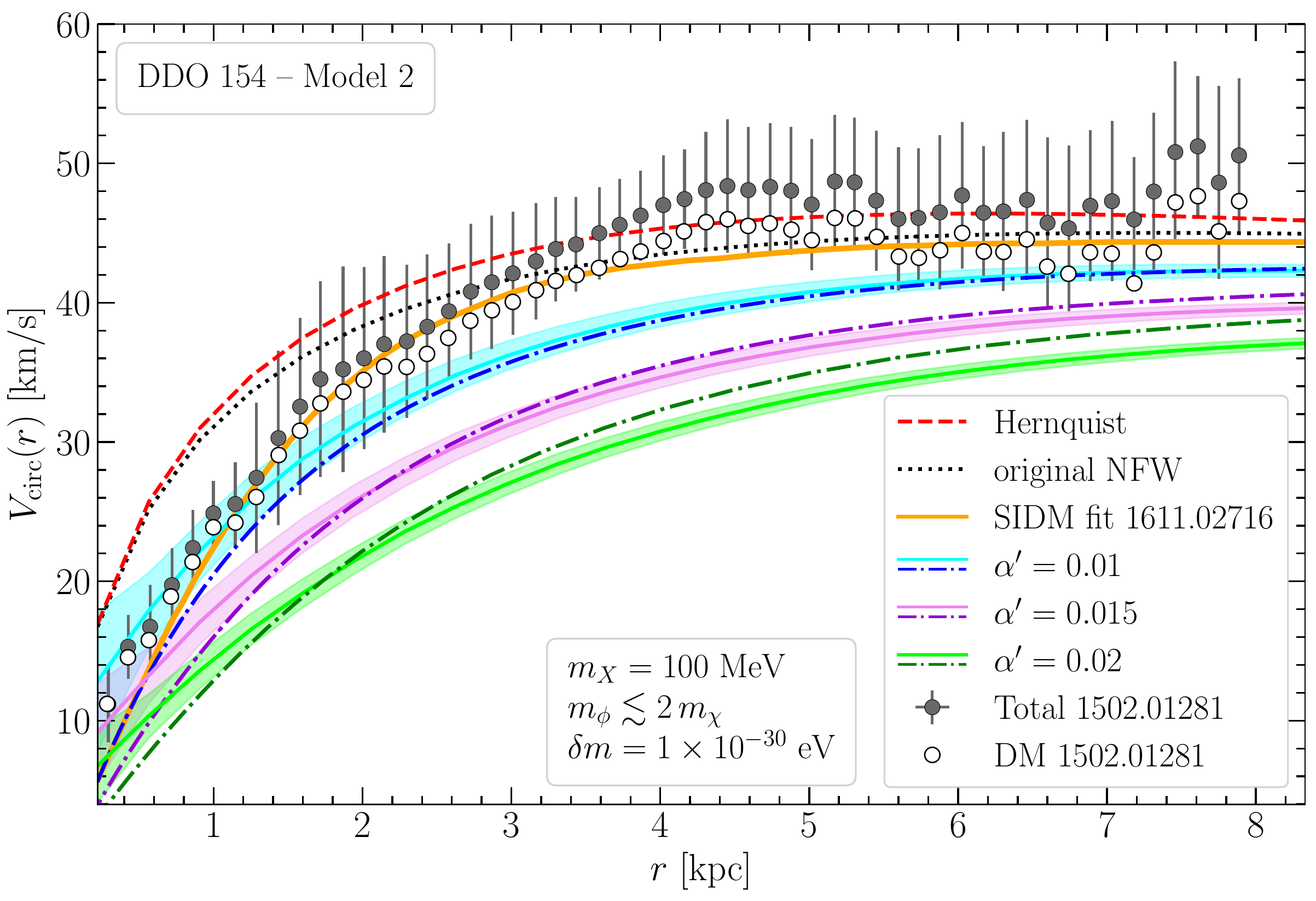}
   \includegraphics[width=0.4925\textwidth]{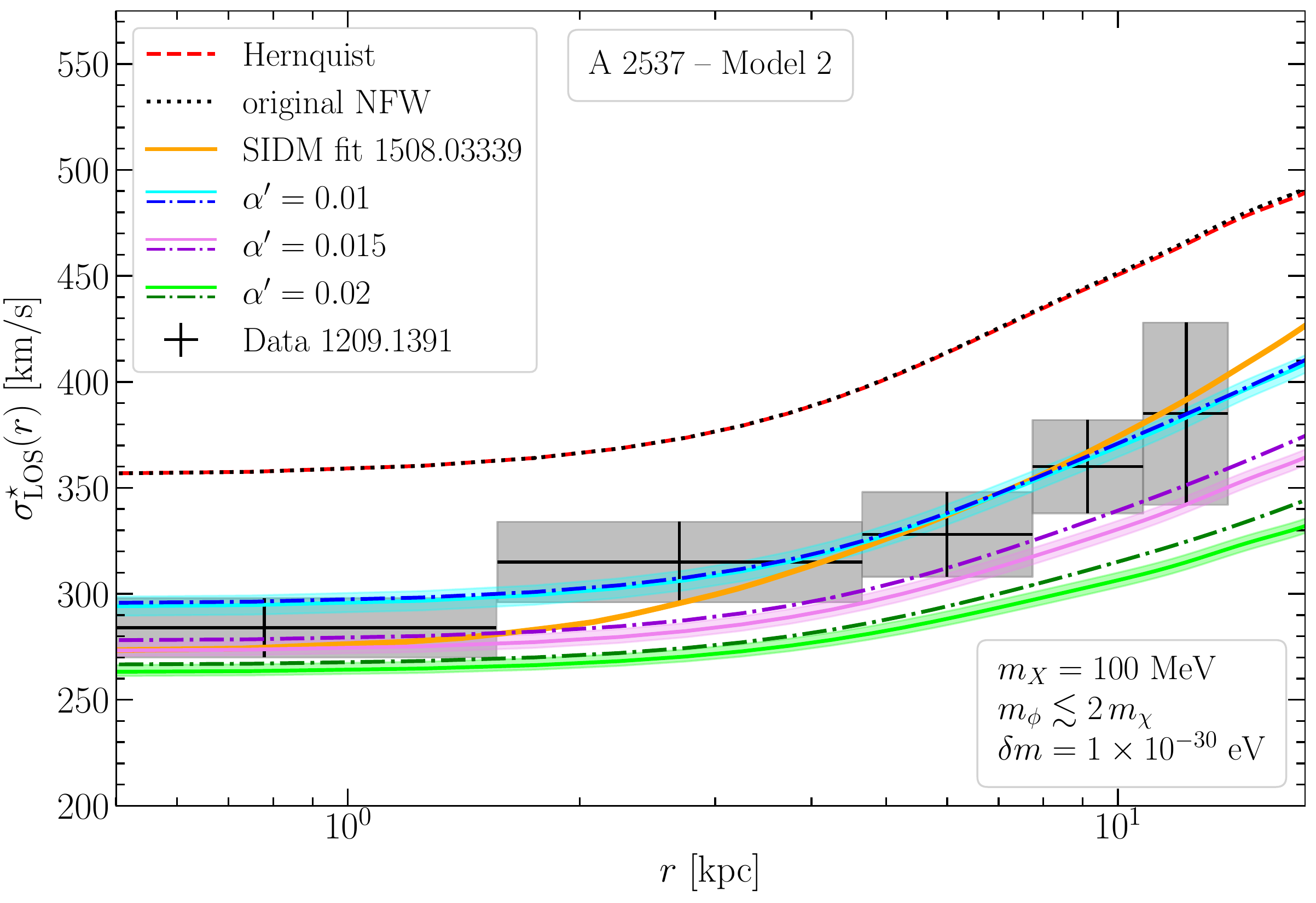}}
  \caption{Comparison between our model predictions and observational data. Left: circular velocity as a function of distance from the galactic center of the dwarf DDO 154. The data points and the corresponding error bars are taken from ref.~\cite{Oh:2015xoa}. In particular, the grey dots show the total effect of DM, gas and stars on the rotation curve, whereas the white dots show just the DM contribution obtained after a careful modelling of stars and gas components (see ref.~\cite{Oh:2015xoa} for details). Right: projected stellar velocity dispersion along the line-of-sight as a function of radial distance for the cluster A2537. The data points and the error bars are taken from ref.~\cite{Newman:2012nv}.
  In all panels, N-body simulation results are shown as solid lines surrounded by the $1\sigma$ uncertainty band, obtained by assuming that the number of particles in each bin is Poisson-distributed. The black dotted curve corresponds to the original NFW profile, whereas the matched Hernquist profile is shown with the red dashed line. The other dot-dashed curves are the results of Fig.\ \ref{model-res}. The orange solid line is the SIDM prediction from ref.~\cite{Kamada:2016euw} for DDO 154 and from ref.~\cite{Kaplinghat:2015aga} for A2537.}
\label{fig:sim_results_with_data}
\end{figure*}

\begin{figure*}[ht]
  \centerline{
  \includegraphics[width=0.5\textwidth]{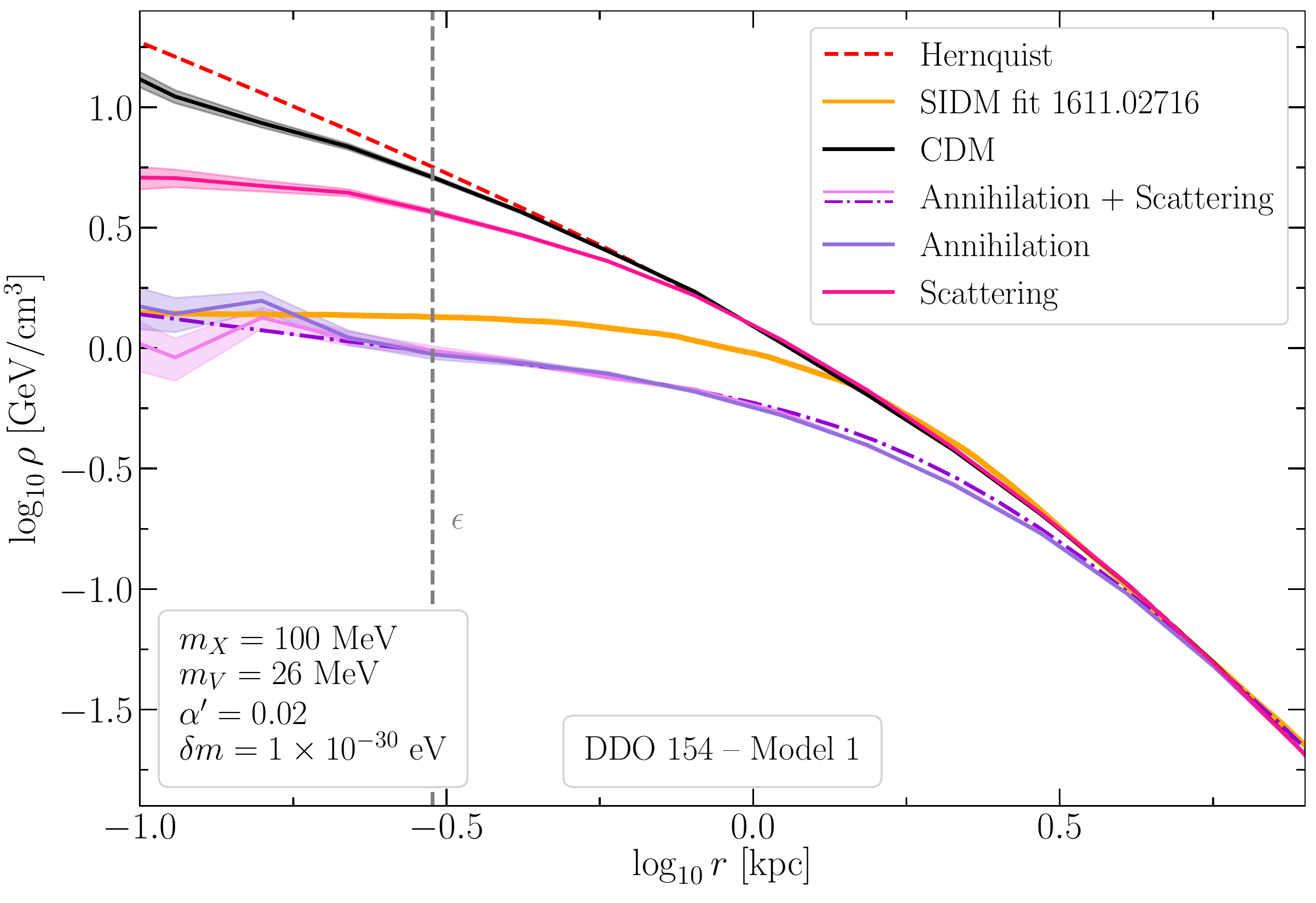}
  \includegraphics[width=0.5\textwidth]{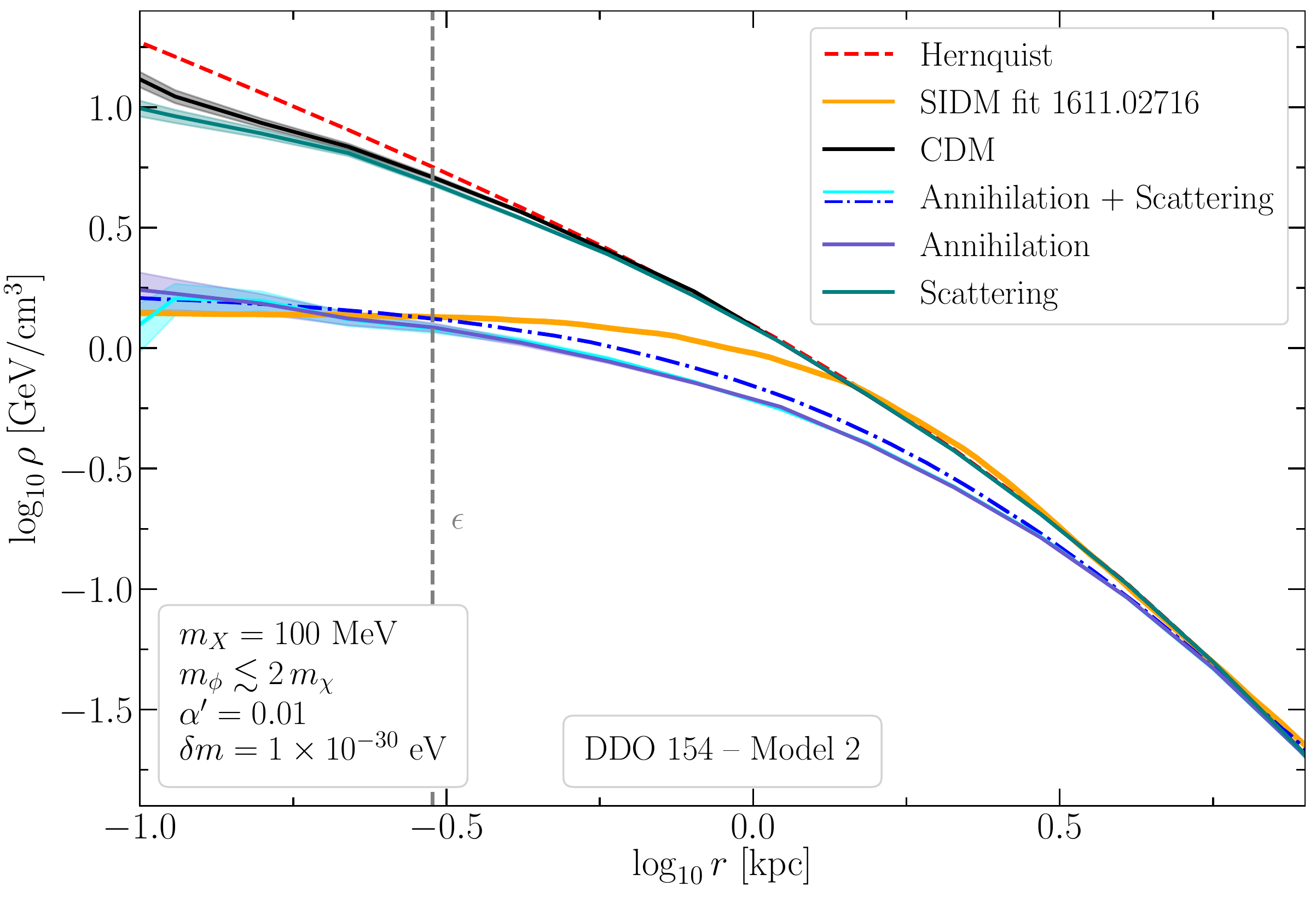}}
  \centerline{
  \includegraphics[width=0.5\textwidth]{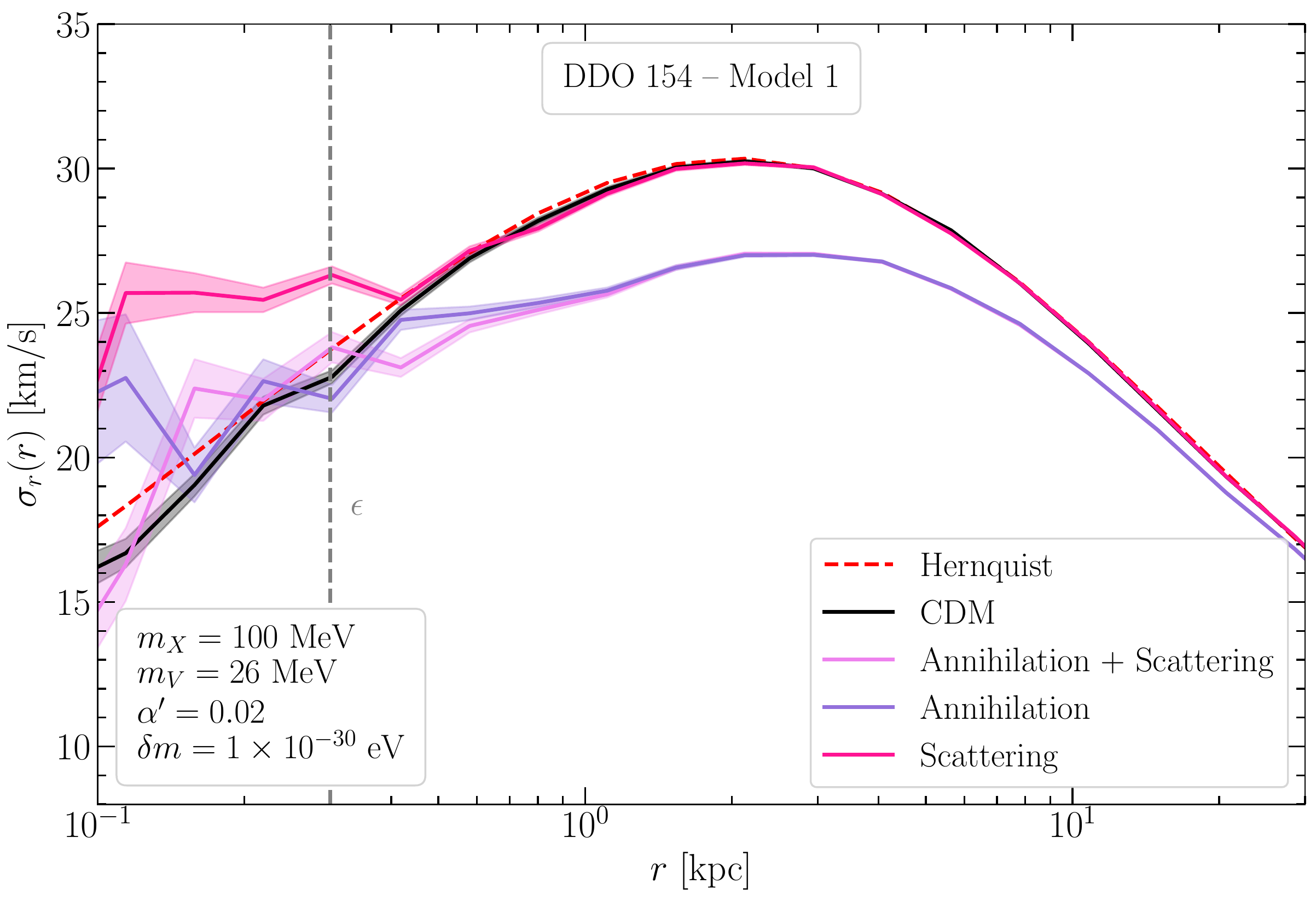}
  \includegraphics[width=0.5\textwidth]{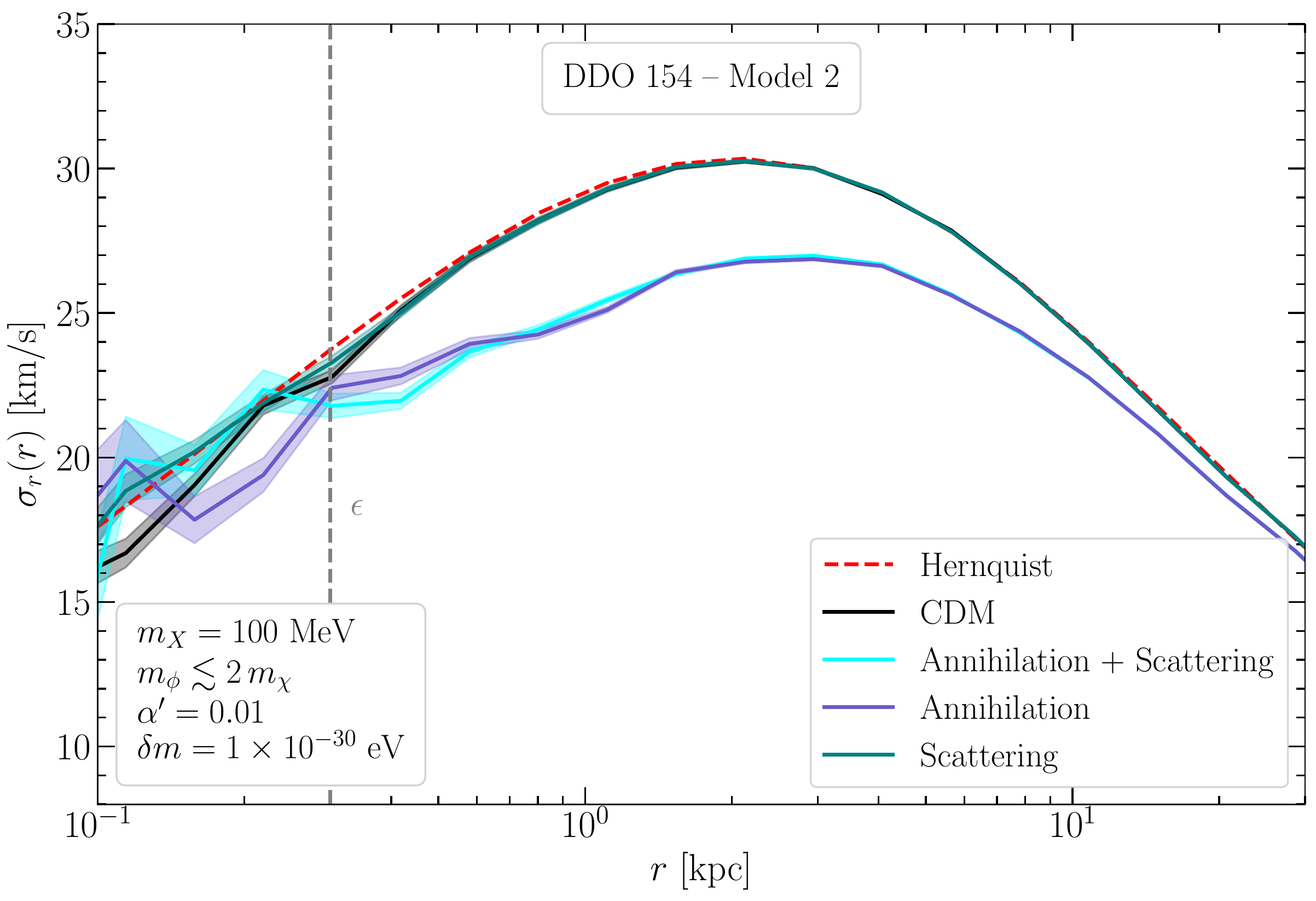}}
  \caption{Top: Radial density profile of the dwarf galaxy DDO 154 for Model 1 with $m_V = 26$ MeV (left) and for Model 2 with $\alpha' = 0.01$ (right) from N-body simulations. The other model parameters are the same as in Fig.~\ref{fig:sim_results}. The contributions of DM scattering and DM annihilation to the total profile are shown separately.
  The black solid curve corresponds to the result with just collisionless cold DM and the Hernquist profile for the initial halo is shown with the red dashed line.
  The gray dashed vertical line shows the position of the gravitational softening length $\epsilon$ used in the simulations.
  Bottom: Corresponding radial velocity dispersion of DDO 154 for the same Model 1 and Model 2 considered in the top row.}
\label{fig:sim_result_separation}
\end{figure*}

As initial conditions for DM halos corresponding to the dwarf galaxy DDO 154 and the galaxy cluster A2537, we took Hernquist profiles~\cite{Hernquist:1990be}, which are described by the total mass $M$ and the scale radius $a$. Unlike NFW, these profiles have finite mass without any need of truncation and they are perfectly stable in time when evolved with collisionless DM~\citep{Springel:2004kf,Robertson:2016xjh}, as we show in appendix~\ref{app:Nbody}.

To match the initial N-body profiles to the ones
assumed in section \ref{sec:struc}, we used the procedure described in ref.~\cite{Springel:2004kf}.
It consists of choosing the value of the Hernquist mass $M$ as the virial mass $M_{200}$ of the  NFW profile and requiring the two density profiles to
coincide in the inner region where $r \ll r_{200}$.
The latter condition gives a relation between the Hernquist scale radius $a$ and the NFW one $r_s$, namely
\be
    a = r_s \sqrt{2\,[\ln{(1 + c_{200})} - c_{200} / (1 + c_{200})]}\,,
\ee
where $c_{200} = r_{200} / r_s$ is the concentration index. The values of $c_{200}$ we use for our examples are displayed in Fig.~\ref{model-res}.
The comparison between the original NFW and the matched Hernquist profile for our simulated halos is shown in Fig.~\ref{fig:sim_results}. The agreement between these two profiles is excellent in the inner halo regions of interest for our study, suggesting that the simulation outcomes should model to a very good approximation the same dynamics as
in our complementary treatment of section \ref{sec:struc}, on the subgalactic or subcluster scales where they are most relevant.

Fig.\ \ref{fig:sim_results} shows the results of the N-body simulations for both Model 1 and Model 2 and their comparison with those obtained previously in Fig.\ \ref{model-res}.
The overall agreement observed for both DM models suggests that the N-body simulations model reasonably well the physics encapsulated in the quantum Boltzmann equation, where the coherence of DM particles plays a decisive role. Differences between the simulation and the approximate approaches are perhaps more evident in the dwarf galaxy because gravitational effects and DM dynamics have a relatively larger effect in small systems than in large ones.

To compare our model predictions with existing data, we converted our results for the DM density into observed quantities, namely the circular velocity for dwarf systems and the projected stellar velocity along the line-of-sight for galaxy clusters.
The former is defined as $V_{\rm circ} (r) = [G\,M(r) / r]^{1/2}$, where $M(r)$ is the enclosed mass at radius $r$. The left panels of Fig.~\ref{fig:sim_results_with_data} show our DM predictions for the rotation curve of DDO 154 dwarf galaxy within the two classes of models considered in this paper compared to current data. The grey points show the total circular velocity of the dwarf as observed by the LITTLE THINGS survey~\cite{Oh:2015xoa}, whereas the white points represent just the DM contribution to $V_{\rm circ} (r)$, obtained by subtracting the gas and star components after carefully modelling their distribution within the galaxy~\cite{Oh:2015xoa}.
The vector model with $m_V = 34$ MeV provides the best fit to data among the models displayed in the top panel of Fig.~\ref{fig:sim_results_with_data}, with a $\chi^2 / \text{d.o.f.} < 1$, comparable to the SIDM curve found in ref.~\cite{Kamada:2016euw}. For the scalar case, a choice of $\alpha'$ somewhat smaller than $0.01$ would provide  good agreement between our model and observations as shown in the bottom left panel of the same figure. 

For relaxed clusters dominated by a central early-type galaxy, such as in A2537, it is possible to measure the stellar line-of-sight velocity dispersion profiles $\sigma^{\star}_{\rm LOS} (r)$ with spatially-resolved spectroscopy~\citep{Newman:2012nv,Newman:2013}.
In order to convert our model predictions into $\sigma^{\star}_{\rm LOS} (r)$, we used the procedure outlined in appendix A of ref.~\cite{Sagunski:2020spe} combined with the information for A2537 cluster contained in ref.~\cite{Newman:2012nv}.
In particular, as done in the latter reference, we modeled the stellar luminosity density $\nu_{\star} (r)$ with a dual pseudo isothermal elliptical profile (dPIE)~\cite{Eliasdottir:2007md} and converted it into a baryonic density via the relation
\be
\label{eq:rhostar}
    \rho_b (r) = \Upsilon_{\star V}\, \nu_{\star} (r)\,.
\ee
Here $\Upsilon_{\star V} \equiv M_{\star} / L_V$ is the stellar mass-to-light ratio in the V-luminosity band, which is usually assumed to be spatially-independent across the cluster ~\citep{Newman:2012nv,Sagunski:2020spe}. The value of $\Upsilon_{\star V}$ could be inferred from the stellar population synthesis (SPS) up to an unknown initial mass function (IMF) and therefore one usually parametrizes this ignorance with the free parameter $\log{(\Upsilon_{\star V} / \Upsilon_{\star V}^{\rm SPS})}$, where $\Upsilon_{\star V}^{\rm SPS}$ is the SPS predicted mass-to-light ratio for a given IMF. We considered a Chabrier IMF~\cite{Chabrier:2003ki} as done in ref.~\cite{Newman:2012nv} and fixed the value of $\log{(\Upsilon_{\star V} / \Upsilon_{\star V}^{\rm SPS})}$ for the A2537 cluster by matching the baryonic density computed by eq.~\eqref{eq:rhostar} with that obtained in ref.~\cite{Kaplinghat:2015aga}.
The results of this procedure are shown in the right panels of Fig.~\ref{fig:sim_results_with_data} for both the vector and scalar models. Good agreement between them and the existing data is obtained for a wide set of parameters in both classes of models because of the large error bars in the observational data.

The N-body approach allows us to distinguish between the complementary effects of ordinary self-interactions by scattering, versus the novel one from annihilations, which we have investigated in both Models 1 and 2. To estimate the annihilation contribution to the total profile, we turned off the elastic scattering processes. Similarly, the scattering contribution can be estimated by turning off the annihilations. 
The top panels of Fig.~\ref{fig:sim_result_separation} show that the major effect in shaping the halo density profile is given by DM annihilation for the choices of parameters both in Model 1 and Model 2 considered in this paper. This verifies that the annihilation mechanism, investigated here for the first time in quantitative detail, has the capacity to alleviate the small-scale structure problems of CDM in the way originally suggested by \cite{Kaplinghat:2000vt}.

The comparison between annihilation and scattering is more evident by looking at their effect on the velocity dispersion of DM particles within the halo. As well-known in standard SIDM,
particle scatterings lead to a net energy transfer between the outer and inner parts of the halo, causing an increase in the velocity dispersion in the central region with respect to the collisionless cold DM case~\citep{Spergel:1999mh,Vogelsberger:2012ku}. 
However, such an effect is absent in the DM annihilation scenario if the annihilation products are not reabsorbed within the halo, as occurs for the choice of parameters for both Model 1 and Model 2 considered in this paper (see discussion at the beginning of section~\ref{sec:struc}).
On the contrary, the halo is expected to become overall colder than that in the collisionless cold DM scenario because high-velocity particles have higher chance to find a partner to annihilate with than low-velocity particles. This is nicely displayed in the bottom panels of Fig.~\ref{fig:sim_result_separation}, where the velocity dispersion of DDO 154 shows a net decrease at intermediate distances from the galactic center because particles there have normally a larger radial velocity.

Using dark matter annihilation to solve the core-cusp problem naturally gives a roughly constant value of the rate of core formation \cite{McDermott:2017vyk}, as is suggested by fits to astrophysical objects spanning five decades in mass \cite{Kaplinghat:2015aga}.  Relying on dark matter dynamics to resolve these issues is potentially under some tension from the measurement of cusps in the centers of classical dwarf spheroidal galaxies \cite{Read:2018pft, Hayashi:2020jze} and from recently discovered ultrafaint galaxies \cite{Hayashi:2020syu}, although out-of-equilibrium dynamics like tidal effects of the host galaxy may play a role in contributing to the diversity of these systems \cite{Kummer:2019yrb, Robles:2019mfq, Sameie:2019zfo, Kahlhoefer:2019oyt, 2020arXiv200702958C}.
Doing self-consistent fits to the observational data across many different systems will be critical for determining if the mechanism we investigate in this paper is as quantitatively successful as the elastic SIDM mechanism has been. Exploring this model in cosmological N-body simulations to compare against the subhalo abundance, for instance, will also be an important route for future work.

\section{Conclusions}
\label{sec:conc}

The long-standing discrepancies between gravitational N-body simulations of structure formation in the $\Lambda$CDM paradigm and observations of cored density profiles continue to motivate exploration of alternative dark matter models and mechanisms.  In this work we have revived one of the earliest such proposals \cite{Kaplinghat:2000vt} by showing that dark matter annihilations in galactic structures can be responsible for erasure of the cusps, using distinctive properties of
asymmetric dark matter (ADM).  The key idea is that very strong annihilations would freeze out early in cosmic history, solving the problem of removing the ``symmetric'' ADM relic density, and are reactivated at late times relevant for structure formation by oscillations of
DM into its antiparticle. 
The preferred annihilation rate per unit mass $\sigma v/m_\chi\sim 100$\,cm$^2$/g\,\,km/s can be fit in our model by dark matter and mediator masses of order $30 \, {\rm MeV} \simeq m_{V,\phi,a} \lesssim m_\chi \simeq 100 \, {\rm MeV}$, a perturbative self-coupling as given in \Eq{eqalphap}, and a Majorana mass term 
$\delta m$ within the range $(10^{-31}-10^{-28})$\,eV.

To obtain a large-enough annihilation cross section while respecting perturbativity of couplings constrains the DM and the mediator of the strong hidden force to be light, typically below 100\,MeV.  We have illustrated the mechanism in two representative models, with vector or scalar mediators respectively, and using two complementary approaches to model the structure formation dynamics.  A fully consistent simulation is challenging because it must incorporate the quantum coherence of the oscillating
dark matter while tracking the spatially-dependent annihilation rates within a DM halo.  Our N-body simulations, which treat the coherence in an approximate way, give relatively close results to a quantum Boltzmann equation approach, which models the structure formation less rigorously. 
We have tested the scenario on two representative
dwarf spheroidal galaxies, as well as a galactic cluster. Both methods lead to significant coring of the density profile, qualitatively similar to the effects of elastic SIDM scattering that have been widely used to 
address the cusp-core problem.  

Like the elastic SIDM paradigm, the new mechanism we propose here does not, in its simplest forms, address the diversity of halo profiles on all scales. In elastic SIDM this is accomplished by assuming velocity-dependent scattering, with a cross section that goes down at larger DM speeds. Within our mechanism, scalar mediators generically have a relatively stronger coring effect on
small halos than larger (less dense) ones, while vector mediators have the opposite behavior.  We presented evidence that the combination of both mediators could provide a good universal fit, leaving a detailed investigation for future study.

\bigskip
{\bf Acknowledgments.}  We thank A.\ Benson, S.\ Tulin and A.\ Robertson for very helpful correspondence.  JC and SDM thank the CERN Department of Theoretical Physics for its hospitality and stimulating environment during the inception of this work.
We acknowledge Calcul Queb\'ec (\url{https://www.calculquebec.ca}) and Compute Canada (\url{https://www.computecanada.ca}) for supercomputing resources. 
JC and GG are supported by NSERC (Natural Sciences and Engineering Research Council, Canada). GG acknowledges support from CNPq grant No.141699/2016-7, FAEPEX grant No. 2039/20, and FAPESP grant No. 2014/19164-6, as well as the support from Global Affairs Canada and McGill Space Institute. MP is supported by the Arthur B. McDonald Institute for Canadian astroparticle physics research. Fermilab is operated by Fermi Research Alliance, LLC under Contract No.~DE-AC02-07CH11359 with the United States Department of Energy.

\begin{appendix}
\renewcommand{\setthesubsection}{\Alph{section}.\arabic{subsection}}
\begin{widetext}

\begin{figure}[hb]
\centering
  \centering
  \includegraphics[width=0.6\textwidth]{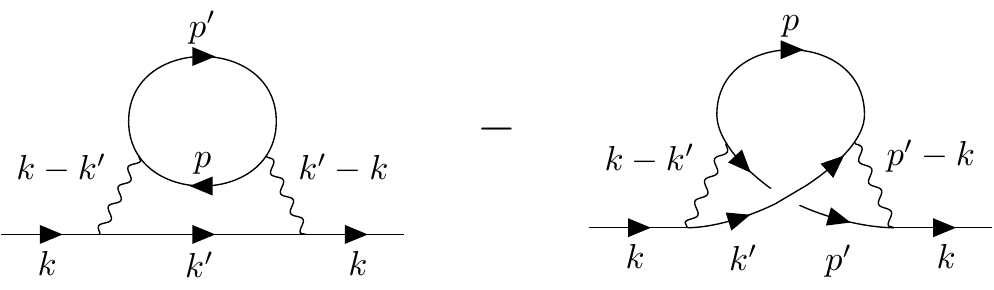}
  \caption{Self-energy diagrams for the vector model}
\label{diags1}
\end{figure}

\section{Scattering term in Model 1 }
\label{appA}
In this appendix we derive the collision term for elastic scattering of $\chi\chi$ or $\chi\bar\chi$ through exchange of the vector boson, needed for the quantum 
Boltzmann equations.  
The  diagrams in Fig.\ \ref{diags1} are the analog of Fig.\ 4b in ref.\ \cite{Tulin:2012re}.  We can read off the imaginary part of the self-energies
$\Sigma^{>,<}$, in
analogy to their eq.\ (A26) of ref.\ \cite{Tulin:2012re},
\bea
	\Sigma^{>,<}(k) &=& i{g'^4\over 4}\int d{k'} d{p'}
	d{p}\, (2\pi)^4 \delta^{(4)}(k + p - k'-p')\,\nn\\
	&\cdot& \Bigg[
	{1\over ((k-k')^2-m_V^2)^2}\, O_- \gamma^\mu S^{>,<}_{k'}
	O_-\gamma^\nu \,{\rm Tr}\left(S^{<,>}_p O_- \gamma_\mu
	S^{>,<}_{p'} O_-\gamma_\nu\right)\nn\\
	&-& {1\over ((k-k')^2-m_V^2)((p-p')^2-m_V^2)}\,
	 O_- \gamma^\mu S^{>,<}_{p'}
	O_-\gamma^\nu S^{<,>}_p O_- \gamma_\mu
	S^{>,<}_{k'} O_-\gamma_\nu\Bigg]\,,
\label{sigmaeq}
\eea
where $dp=d^{\,4}p/(2\pi)^4$ and the Green's
functions are given by
\bea
	S^<_k &=& -2\pi\delta(k^2-m^2)(\slashed{k}+m_\chi)\left[
	\theta_{k^0}{\cal F}_k - \theta_{-k^0}(1-\bar{\cal F}_k)
	\right]\,,\nn\\
	S^>_k &=& +2\pi\delta(k^2-m^2)(\slashed{k}+m_\chi)\left[
	\theta_{k^0}(1-{\cal F}_k) - \theta_{-k^0}\bar{\cal
F}_k\right]\,.
\eea
Here $\bar{\cal F}$ is the matrix with the diagonal entries
interchanged, as in (11) of \cite{Tulin:2012re}, and $O_- = {\rm diag}(1,-1)$.
The trace is over both Dirac and flavor indices. We also define
\be
	\tilde{{\cal F}} = O_- {\cal F} O_-\,, \quad 
 \tilde{\bar{{\cal F}}} = O_-\bar{{\cal F}} O_-\,.
\ee
This has the effect of reversing the signs of the off-diagonal
elements.

Considering the relevant physical processes,
it is not necessary to take account of all eight of the terms that arise from each diagram, from the products of the $S^{>,<}$ functions.  First, 
since annihilation diagrams are suppressed while $k^0>0$, we can
ignore $k'^0<0$, which would give the $s$-channel diagram.  Second, by energy conservation, we must have either
$p^0>0$ and $p'^0>0$, representing $\chi \chi$ scattering, or 
$p^0<0$ and $p'^0<0$, representing $\chi \bar{\chi}$ scattering.
Let us first write the  
terms that arise from the middle line of (\ref{sigmaeq}),
apart from the factors of $2\pi\delta(\dots)$
\bea
{1\over ((k-k')^2-m_V^2)^2}\, &&\gamma^\mu(\slashed{k}'+m_\chi)\gamma^\nu
\tr\left((\slashed{p}+m_\chi)\gamma_\mu (\slashed{p}'+m_\chi)\gamma_\nu\right)\,
\times \nn\\
\Sigma^>_k: \qquad &&	(1-\tilde{\cal F}_{k'})\left[\theta_{p^0}\theta_{p'^0}
	\tr\left((-{\cal F}_p)(1-\tilde{\cal F}_{p'})\right)
	+ \theta_{-p^0}\theta_{-p'^0}
\tr\left((1-\bar{{\cal F}}_p)(-\tilde{\bar{{\cal F}}}_{p'})\right)\right]\,,
	\nn\\
\Sigma^<_k: \qquad &&	\phantom{(1)}-\tilde{\cal F}_{k'}
\left[\theta_{p^0}\theta_{p'^0}
	\tr\left((1-{\cal F}_p)(-\tilde{\cal F}_{p'})\right)
	+ \theta_{-p^0}\theta_{-p'^0}
\tr\left((-\bar{{\cal F}}_p)(1-\tilde{\bar{{\cal F}}}_{p'})\right)\right]\,.
	\nn\\
\eea
Similarly the last line of (\ref{sigmaeq}) contributes
\bea
-{1\over ((k-k')^2-m_V^2)((p-p')^2-m_V^2)}\ &&  
	\gamma^\mu(\slashed{p}'+m_\chi)\gamma^\nu
(\slashed{p}+m_\chi)\gamma_\mu (\slashed{k}'+m_\chi)\gamma_\nu\, \times	\nn\\
\Sigma^>_k: \qquad &&	(1-\tilde{\cal F}_{p'})\left[\theta_{p^0}\theta_{p'^0}
	\left((-{\cal F}_p)(1-\tilde{\cal F}_{k'})\right)
	+ \theta_{-p^0}\theta_{-p'^0}
\left((1-\bar{{\cal F}}_p)(-\tilde{\bar{{\cal F}}}_{k'})\right)\right]\,,
	\nn\\
\Sigma^<_k: \qquad &&	\phantom{(1)}-\tilde{\cal F}_{p'}
\left[\theta_{p^0}\theta_{p'^0}
	\left((1-{\cal F}_p)(-\tilde{\cal F}_{k'})\right)
	+ \theta_{-p^0}\theta_{-p'^0}
\left((-\bar{{\cal F}}_p)(1-\tilde{\bar{{\cal F}}}_{k'})\right)\right]\,.
	\nn\\
\eea

The collision term comes from $\Sigma^{>,<}$ by 
\be
	{\cal C}_s = i\int {d^{\,4}k\over (2\pi)^4}
	\,{\rm tr}\left[\left({\slashed{k}+m_\chi\over 4 m_\chi}\right)
	\left(\{\Sigma_k^<,S_k^>\} - \{\Sigma_k^>,S_k^<\}\right)
	\right]
\ee
where unlike $\tr$ above, $\rm{tr}$ denotes only the trace over 
Dirac matrices.  Since we are interested in low densities, we can
neglect terms of order ${\cal F}^3$, which means that we need only
keep terms of order
\bea 
	\Sigma_k^<: O({\cal F}^2)\,, \quad
S_k^>: O(1)\,,\quad  \Sigma_k^>: O({\cal F})\,, \quad
S_k^<: O({\cal F})\,.
\eea
After carrying out the Dirac traces and combining like terms, we find that
the respective contributions from the two diagrams are
\bea
{\cal C}_1 &=& -4 g'^4\int d\Pi_k d\Pi_{k'} d\Pi_p d\Pi_{p'}\,{ (2\pi)^4
	\delta^{(4)}(k+p-k'-p')\over ((k-k')^2-m_V^2)^2}
\left[(k\cdot p)(k'\cdot p') + (k\cdot p')(k'\cdot p) -m_\chi^2(k\cdot
k'+p\cdot p') +  2 m_\chi^4\right] 
\nn\\
&\times&\theta_{k^0}\theta_{k'^0}\Big\{
	\theta_{p^0}\theta_{p'^0} \Big[
	\tilde{\cal F}_{k'}\tr\tilde{{\cal F}}_{p'}
	-{\cal F}_k \tr{{\cal F}}_{p}
	\Big] + 
 \theta_{-p^0}\theta_{-p'^0} \Big[
	\tilde{\cal F}_{k'}\tr\bar{{\cal F}}_{p}
	-{\cal F}_k \tr{\bar{\tilde{{\cal F}}}}_{p'}
		\Big] \Big\}\,,\nn\\
{\cal C}_2 &=& -4 g'^4\int d\Pi_k d\Pi_{k'} d\Pi_p d\Pi_{p'}\,{ (2\pi)^4
	\delta^{(4)}(k+p-k'-p')\over ((k-k')^2-m_V^2)((p-p')^2-m_V^2)}
\nn\\
&\times&\left[(k\cdot p)(k'\cdot p') -\sfrac12m_\chi^2(k\cdot k'+p\cdot p'+
k\cdot p + k\cdot p' + k'\cdot p + k'\cdot p'
) +  m_\chi^4\right]
\nn\\
&\times&\theta_{k^0}\theta_{k'^0}\Big\{\theta_{p^0}\theta_{p'^0} \Big[
	\tilde{{\cal F}}_{p'}\tilde{{\cal F}}_{k'} -
	 \sfrac12\{{\cal F}_p,{\cal F}_k\}
	\Big]+
\theta_{-p^0}\theta_{-p'^0}\Big[
\tilde{{\cal{F}}}_{p'}\bar{{\cal F}}_p - 
\sfrac12\{\bar{\tilde{{\cal F}}}_{k'},{\cal F}_k\}
	\Big]\Big\}\,,
\eea
where $d\Pi_p = d^{\,4}p\, \delta(p^2-m_\chi^2)/(2\pi)^3$.

In the non-relativistic limit, it further simplifies
since the squared matrix element in brackets is equal to $2\,m_\chi^4$ for
${\cal C}_1$, while for ${\cal C}_2$ it depends on which of the theta
functions are taken: $[\dots] = -m_\chi^4$ for positive energies and $+m_\chi^4$ for
negative energies.  The resulting collision term is
\bea
	{\cal C}_s &=& -{g'^4\over 4(2\pi)^8 m_V^4}
	\int d^{\,3}k\cdots d^{\,3}p'\,\delta^{(4)}(\cdots)\,
	\Big[ 4\left(\tilde{{\cal F}}_{k'}\tr{\cal F}_{p'}
-{\cal F}_{k}\tr {\cal F}_p\right)\nn\\
&-& \tilde{\cal F}_{p'}\tilde{\cal F}_{k'}
+ 
 \sfrac12\{{\cal F}_p,{\cal F}_k\}
+\tilde{\cal F}_{p}\bar{\cal F}_{p'}  
-\sfrac12\{\bar{\tilde{{\cal F}}}_{k'},{\cal F}_k\} \Big]\,.
\eea
Here, we used the identities $\tr\tilde{\cal F}_{p} = \tr\bar{{\cal F}}_{p} = 
\tr\bar{\tilde{{\cal F}}}_{p} = \tr{\cal F}_{p}$, as well as 
the fact that any terms with negative energies can be
transformed to the corresponding phase space integrals with positive energy
by changing $p\leftrightarrow p'$.

The next step is to make the ansatz
\be
	{\cal F}_k = e^{-\beta\omega_k} {n\over n_{\rm eq}}\,,
\ee
where $\omega_k \cong m_\chi + k^2/2m_\chi\equiv m_\chi + E_k$, 
\be
	n = \left({n_{11}\atop n_{21}} \, {n_{12}\atop n_{22}}
	\right)\,,
\ee
and $n_{\rm eq}$ is the equilibrium number density.  Then the
momentum integrals can be carried out to get collision terms as a
function of the density matrix $n$
\bea
{\cal C}_s  &=& -{g'^4\, e^{-2\beta m_\chi}\over 4(2\pi)^8
m_V^4\, n_{\rm eq}^2}
	\int d^{\,3}k\cdots d^{\,3}p'\,\delta^{(4)}(\cdots)\,
	\Big[e^{-\beta( E_{k}+E_p)}\left[
	 4(\tilde n - n)\tr n -\tilde n^2 + n^2\right)\nn\\
	&+& 
	e^{-\beta(E_p+E_{p'})}\tilde n\bar n -
e^{-\beta(E_k+E_{k'})}\sfrac12\{\tilde{\bar{n}},n\}\Big]\,.
\eea
The integrals are all equal to $(m_\chi T)^{9/2}/T$ times a dimensionless
number, and there are only two different possibilities, depending upon
whether the two energies in the Boltzmann factors are both
initial/final state, or one initial and one final.  We get
\bea
{\cal C}_s  &=& -{g'^4\,m_\chi^{3/2} T^{1/2}\,\over 16 \pi\,
m_V^4}\Big[ I_s\,\left(4(\tilde n -n)\tr n
-\tilde n^2 + n^2\right) + 
I_d\left(\tilde n\bar n -\sfrac12\{\bar{\tilde n},n\}\right)
\Big]\,,
\eea
where the two dimensionless integrals are
\bea
	I_s &=& {1\over 8\pi^4}\int d^{\,3}p\,d^{\,3}k\,d^{\,3}p'\,d^{\,3}k'\,
	\delta^{(4)}(p+k-p'-k')\, e^{-(p^2+k^2)/2}\nn\\
	&=& {1\over 8\pi^4}\int d^{\,3}p\,d^{\,3}k\,d^{\,3}p'
	\,\delta(\vec p\cdot \vec k)\, e^{-(\vec p+\vec p')^2/2 - (\vec k+\vec p')^2/2}
	\cong 
 2.26\,,\nn\\
	I_d &=& {1\over 8\pi^4}\int d^{\,3}p\,d^{\,3}k\,d^{\,3}p'\,d^{\,3}k'\,
	\delta^{(4)}(p+k-p'-k')\, e^{-(p^2+p'^2)/2}\nn\\
	&=& {1\over 8\pi^4}\int d^{\,3}p\,d^{\,3}k\,d^{\,3}p'
	\,\delta(\vec p\cdot \vec k)\, e^{-(\vec p+\vec p')^2/2 - p'^2/2} = \infty\,,
\eea
and the zeroth component of the delta function is in terms of the non-relativistic dimensionless energies.
The second forms of the integrals, in which the delta function of energies simplifies, are obtained by shifting $\vec p\to \vec p+\vec p'$ and $\vec k\to \vec k + \vec p'$.  
We evaluated $I_s$ numerically.  The divergent integral is
inconsequential because it multiplies 
$\tilde n\bar n -\sfrac12\{\bar{\tilde n},n\}\equiv 0$, which
vanishes identically.  In retrospect we understand that this term is
unphysical, since it corresponds to the interference of the $t$- and $u$-channel scattering diagrams, which vanishes for scattering of $\chi$
with $\bar{\chi}$.  The relevant matrix evaluates to
\be
4(\tilde n -n)\tr n
-\tilde n^2 + n^2 = -6(n_{11} + n_{22})\left({0\atop n_{21}}\
	{n_{12}\atop 0}\right)\,,
\ee
so the collision term from scattering is
\be
	{\cal C}_s = {3 I_s g'^4\,m_\chi^{3/2} T^{1/2}\,\over 8 \pi m_V^4 }
\,(n_{11} + n_{22})\left({0\atop n_{21}}\
	{n_{12}\atop 0}\right) \equiv \sfrac32\langle\sigma v\rangle_s (n_{11} + n_{22}) \left({0\atop n_{21}}\
	{n_{12}\atop 0}\right)\,,
\ee
which would appear in eq.\ (20) of ref.\ \cite{Tulin:2012re}.  The normalization of $\langle\sigma v\rangle_s$ is chosen to agree with the usual definition, in which the low-energy
cross section is $\sigma \approx g^4 m_\chi^2/(4\pi m_v^4)$,
and the thermal averaging is done as in ref.\ \cite{Gondolo:1990dk}.

\section{Thermal decoherence in the Boltzmann equation}
\label{thdec}
For the vector model, we have simulated the  effect of thermal
decoherence due to the oscillation rate depending on the momentum
in the quantum Boltzmann equation for the density matrix 
\be
    {d{\cal F}_k\over dt} - H k {d{\cal F}_k\over dk} = -i[{\cal H}_k,{\cal F}] + {\cal C}[{\cal F}], \qquad {\cal H}_k = 
    \omega_k\mathbb{1} + {m_\chi \delta m\over \omega_k}\left({0\atop 1}\,{1\atop 0}\right)\,,
\label{qbe}
\ee
where $H$ is the Hubble rate and $\omega_k=\sqrt{k^2 + m_\chi^2}$.
The $k$-dependence in the second term of the Hamiltonian implies that high-$k$ parts of the distribution oscillate with slightly lower frequency than low-$k$ parts, which is an additional source of decoherence that is neglected by integrating over momenta to reduce eq.\ (\ref{qbe}) to an equation for the
number density matrix $n$.  Our goal is to verify that this neglect is justified.
For the scalar model, this issue is less important since decoherence is not a requirement for annihilations to occur.

To model the effect one would like to  divide
the particle distribution into several momentum bins.  We will
be content to take
just two, labeled by $s,l$ for small and large momenta, relative to
the midpoint of the distribution.  Accordingly, we split the density
matrix into
\be
	n_t = n_s + n_l
\ee
and one finds separate Boltzmann equations for each component, that are
coupled to each other through the collision terms.  The Boltzmann
equations take the form
\bea	
 	\dot n_s + 3H n_s = -i[H_s,n_s] - {\langle\sigma v\rangle_s\over 8} \left(S_s + S\right) - {\langle\sigma v\rangle_a\over 2}  \left(A_s- n_{\rm eq}^2\right)\,,\nn\\
 	\dot n_l + 3H n_l = -i[H_l,n_l] - {\langle\sigma v\rangle_s\over 8} \left(S_l + S\right) - {\langle\sigma v\rangle_a\over 2}  \left(A_l- n_{\rm eq}^2\right)\,,
 \label{boltz-2bin}
\eea
where $\langle\sigma v\rangle_{s,a}$ are the scattering and annihilation cross sections, the 
matrices $S_i$, $S$, $A_i$ are defined as
\be
    S_i = \left(\begin{array}{cc} n_{i,11}(6n_{11} + 8 n_{22}) - n_{i,12}n_{21} - n_{i,21}n_{12}
    &  7n_{i,12} n_t -n_{i,t}n_{12}\\  7n_{i,21} n_t -n_{i,t}n_{21} & 
    n_{i,22}(8n_{11} + 6n_{22}) - n_{i,12}n_{21} - n_{i,21}n_{12}\end{array}\right)\,,
\ee
\be
    S = \left(\begin{array}{cc} 
    -3 n_{11}^2 - 4 n_{11}n_{22} + n_{12}n_{21}
    &  3 n_{12} n_t \\  3 n_{21} n_t & -3 n_{22}^2 - 4 n_{11}n_{22} + n_{12}n_{21}
    \end{array}\right)\,,
\ee
\be
    A_i = \left(\begin{array}{cc} 
    2\, n_{i,11}n_{22} - (n_{i,12}n_{21} + n_{i,21}n_{12} )   &  
    (n_{i,12}n_t - n_{i,t}n_{12})\\  (n_{i,21}n_t - n_{i,t}n_{21})  & 2\, n_{i,22}n_{11} -(n_{i,12}n_{21} + n_{i,21}n_{12} )     \end{array}\right)\,,
\ee
and we defined $n_{ij} = n_{s,ij} + n_{l,ij}$, $n_t = n_{11} + n_{22}$, and $n_{i,t} = n_{i,11}+ n_{i,22}$.  These expressions can be read from the form of
the collision and annihilation terms in terms of  the ${\cal F}$ matrices, before doing the final integral over the momentum $k$ of the particle whose distribution is being tracked in the Boltzmann equation. 
If one adds the two equations together, they revert to 
the standard equation in terms of $n_t$ alone.  The decoherence
effect comes from the fact that the free Hamiltonians ${\cal H}_{s,l}$ are
slightly different for the two components, which for non-relativistic particles is
\be
	{\cal H}_{s,l} \cong  m_\chi\left({1\atop 0}\ {0\atop 1}\right) 
	+ \delta m \left({0\atop 1}\,{1\atop 0}\right)+
	{\langle k^2\rangle_{s,l}\over 2m_\chi}
	\left[1 - {\delta m\over m_\chi}\left({0\atop 1}\,{1\atop 0}\right)\right]\,.
\label{Hsleq}
\ee
The important feature is the difference between 
$\langle k^2\rangle_{l}$ and $\langle k^2\rangle_{s}$, so for
simplicity one could take, for example, $\langle k^2\rangle_{s}=\sfrac12\langle k^2\rangle$ and
$\langle k^2\rangle_{l} = \sfrac32\langle k^2\rangle$, which is a
temperature-dependent split.  For temperatures such that scattering is 
still in equilibrium, we can estimate $\langle k^2\rangle \sim 3
m_\chi^2/x$, where $x=m_\chi/T$.  After scatterings freeze out, the wavenumber redshifts
as $1/a$, so $\langle k^2\rangle \sim 3
m_\chi^2 \,x_f/x^2$.

This effect can be important only in the early
universe when the momenta are sufficiently large that $k^2/m_\chi^2$ is not negligible.
We have applied this formalism to check the
early-universe solutions of section \ref{sec:early}.  We found no appreciable effect from this extra source of decoherence.

\end{widetext}

\section{N-body simulation details}\label{app:Nbody}
In this appendix, we describe the scattering and annihilation algorithms used in the simulations presented in this paper. Validation tests of the code are also presented and discussed.

\begin{subappendices}
\subsection{Scattering}\label{subsec:scatt}
Elastic scattering between DM particles has been implemented stochastically on top of \texttt{GADGET-2} in the same way as done by ref.~\cite{Robertson:2016xjh}, which was derived directly from the classical Boltzmann equation~\cite{Rocha:2012jg}.
We summarize below the main relevant information and refer the reader to ref.~\cite{Robertson:2016xjh} for a more detailed description.

The scattering rate for a DM particle of mass $m_{\chi}$ at position $\vec{r}_i$ and velocity $\vec{v}_i$, moving in an equal-mass particle background characterized by a normalized velocity distribution $f_v (\vec{r}, \vec{v})$ and a local density $\rho (\vec{r})$, is
\be
\label{eq:scattrate}
    \Gamma_{i,\,\text{scatt}} = \int f_v (\vec{r}_i, \vec{v})\,\rho (\vec{r}_i) \,\frac{\sigma}{m_{\chi}}\, |\vec{v}_i - \vec{v}|\,d\vec{v}\,.
\ee
Here $\sigma / m_{\chi}$ is the DM scattering cross section per unit particle mass, which can be velocity-dependent. In simulations, individual physical particles cannot be resolved and all the properties of the latter should be translated to those of the simulation particles. For instance $\sigma / m_{\chi}$ should be replaced by $\sigma_p / m_p$, where $\sigma_p$ and $m_p$ are the scattering cross section and the mass of the simulation particles, respectively. In a similar manner, the quantities $f_v (\vec{r}, \vec{v})$ and $\rho (\vec{r})$ should be estimated from the volume within a sphere of radius $h_S$, called \textit{scatter search radius}, centered at the DM particle position $\vec{r}_i$.
Assuming all the simulation particles within the scattering volume contribute equally, independently of their location (top-hat kernel), eq.~\eqref{eq:scattrate} can be written as a sum over the $N_p$ neighboring particles~\cite{Robertson:2016xjh}
\be
    \Gamma_{i,\,\text{scatt}} = \sum_{j=1}^{N_p} \frac{\sigma_p\,|\vec{v}_i - \vec{v}_j|}{\sfrac{4}{3} \pi h_S^3}\,.
\ee
Hence the probability for particles $i$ and $j$, separated by a distance smaller than $h_S$, to scatter within the next time step of size $\Delta t$, is given by
\bea
\label{eq:Pscatt}
    P_{ij,\,\text{scatt}} &=& \bigg(\frac{\sigma}{m_{\chi}}\bigg)\,\rho_{ij} \,|\vec{v}_i - \vec{v}_j|\,\Delta t\,,\nn\\ 
    \rho_{ij} &=& 
\begin{cases}
    \frac{3\,m_p}{4\, \pi h_S^3} \qquad 0 \leq \tilde{r} \leq h_S \\
    0 \qquad \qquad \quad\,\,\,\, \tilde{r} > h_S
\end{cases}
\eea
where $\tilde{r} \equiv |\vec{r}_i - \vec{r}_j|$ and $\rho_{ij}$ is the target density, which is constant in this case because it is estimated using a top-hat kernel function. This is the simplest choice, but not the most common one used to implement DM self-interaction in N-body simulations. For instance, refs.~\citep{Rocha:2012jg,Vogelsberger:2012ku,Vogelsberger:2018bok} used a cubic spline kernel $W(r, h_S)$ like the one already used in \texttt{GADGET} to compute the gravitational force in the context of smoothed-particle hydrodynamics (SPH)~\cite{Monaghan1985}. Such a kernel allows nearby particles separated by a distance less than $h_S$ to have higher scatter probability than those further apart because $W$ is a smoothing function peaked at $\tilde{r} = 0$.
For appropriate choices of $h_S$, both the top-hat and cubic spline kernels provide results in agreement with analytical expectations~\citep{Rocha:2012jg,Vogelsberger:2012ku,Vogelsberger:2018bok,Robertson:2016xjh} and with each other, within numerical uncertainties~\cite{Robertson:2016xjh}. Therefore, we preferred using the simple and intuitive top-hat kernel with $\rho_{ij}$ given by eq.~\eqref{eq:Pscatt} and with a fixed value of $h_S$ of the same order of the gravitational softening length $\epsilon$. In particular, we consider $h_S = \epsilon$ as chosen by ref.~\cite{Robertson:2016xjh} because it gives the expected scattering rate, as discussed in section~\ref{subsec:tests}.

To see which particles do actually scatter at each time step, the probability in eq.~\eqref{eq:Pscatt} is computed for each pair of nearby particles and compared with a random number drawn from a uniform distribution. For isotropic scattering and equal-mass particles, the post-scatter velocities are computed as $\vec{v}'_{i,j} = \vec{v}_{\text{CM}} \pm (v_{\text{rel}}/2)\,\hat{e}$, where $\vec{v}_i$ and $\vec{v}_j$ are the initial velocities, $\vec{v}_{\text{CM}}$ is the center-of-mass velocity between particle $i$ and $j$, $v_{\text{rel}}$ is the magnitude of their relative velocity and $\hat{e}$ is a randomly oriented unit vector.

This scattering algorithm is very similar to that used in several SIDM cosmological simulations~\citep{Kochanek:2000pi,Yoshida:2000uw,Dave:2000ar,Koda:2011yb,Vogelsberger:2012ku,Fry:2015rta,Elbert:2016dbb,Kim:2016ujt}, which mainly differ in the number of neighbors within the scattering volume.
The majority of these simulations have treated the DM scattering as isotropic, which usually results from a short-range interaction mediated by a massive particle. In particular the mediator mass $m_{V, \phi}$ should be much heavier than the DM particle momenta, which translates into $m_{V, \phi} \gg 10^{-3}\, m_{\chi}$ for DM particles moving in Milky-Way-like galaxies today.
If this is not the case, the cross section will depend on the momentum exchange, which increases with the collision velocity or the scattering angle, leading typically to velocity-dependent anisotropic scatterings. The latter are common in long-range interactions via light or massless mediators, which occur in several motivated DM scenarios such as mirror~\citep{Blinnikov:1983gh,Berezhiani:1995am,Foot:2004pa}, atomic~\citep{Cline:2012is,Cline:2013pca,Cline:2013zca,CyrRacine:2012fz} and hidden sector DM models~\citep{Feng:2009mn,Foot:2014uba,Foot:2016wvj,Boddy:2014yra,Boddy:2016bbu}.
The simplest way to take them into account is to assume the scattering is still isotropic but the cross section $\sigma$ in eq.~\eqref{eq:Pscatt} is replaced by the \textit{momentum transfer cross section} $\sigma_{T}$, defined as~\citep{Feng:2009hw,Buckley:2009in,Tulin:2013teo}
\be
\label{eq:sigmaTdef}
    \sigma_T = \int_0^{\pi} \frac{d\sigma}{d\Omega}\,(1-\cos{\theta})\,d\Omega\,,
\ee
where $\theta$ and $d\Omega$ are the scattering and solid angles, respectively.
Here, the differential cross section $d\sigma/d\Omega$ can be derived within the Born approximation~\cite{Born:1926yhp} from particle scattering mediated by a Yukawa interaction~\citep{Ibe:2009mk,Tulin:2017ara}
\be
\label{eq:scattdiff}
    \frac{d\sigma}{d\Omega} = \frac{\sigma_s}{4\pi} \bigg(1 + \frac{v_{\text{rel}}^2}{\omega^2}\,\sin^2{\frac{\theta}{2}}\bigg)^{-2}\,,
\ee
where $\sigma_s$ is given by eq.~\eqref{eqsigmas} and $\omega = r_m\,c$ with $r_m = m_V / m_{\chi}$ or $r_m = m_{\phi} / m_{\chi}$ depending on the model under consideration. 
In the limit $v_{\text{rel}} \ll \omega$, $d\sigma/d\Omega$ becomes velocity-independent, $\sigma = \int (d\sigma / d\Omega)\,d\Omega \approx \sigma_s$ and the scattering is isotropic. In the opposite regime, eq.~\eqref{eq:scattdiff} scales as $\propto v_{\text{rel}}^{-4}$ like in Rutherford scattering which is mediated by the Coulomb potential.
The approximation of using $\sigma_T$ instead of $\sigma$ as the scattering cross section captures most of the effects of more complicated scattering dynamics where the DM velocity distribution is close to isotropic, which is the case of dwarf galaxies~\cite{Tulin:2013teo}~\footnote{Although in most astrophysical systems DM is expected to have an approximately isotropic velocity distribution, this is not the case of colliding galaxy clusters where there is a preferred direction along which DM particles collide~\citep{Kahlhoefer:2013dca,Robertson:2016qef}.}. This is because $\sigma_{T}$ estimates the average forward momentum lost during the collision, and several simulations used it to model DM long-range interactions~\citep{Vogelsberger:2012ku,Zavala:2012us,Vogelsberger:2013,Vogelsberger:2014pda,Vogelsberger:2015gpr}.
Ref.~\cite{Kahlhoefer:2013dca} proposed an alternative definition of $\sigma_{T}$, namely 
\be
\label{eq:sigmaTpdef}
    \sigma_{\tilde{T}} = 2 \int \frac{d\sigma}{d\Omega}\, (1 - |\cos{\theta}|)\,d\Omega\,,
\ee
to  account for particle indistinguishability, which implies that $d\sigma/d\Omega$ is invariant under $\cos\theta\to-\cos\theta$. The overall factor of $2$ is used to give $\sigma_{\tilde{T}} \approx \sigma$ in the isotropic regime, as done in ref.~\cite{Robertson:2020pxj}.
This new definition of the scattering cross section has been shown to provide better results than $\sigma_T$ in simulations of isolated DM halos~\cite{Robertson:2016qef}.
An even better description of anisotropic scattering within the isotropic approximation seems to be given by the replacement of $\sigma$ with the \textit{viscosity (or conductivity) cross section}~\cite{Tulin:2013teo}
\be
\label{eq:sigmaVdef}
    \sigma_{V} = \frac{3}{2} \int \frac{d\sigma}{d\Omega}\,(1-\cos^2{\theta})\,d\Omega
\ee
because $\sigma_{V}$ takes into account of both forward and backward scatterings at the same time in addition to particle indistinguishability. Again, in order to match $\sigma_{V} \approx \sigma$ in the isotropic regime, the original expression is multiplied by an overall factor of $3/2$.
In the present case, the DM particles are treated as being distinguishable since they have an associated oscillation phase $\varphi$, which might evolve differently during the simulation from particle to particle as discussed in section~\ref{sec:sim}. 

As noticed by ref.~\cite{Tulin:2017ara}, $\sigma_{V}$ differs just by a $\mathcal{O}(1)$ factor from $\sigma_{T}$ and $\sigma_{\tilde{T}}$ for distinguishable particles and apart from the rescaling factor, any of them can be taken as good measures of DM self-interactions, since systematic uncertainties in astrophysical observations are still too big to allow for discrimination between the different prescriptions.
They become identical in the isotropic regime where $v_{\text{rel}} \ll \omega$, which is well satisfied for the choice of model parameters and DM halos considered in this paper (see section~\ref{sec:struc}). We have chosen $\sigma_T$ because of its wide use in SIDM simulations, and we checked \textit{a posteriori} that the use of the other prescriptions lead to the same simulation results for the DM models considered in this paper.

With eq.~\eqref{eq:scattdiff}, which is valid within the Born approximation for $r_m \gg \alpha'$ (recall $\alpha'$ is the dark fine-structure constant), the momentum transfer cross section in eq.~\eqref{eq:sigmaTdef} and its modified version in eq.~\eqref{eq:sigmaTpdef} turn out to be~\citep{Feng:2010zp,Khrapak:2004}
\be
\begin{split}
    \sigma_T &= \sigma_s\,\frac{2\,\omega^4}{v_{\text{rel}}^4}\,\bigg[\ln{\bigg(1 + \frac{v_{\text{rel}}^2}{\omega^2}\bigg)} - \frac{v_{\text{rel}}^2}{\omega^2 + v_{\text{rel}}^2}\bigg]\,, \\
    \sigma_{\tilde{T}} &= \sigma_s\,\frac{4\,\omega^4}{v_{\text{rel}}^4}\,\bigg[2\,\ln{\bigg(1 + \frac{v_{\text{rel}}^2}{2\,\omega^2}\bigg)} - \ln{\bigg(1 + \frac{v_{\text{rel}}^2}{\omega^2}\bigg)}\bigg]\,,
\end{split}
\ee
and, analogously, the viscosity cross section in eq.~\eqref{eq:sigmaVdef} becomes
\be
\label{eq:sigmaV}
    \sigma_V = \sigma_s\,\frac{6\,\omega^4}{v_{\text{rel}}^4}\,\bigg[\bigg(1 + \frac{2\,\omega^2}{v_{\text{rel}}^2}\bigg)\,\ln{\bigg(1 + \frac{v_{\text{rel}}^2}{\omega^2}\bigg)} - 2 \bigg]\,.
\ee
The tests for the code implementations of DM scattering are presented in section~\ref{subsec:tests}.

\subsection{Annihilation}\label{subsec:ann}
We have implemented DM annihilation in a stochastic way similar to what was done for scattering. To the best of our knowledge, the first self-consistent implementation of DM annihilation in N-body simulations was performed in ref.~\cite{Iwanus:2017mue}, followed by ref.~\cite{List:2019jrl}, where the energy released from annihilations to the surrounding gas particles was properly accounted for throughout the simulation.  
We followed a simplified version of this annihilation algorithm, without including the energy transfer between different particle species, because our simulation considers only DM particles $\chi$ and the annihilation products  escape the galaxies without affecting their environment and therefore the observable quantities. 

In detail, the annihilation rate for a DM particle of mass $m_{\chi}$ at position $\vec{r}_i$, moving in an equal-mass particle background of density $\rho (\vec{r})$, is
\be
\label{eq:annrate}
    \Gamma_{i,\,\text{ann}} = \rho (\vec{r}_i) \,\frac{\langle \sigma_{\text{ann}} v \rangle}{m_{\chi}}\,,
\ee
where $\langle \sigma_{\text{ann}} v \rangle$ is the velocity-averaged annihilation cross section. 
Depending on the particle nature of DM, the estimation of $\rho (\vec{r})$ can include all the DM particles in the system (for Majorana fermions or real scalars) or just antiparticles (for Dirac fermions or complex scalars). 
Following the same steps as in section~\ref{subsec:scatt} and introducing the \textit{annihilation search radius} $h_A$, we can generalize eq.~\eqref{eq:annrate} and write the probability for DM particles $i$ and $j$, separated by a distance smaller than $h_A$, to annihilate within the next time step of size $\Delta t$ as
\bea
\label{eq:Pann}
    P_{ij,\,\text{ann}} &=& \bigg(\frac{\langle \sigma_{\text{ann}} v \rangle_{ij}}{m_{\chi}}\bigg)\,\rho_{ij}\,\Delta t\,, \nn\\
    \rho_{ij} &=& 
\begin{cases}
    \frac{3\,m_p}{4\, \pi h_A^3} \qquad 0 \leq \tilde{r} \leq h_A \\
    0 \qquad \qquad \quad\,\,\,\, \tilde{r} > h_A
\end{cases}
\eea
where $\tilde{r} \equiv |\vec{r}_i - \vec{r}_j|$ and $\rho_{ij}$ is the target density, which is estimated using a top-hat kernel function.
The velocity-averaged annihilation cross section $\langle \sigma_{\text{ann}} v \rangle_{ij}$ depends generally on the relative velocity $v_{\text{rel}} = |\vec{v}_i - \vec{v}_j|$ between particles $i$ and $j$. In the non-relativistic limit, valid in the context of galaxies, it is usually expanded in powers of $v_{\text{rel}}$ as
\be
\langle \sigma_{\text{ann}} v \rangle_{ij} \simeq \sigma_{\text{ann},\,s} + \sigma_{\text{ann},\,p}\,v_{\text{rel}}^2 + \mathcal{O}(v_{\text{rel}}^4)\,,
\ee
where $\sigma_{\text{ann},\,s}$ and $\sigma_{\text{ann},\,p}$ are constants corresponding to the $s$-wave and the $p$-wave annihilation terms, respectively. For the models considered in this paper, only $\sigma_{\text{ann},\,s}$ contributes to $\langle \sigma_{\text{ann}} v \rangle_{ij}$, which is given by what we refer to as $\langle \sigma v \rangle_a$ in eq.~\eqref{eqxsects}. This makes the annihilation probability in eq.~\eqref{eq:Pann} velocity-independent. 
As in the scattering case, the annihilation search radius $h_A$ entering the density $\rho_{ij}$ is in principle a free parameter that should be chosen to reproduce analytical results. As we will discuss in section~\ref{subsec:tests}, $h_A = \epsilon$ turns out to be the best choice (recall $\epsilon$ is the gravitational softening length). 
To see which particles annihilate at each time step, the probability in eq.~\eqref{eq:Pann} is computed for each pair of nearby simulation particles and compared to a random number. If the latter is below than the former, annihilation happens and the two particles in the event are removed from the system.

\subsection{Validation tests}
\label{subsec:tests}
The simplest test for both our scattering and annihilation algorithms is a uniform cube of $N_c$ particles moving through a background of stationary particles with constant number density $n_b$. All the particles making up the simulated system have the same mass $m_{p}$. To allow simple predictions, we impose that the cube particles move with constant speed $v_0$ along the same axis, they can scatter with constant cross section $\sigma_p$ at most once, and gravity is turned off~\cite{Robertson:2016xjh}.

The scattering or annihilation rate for each simulation particle in the cube is $\Gamma = n_b\, m_p (\sigma v_0/m_{\chi})$, where $m_{\chi}$ and $\sigma$ are respectively the mass and the cross section of the physical particles, whose properties have been translated into those  of the simulation particles via $\sigma_p = m_p\,(\sigma/m_{\chi})$. Naively, we can estimate the expected number of interactions after a time $t$ as
\be
    N_{\text{exp}} = N_c\, \Gamma\, t = N_c\,n_b\, m_p (\sigma v_0/m_{\chi})\,t\,.
\ee
This expression has however several limitations because not only it does not take into account that not all the particles in the cube have interacted between zero and time $t$ but also that those that have scattered or annihilated could do it just once.
These effects are properly captured by considering how the number of cube particles changes with time, which can be described by
\be
    \frac{d N_c}{dt} = - \Gamma\,t\,.
\ee
With this in mind, the expected number of scattering or annihilating particles in the simulation after a time $t$ can be better expressed by
\be
\label{eq:Nexp}
    N_{\text{exp}} = N_c (0)\,\{1 - \exp{[-n_b\,m_p\,(\sigma v_0/m_{\chi})\,t]}\}\,,
\ee
where $N_c (0)$ is the initial number of particles in the cube.

\begin{figure*}[t]
  \centerline{\includegraphics[height=0.35\textwidth]{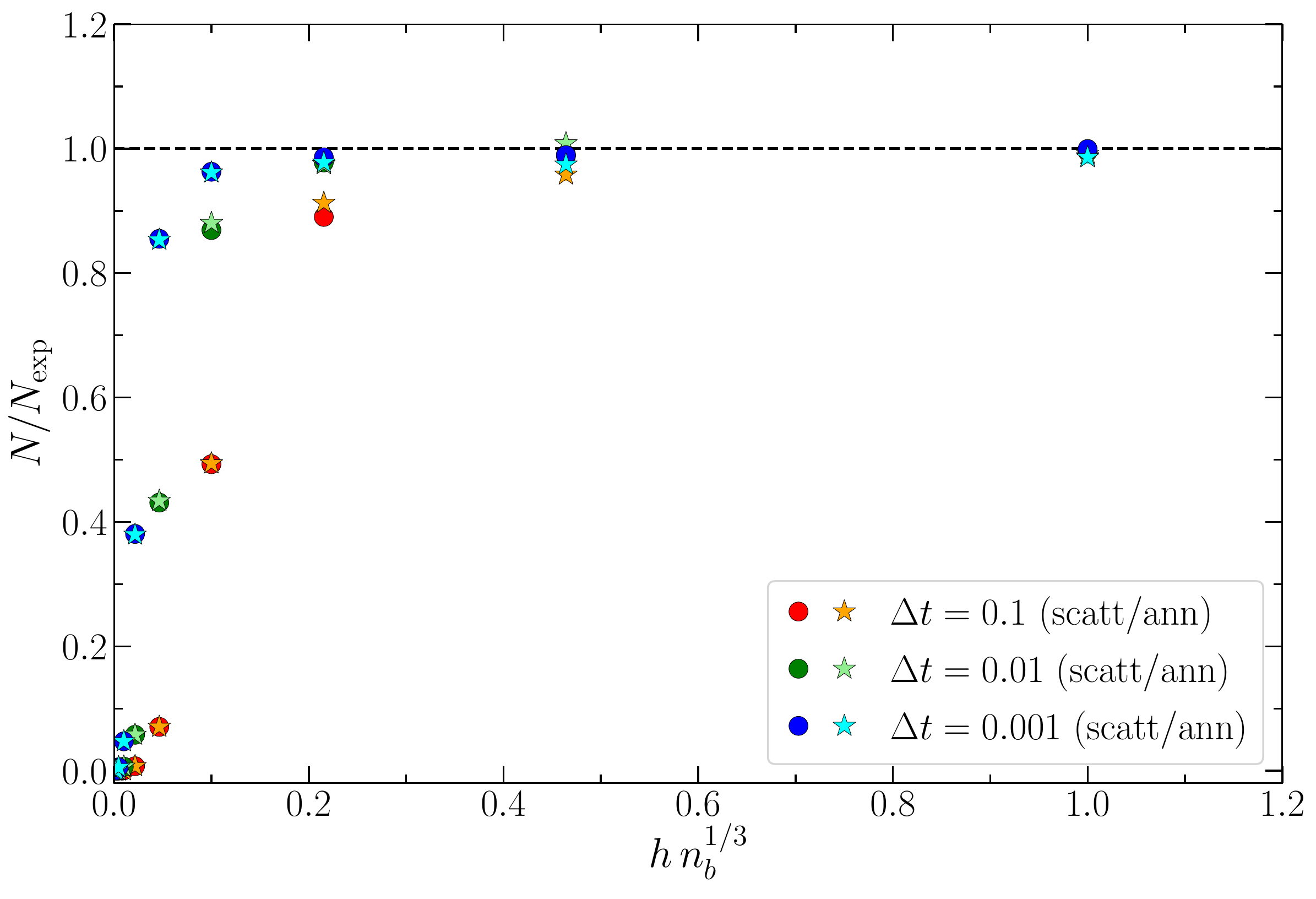}
  \includegraphics[height=0.345\textwidth]{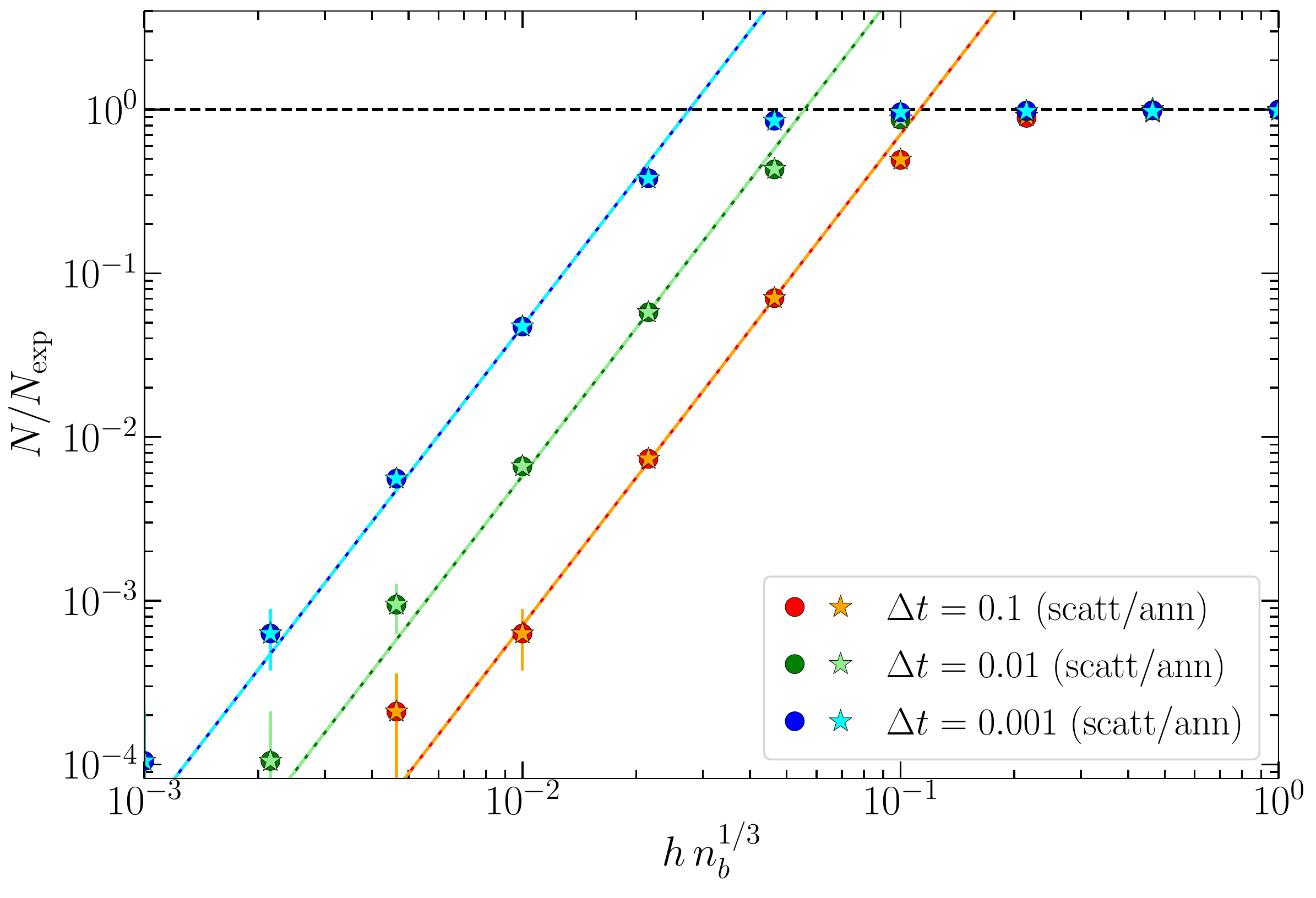}}
  \caption{Ratio of the number of cube particles scattering (annihilating) in our test simulations to the expected number of the same events given by eq.~\eqref{eq:Nexp}, as a function of the scatter (annihilation) search radius $h$. Points correspond to the results of simulations where only scattering was turned on, whereas stars are used for simulations in which only annihilation took place.
  Different color points represent different choices of the simulation time step $\Delta t$, which is measured in units of $\ell/v_0$ with $\ell$ being the side of the cube.
  The left and right plots show the same data with different axis scales, linear on the left and logarithmic on the right. The solid (dashed) lines in the right panel show $N \propto h^3$, which is the result expected from ``probability saturation,'' as originally noticed by ref.~\cite{Robertson:2016xjh}. The error bars show the $1\sigma$ uncertainty assuming $N$ is Poisson distributed.}
\label{fig:test1cmp}
\end{figure*}

The comparison between $N_{\text{exp}}$ and the number $N$ of cube particles that have scattered (annihilated) in our test simulations is plotted in Fig.~\ref{fig:test1cmp} as a function of the scatter (annihilation) search radius $h$. Here, each point corresponds to a simulation where only scattering was turned on, while stars are used for simulations in which only annihilation was active. 
As already noticed in refs.~\citep{Robertson:2016xjh,Rocha:2012jg}, $N$ falls below than expected only for $h$ smaller than $20\%$ of the mean background interparticle separation, but this minimum value depends heavily on the chosen time step $\Delta t$~\cite{Robertson:2016xjh}, as confirmed by Fig.~\ref{fig:test1cmp}.
The $h$-dependence of the number $N$ of scattering or annihilating particles in simulations arises in the ``probability-saturated'' regime, where the probability for a particle to scatter or annihilate within a time step, given by eqs.~\eqref{eq:Pscatt} or~\eqref{eq:Pann}, becomes greater than unity.
The latter probability enters the definition of $N$, which depends on it ($\propto h^{-3}$) and on the number of neighbouring particles that a particle finds at each time step ($\propto h^3$). This makes $N$ generally insensitive to $h$, except in the probability-saturated regime, where $N \propto h^3$ as shown by the solid and dashed lines in the right panel of Fig.~\ref{fig:test1cmp}.
In order to avoid probability saturation, a shorter time step should be used when using smaller $h$, since the probability for a pair of particles to scatter or annihilate is proportional to $\Delta t / h^3$.

Although ref.~\cite{Robertson:2016xjh} suggested that probabilities exceeding unity within each time step should not appear in SIDM simulations powered by \texttt{GADGET} for reasonable values of $\sigma / m_{\chi}$, we decided to implement a time-step criterion. 
In particular, similar to what was done in ref.~\cite{Koda:2011yb}, an individual particle time step $\Delta t$ is modified by rearranging eq.~\eqref{eq:Pscatt} as
\be
    \Delta \tilde{t} = \frac{4\,\pi h_S^3}{3\, m_p}\,\frac{m_{\chi}}{\sigma \,v_{\text{rel}}}\,P_{\text{max}}
\ee
if the probability of interaction for any pair involving such a particle was greater than $P_{\text{max}} = 0.1$ during the last tree-walk. This restriction is important only for scattering particles, because annihilating particles are removed from the system as soon as the interaction takes place and therefore their time step becomes meaningless. Although limiting the individual time step makes the Monte Carlo method more computationally costly, it allows for suppression of the probability of having multiple scatterings in the same $\Delta t$, which is an important issue in SIDM simulations.

\begin{figure*}[t]
  \centerline{\includegraphics[height=0.35\textwidth]{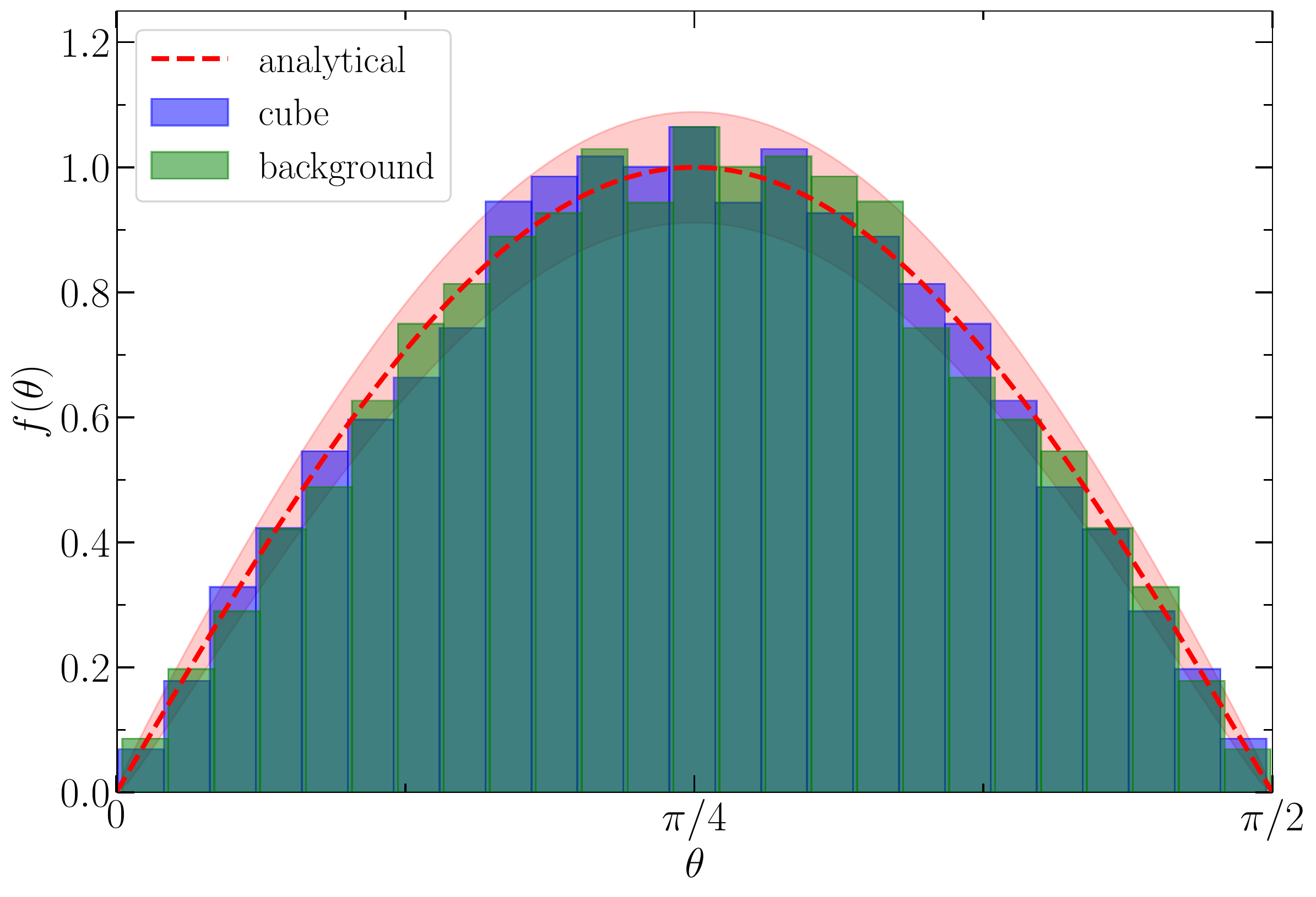}
  \includegraphics[height=0.35\textwidth]{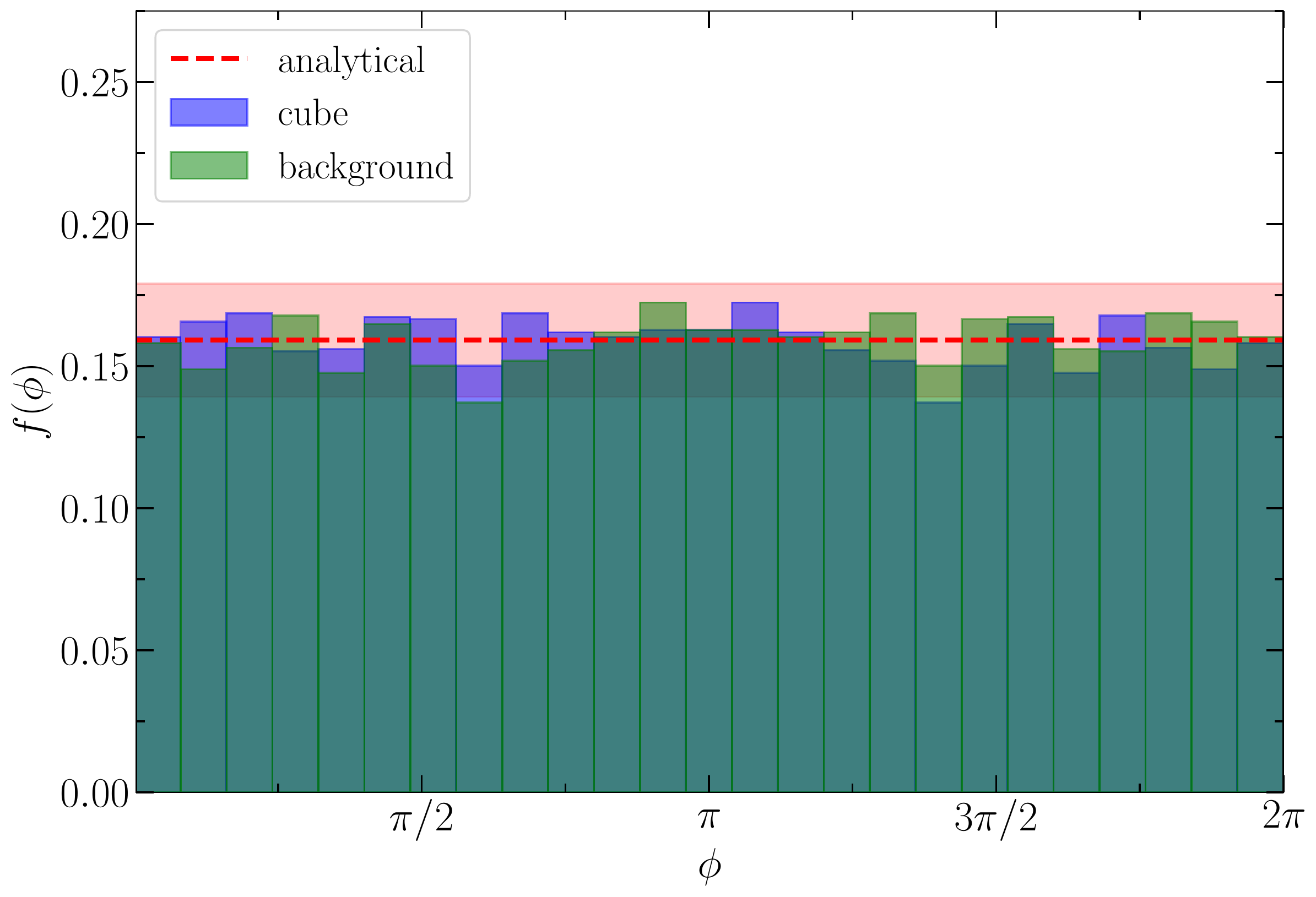}}
  \includegraphics[height=0.35\textwidth]{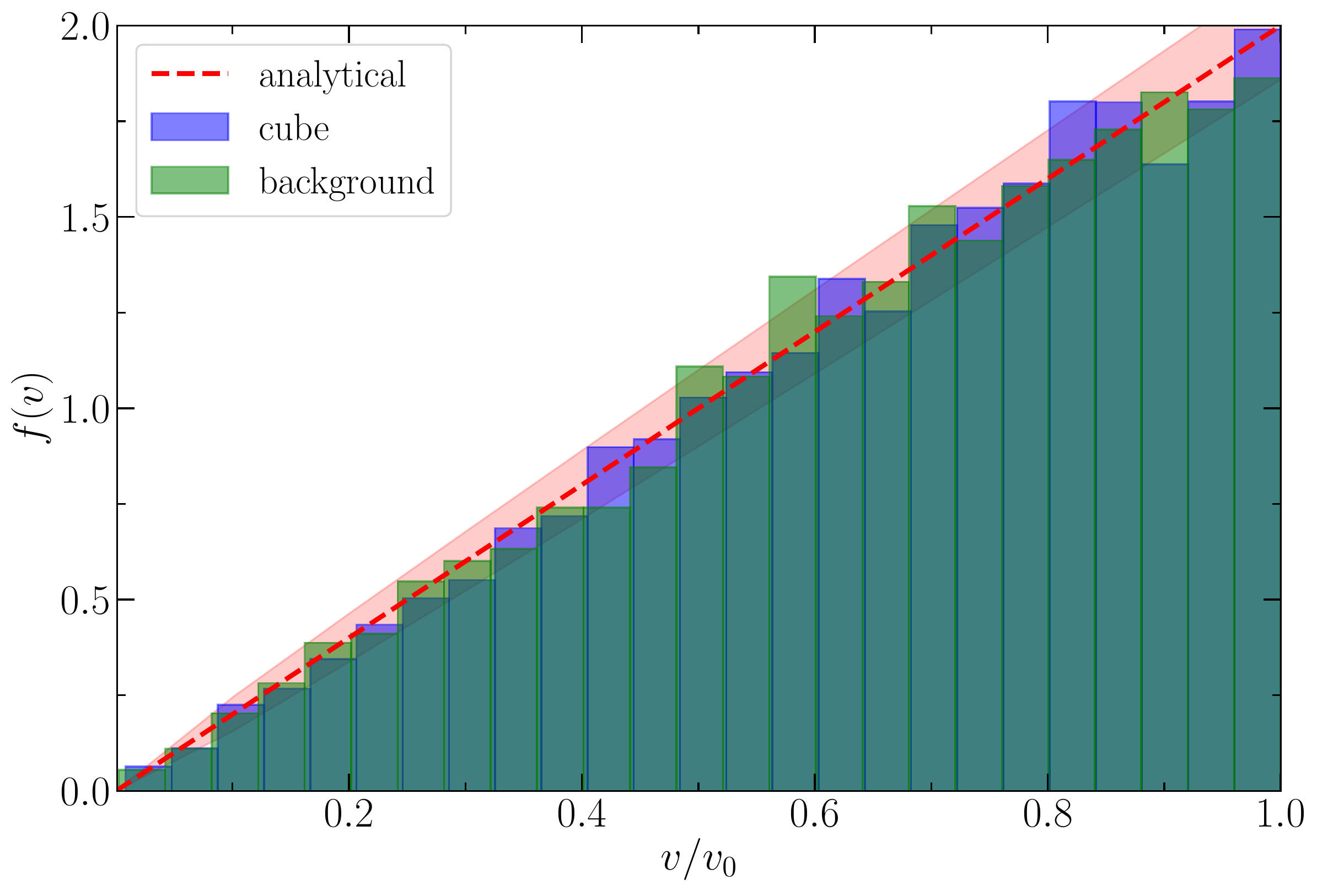}
  \caption{Distributions of polar and azimuthal angles (top) and velocity magnitude (bottom) of scattered particles in one of our test simulations. The expected results are the red dashed lines, and their $1\sigma$ uncertainty regions are shaded red, computed assuming that the number of particles in each bin is Poisson distributed.}
\label{fig:test1kin}
\end{figure*}

For simulations with only scattering, the directions and velocities of the scattered particles can also be compared to the expected normalized distributions. The latter can be obtained by transforming the differential cross section from the centre-of-mass frame of the collision to the simulation frame. 
For isotropic scatterings, these distributions turn out to be the same for both background and cube particles and take a simple form, with $f(\theta) = \sin{2\theta}$, $f(\phi) = (2\pi)^{-1}$ and $f(v) = 2 v$, where the latter is valid for $v \leq v_0$, and becomes  zero otherwise.
Fig.~\ref{fig:test1kin} shows the comparison between the expected distributions and those reconstructed from one of our test simulations, which agree to within $1\sigma$ uncertainty.

Another important test of our code is to check whether particle scattering and annihilation are well modeled in isolated DM halos. 
We focus on halos with Hernquist profile~\cite{Hernquist:1990be}, whose density distribution can be written as
\be
\label{eq:Hernquist}
    \rho (r = x\,a) = \frac{\rho_H}{x\,(1 + x)^3}\,,
\ee
where $\rho_H \equiv M / (2\pi a^3)$, $M$ is the total halo mass and $a$ the scale radius. We have chosen this type of halo because it has a finite mass without need of truncation and the phase-space distribution function $f(E)$, with $E$ being the particle energy, has an analytic form~\cite{Hernquist:1990be}.
The latter property allows to easily generate equilibrium initial conditions for N-body simulations and compute scattering or annihilation quantities analytically.
The initial conditions for particle positions and velocities making up our test halos have been generated randomly from the Hernquist distribution function $f(E)$ using the von Neumann rejection method~\cite{Neumann:1951}, as originally done in ref.~\cite{Aarseth:1974}. To prevent centroid motion of the generated DM halo during the simulation, we set its initial centre-of-mass position and velocity to zero by an overall boost.

\begin{figure*}[t]
  \centerline{\includegraphics[height=0.35\textwidth]{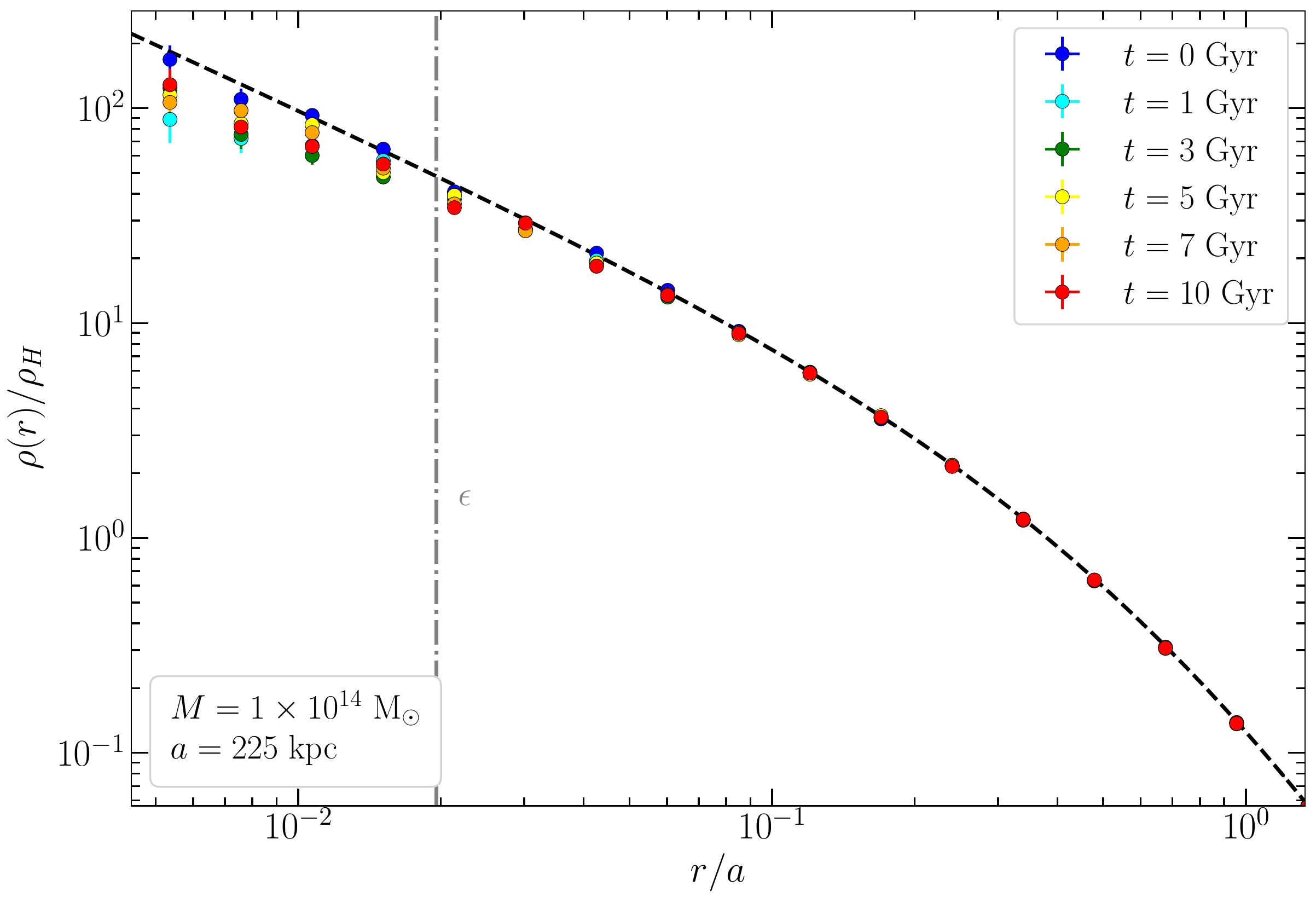}
  \includegraphics[height=0.35\textwidth]{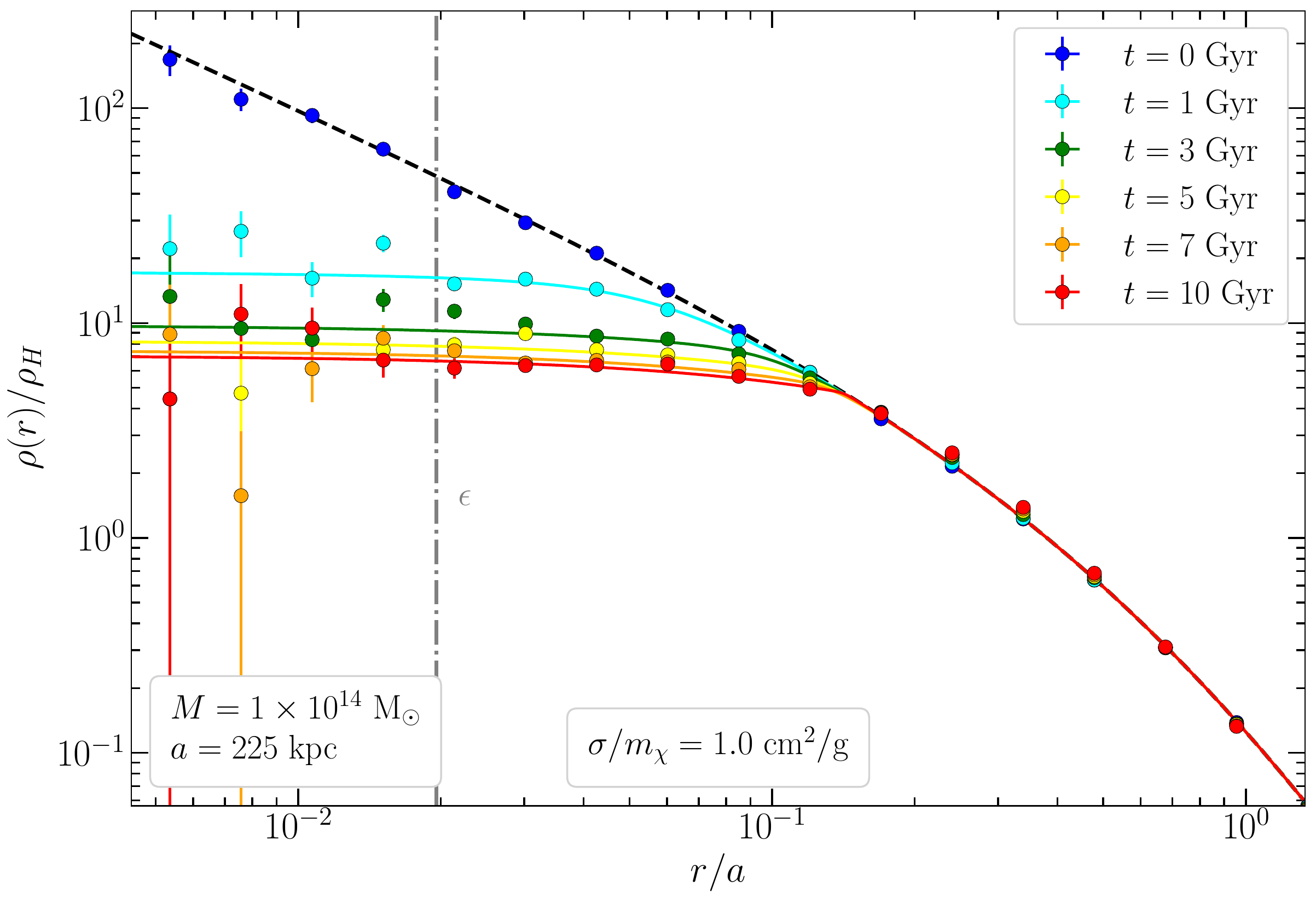}}
  \includegraphics[height=0.35\textwidth]{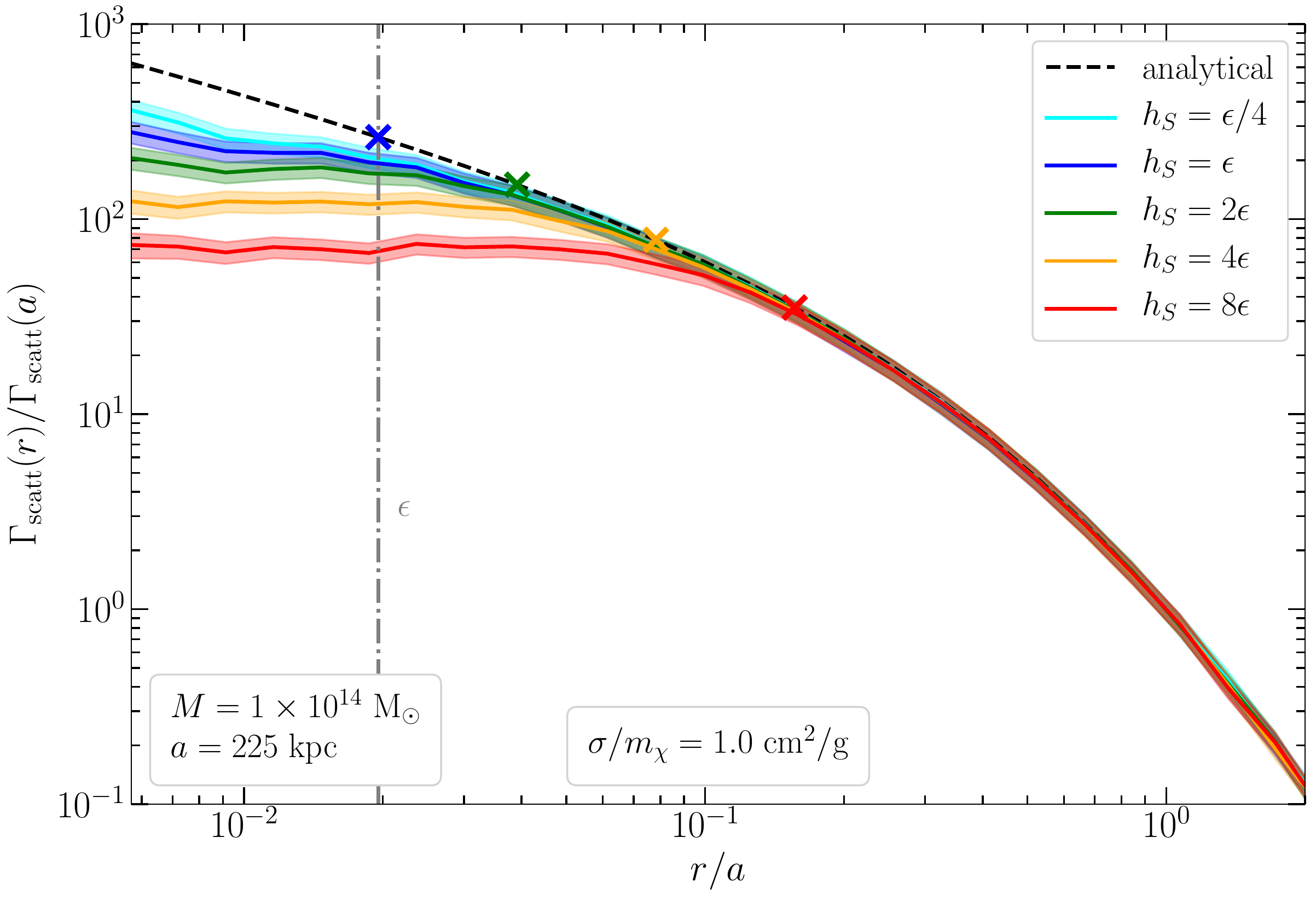}
  \caption{Top: Radial density profile of a DM halo with Hernquist mass $M = 10^{14}\,\,\text{M}_{\odot}$ and radius $a = 225\,\,\text{kpc}$ as a function of the distance $r$, in units of $\rho_H = M / (2\pi a^3)$ and $a$ respectively. The black dashed line displays the original density profile in eq.~\eqref{eq:Hernquist}.
  The left panel shows the halo stability across a time window of $10$ Gyr. 
  The right panel shows how the density profile evolves with time assuming particle scattering with constant $\sigma / m_{\chi} = 1.0\,\,\text{cm}^2/\text{g}$. Here we chose the scatter search radius as $h_S = \epsilon$. The solid lines correspond to the best-fit cored-Hernquist profiles, given by eq.~\eqref{eq:coredHernquist}, where $r_c$ and $\beta$ are left as free parameters.
  The $1\sigma$ error bar for each data point is computed assuming that the number of particles in each bin is Poisson distributed.
  Bottom: Extracted scattering rate per particle for the same Hernquist profile DM halo after $3$ Gyr, for different choices of the scatter search radius $h_S$, and its comparison with the theoretical expectation. Although the simulations were run with $\sigma/m_{\chi} = 1.0\,\,\text{cm}^2/\text{g}$, the result is independent of the scattering cross section since a ratio is considered.
  The colored crosses along the analytical curve correspond to the radius equal to $h_S$. The $1\sigma$ uncertainty for each colored line is computed assuming that the number of particles in each bin is Poisson distributed and displayed with a same-color shaded region.
  In all three panels, the gray dot-dashed vertical line shows the position of the gravitational softening length $\epsilon$ used in all simulations for the considered halo. We used a time-integration parameter of $\eta = 0.005$ and tree-force accuracy of $\alpha = 0.0012$ in each simulation run.}
\label{fig:test2scatt}
\end{figure*}

Since we consider a large range of halo masses in this paper, ranging from $10^{10}$ to $10^{15}\,\,M_{\odot}$, we tested our code with simulated Hernquist halos having different values of $M$ and $a$, finding good agreement between the analytical expectations and the simulation outcomes for all of them. 
As a prototype example, we focus here just on an isolated Hernquist halo with total mass $M = 10^{14}\,\,M_{\odot}$ and scale radius $a = 225\,\,\text{kpc}$.
The simulations for such a halo were run with $N = 128^3$ particles, each having mass $m_p \simeq 4.8 \times 10^8\,\,M_{\odot}$, and the Plummer-equivalent gravitational softening length was set to $\epsilon = 4.4\,\,\text{kpc}$. 
We evolved the generated halos with collisionless DM first to study their stability and to check whether cores can form as numerical artifacts. The results for our prototype halo as a function of the simulation time are shown in the top left plot of Fig.~\ref{fig:test2scatt}. 
For a suitable choice of time-integration and tree-force accuracy parameters, $\eta = 0.005$ and $\alpha = 0.0012$ respectively, the density and velocity distributions remained unchanged except for the formation of a small constant core with size similar to the gravitational softening length $\epsilon$, as observed by ref.~\cite{Robertson:2016xjh}.  

When DM scattering is turned on, these cores quickly become larger because particles scatter mostly in high density regions until the cores settle to a size that is independent of the value of $\sigma/m_{\chi}$~\cite{Kochanek:2000pi}. The right panel of Fig.~\ref{fig:test2scatt} shows the results for the density profile of our prototype halo at different simulation times for a constant scattering cross section per unit DM mass of $1.0\,\,\text{cm}^2/\text{g}$.
The resulting profiles are well fitted by a cored-Hernquist profile of the form~\cite{Robertson:2016qef}
\be
\label{eq:coredHernquist}
    \rho(x) = \frac{\rho_H}{(1 + x)^3} \frac{1}{[x^{\beta} + (r_c / a)^{\beta}]^{1/\beta}}\,,
\ee
where $x \equiv r /a$, $r_c$ is the core-radius and $\beta$ is an index controlling the sharpness of the transition from $\rho \propto x^{-1}$ to a constant density.
Leaving $r_c$ and $\beta$ as free parameters in the fit, we found that their best-fit values across a time evolution of $10$ Gyr are $r_c / a \in (0.06, 0.14)$ and $\beta \in (3.5, 62.0)$, in agreement with  refs.~\citep{Kochanek:2000pi,Robertson:2016qef}.

The results shown in the right panel of Fig.~\ref{fig:test2scatt} have been obtained by choosing the scatter search radius equal to the gravitational softening length in the computation of the scattering probability, given by eq.~\eqref{eq:Pscatt}. This was done in light of the result shown in the bottom panel of Fig.~\ref{fig:test2scatt}, where the scattering rate extracted from simulations is compared to the analytical prediction.
In particular, the latter can be computed by integrating eq.~\eqref{eq:scattrate} over the velocity distribution function, obtaining the average rate for particles at position $\vec{r}$~\cite{Robertson:2016xjh}
\be
\label{eq:scattrate_sim}
   \Gamma_{\text{scatt}} (\vec{r}) = \frac{\langle \sigma\,v_{\text{pair}} \rangle (\vec{r}) \,\rho(\vec{r})}{m_{\chi}}\,,
\ee
where $\rho(\vec{r})$ is given by eq.~\eqref{eq:Hernquist} and $\langle \sigma\,v_{\text{pair}} \rangle = \sigma\, \langle v_{\text{pair}} \rangle$ if the scattering cross section is velocity-independent. The mean pairwise particle velocity $\langle v_{\text{pair}}\rangle$ can be computed from the one-dimensional velocity dispersion $\sigma_{1D}$ for Hernquist halos~\cite{Hernquist:1990be}
\bea
\sigma_{1D}^2 &=& \frac{G\, M}{12\, a} \bigg\{12\, x\, (1+x)^3\,\ln{\bigg(\frac{1 + x}{x}\bigg)}\nn\\
&-& \frac{x}{1+x} \Big[25 + 52\,x + 42\,x^2 + 12\,x^3 \Big] \bigg\}
\eea
as $\langle v_{\text{pair}}\rangle = (4/\sqrt{\pi})\,\sigma_{1D}$, where we have assumed the velocities are isotropic and follow a Maxwell-Boltzmann distribution.
From simulations, the scattering rate per particle as a function of the radius $r$ can be estimated as the ratio between the number of
particles that have scattered within the radial bin including $r$, and the time averaged number of particles within the same bin~\cite{Robertson:2016xjh}.
To avoid any modification of the density profile and velocity distribution due to DM self-interactions that would lead the scattering rate not to follow the analytic prediction, we turned off the change in the momenta of the scattered particles in the simulations used in the bottom panel of Fig.~\ref{fig:test2scatt}. 

Such a plot clearly shows that our code accurately reproduces the scattering rate within the halo except for distances less than $h_S$, where the extracted rate falls below the analytic result. This can be explained by the fact that the scatter search radius acts as a scale below which the particle density entering the scattering probability becomes smooth, preventing a faithful reconstruction of the real density. 
Although it suggests choosing a small value of $h_S$ in order to correctly capture the scattering dynamics in small high-density regions, there is a natural lower bound for $h_S$ set by the gravitational softening length $\epsilon$. 
As we have found in simulation runs for collisionless DM, the density gets affected by a small core of size of the order of the softening length, which arises from smoothing the gravitational potential at $r < \epsilon$ and thus the particle distribution at these scales. 
This explains why $\Gamma_{\text{scatt}}$ ceases to change for scales smaller than $\epsilon$ in simulations where $h_S < \epsilon$, as displayed in the bottom panel of Fig.~\ref{fig:test2scatt}.
Therefore, reducing $h_S$ to values below $\epsilon$ cannot improve the agreement between the extracted rate and the analytic result, but rather it tends to create probability saturation problems that can be solved by choosing smaller time steps or a time-step delimiter, as already discussed above.
Considering all these factors and the results in the bottom panel of Fig.~\ref{fig:test2scatt}, we find that setting $h_S = \epsilon$ provides the best reconstruction of scattering dynamics at low scales, and it limits the occurrence of events where the scattering probability becomes greater than unity.

\begin{figure*}[t]
  \centerline{\includegraphics[height=0.35\textwidth]{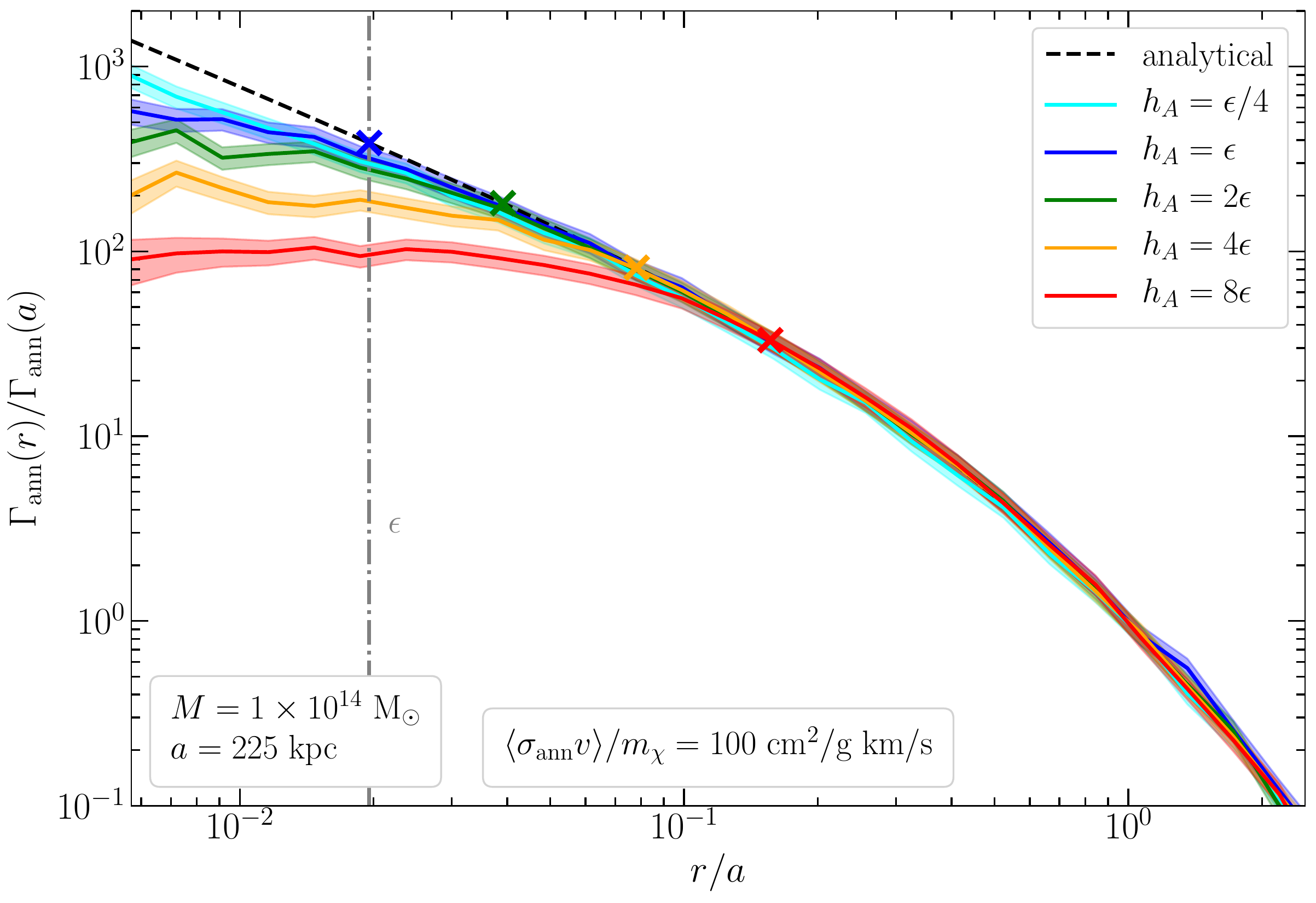} 
  \includegraphics[height=0.35\textwidth]{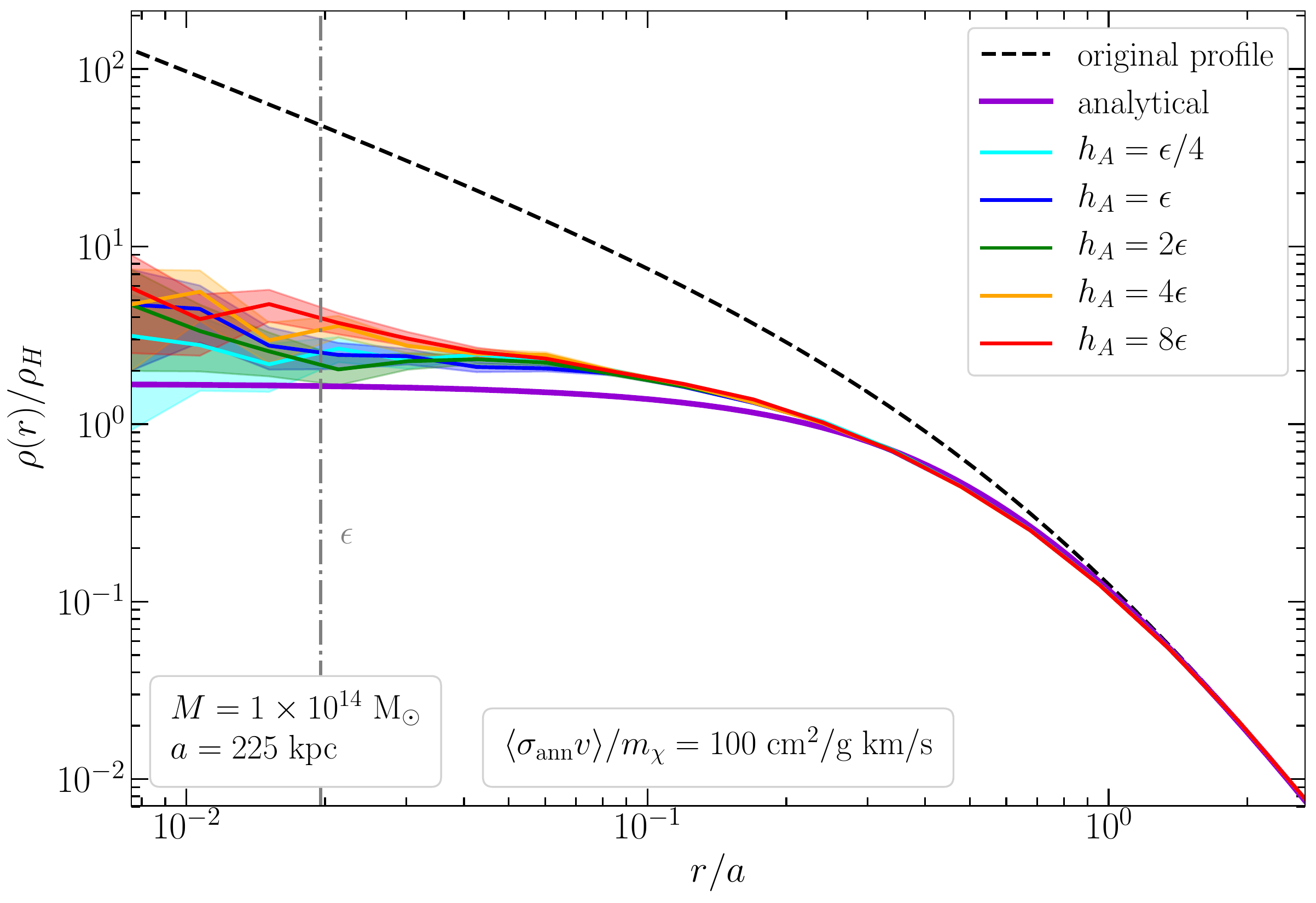}}
  \includegraphics[height=0.35\textwidth]{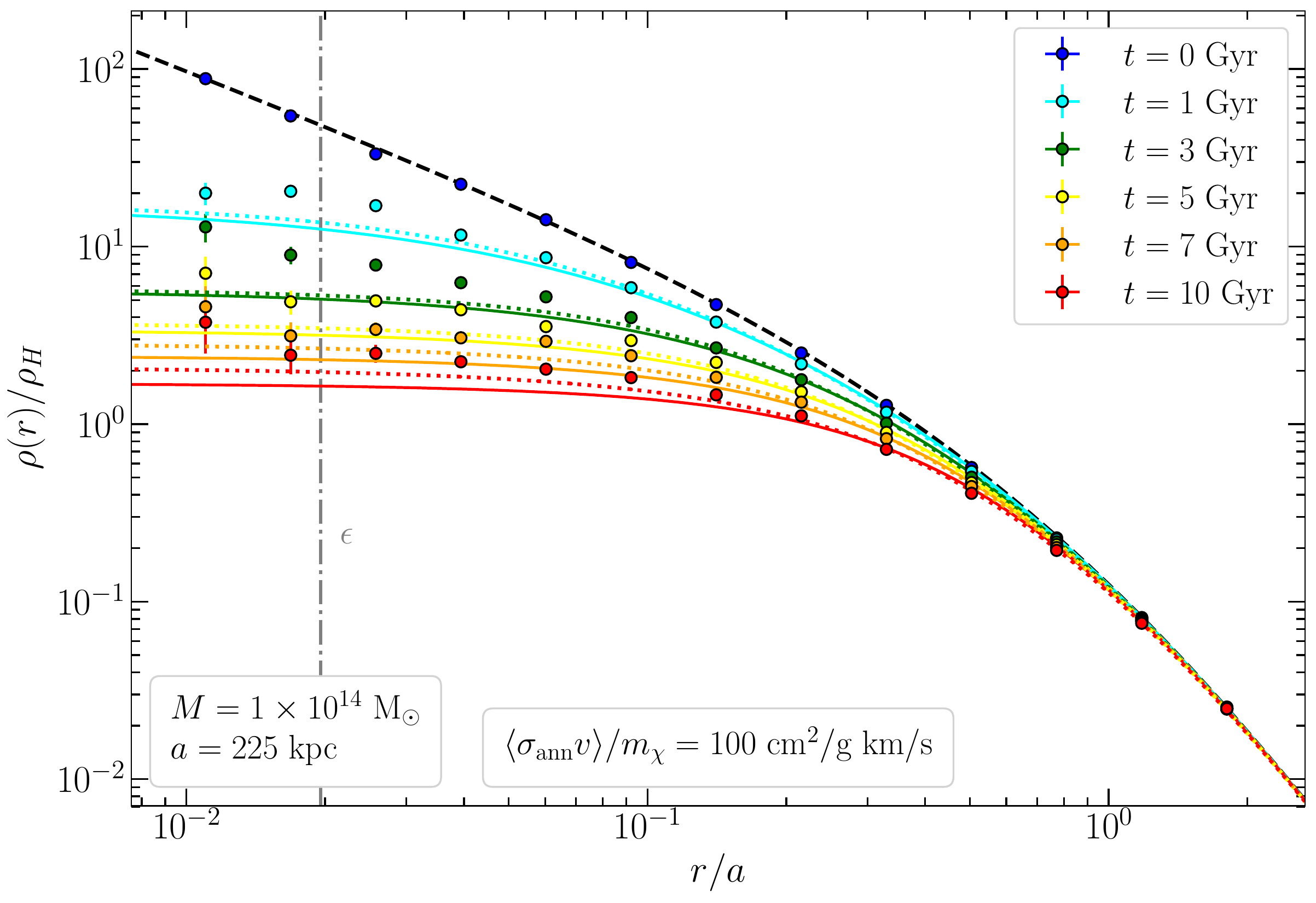}
  \caption{Top left: Similar to the bottom panel of Fig.~\ref{fig:test2scatt}, but for the extracted annihilation rate per particle and different choices of the annihilation search radius $h_A$. The simulations for the prototype Hernquist halo were run with $\langle \sigma_{\text{ann}} v \rangle/m_{\chi} = 100\,\,\text{cm}^2/\text{g}\,\,\text{km}/\text{s}$.
  Top right: Radial density profile of the prototype Hernquist halo undergoing DM annihilation for $10$ Gyr, for different choices of $h_A$. The black dashed line displays the original density profile in eq.~\eqref{eq:Hernquist}. The violet solid line corresponds to the theoretical prediction given by eq.~\eqref{eq:rhoann_time} at $(t - t_{\text{ini}}) = 10$ Gyr.
  Bottom: Similar to the right panel of Fig.~\ref{fig:test2scatt}, but for particle annihilation with the same constant velocity-averaged cross section considered above.
  Here we chose the annihilation search radius as $h_A = \epsilon$. Each colored solid curve shows the analytical expectation given by eq.~\eqref{eq:rhoann_time} at the time of the corresponding same-color data points.
  The dotted lines represent the best-fit cored-Hernquist profiles, given by eq.~\eqref{eq:coredHernquist}, where $r_c$ and $\beta$ are left as free parameters.}
\label{fig:test2ann}
\end{figure*}

With the same Hernquist halos it is possible to also test the goodness of our annihilation algorithm. 
For instance, the annihilation rate extracted from simulations can be compared to the analytic result, which is given by eq.~\eqref{eq:scattrate_sim} with the replacement of $\langle \sigma\,v_{\text{pair}} \rangle$ with $\langle \sigma_{\text{ann}}\,v \rangle$. The extracted annihilation rate as a function of the radius $r$ has been obtained in the same way as done for scattering events, namely as the ratio between the number of annihilated particles located in the radial bin including $r$ and the time averaged number of particles within the same bin. To allow for a direct comparison to the analytical result we did not remove the annihilated particles from the system, but the annihilation algorithm was used to count and localize the particles that actually annihilate. 
The results from this comparison are shown in the top left panel of Fig.~\ref{fig:test2ann}, where we have chosen a constant ($s$-wave) velocity-averaged annihilation cross section per unit DM mass of $ 100\,\,\text{cm}^2/\text{g}\,\,\text{km}/\text{s}$.
Similarly to what was observed in the scattering case, choosing an annihilation search radius $h_A$ equal to the gravitational softening length $\epsilon$ provides the best reconstruction of the annihilation rate. This is because the particle distribution at scales below $h_A$ and $\epsilon$ are smoothed, leading to a decreased annihilation rate compared to the true unsmoothed one. 

The effect that DM annihilation has on the time evolution of the halo density profile was first studied by ref.~\cite{Kaplinghat:2000vt}, which proposed the following analytic formula for $\rho (r, t)$
\be
\label{eq:rhoann_time_eq}
    \frac{d}{dt} \bigg(\frac{\rho(r, t)}{\rho_A}\bigg) = - \frac{1}{t_0}\,\bigg(\frac{\rho (r, t)}{\rho_A}\bigg)^2\,,
\ee
where $\rho_A \equiv m_{\chi} / (\langle \sigma_{\text{ann}}\,v \rangle\,t_0)$ and $t_0$ is the age of the universe today.
This equation comes directly from the definition of the DM annihilation rate and can be easily adapted to our simulation setup, where the annihilation probability is given by eq.~\eqref{eq:Pann}, by replacing $\rho_A$ with $\rho_A / 2$. The factor $1/2$ arises because the halo density is reduced by two units of the simulation particle mass in each annihilation event.
The general solution of eq.~\eqref{eq:rhoann_time_eq} is given by
\bea
\label{eq:rhocore_time}
    \rho (r, t) &=& \bigg[\frac{1}{\rho_{\text{core}} (t)} + \frac{1}{\rho (r, t_{\text{ini}})} \bigg]^{-1}\,,\nn\\
    \rho_{\text{core}} (t) &\equiv& \frac{m_{\chi}}{2\,\langle \sigma_{\text{ann}}\,v \rangle\,(t - t_{\text{ini}})}\,,
\eea
where the latter definition is valid for our simulations and $t_{\text{ini}}$ is the initial time, which can be taken as the time when the simulation starts and the density profile is $\rho (r, t_{\text{ini}})$. One observes  that the core density $\rho_{\text{core}} (t)$ is generally greater than $\rho_A$ and the equality is reached when $(t - t_{\text{ini}}) = t_0$.

Focusing on just DM halos with initial Hernquist profile and taking $t_{\text{ini}} = 0$, we find that the density profile of halos undergoing DM annihilation should be described after a time $t$ by
\be
\label{eq:rhoann_time}
    \rho (x, t) = \frac{\rho_H}{x\,(1 + x)^3 + \rho_H / \rho_{\text{core}} (t)}
\ee
with $x \equiv r /a$, which has been obtained by substituting eq.~\eqref{eq:Hernquist} into eq.~\eqref{eq:rhocore_time}. The halo density is therefore characterized by a core of constant density $\rho_{\text{core}} (t)$ at small scales.
The bottom panel of Fig.~\ref{fig:test2ann} shows the evolution of the density profile of the prototype Hernquist halo in the simulation at different times $t$ and its comparison with the theoretical prediction. 

Although the data points and the solid lines given by eq.~\eqref{eq:rhoann_time} match very well at large radii, there is some disagreement at small distances. These differences are not eliminated by changing the annihilation search radius, as shown by the right panel of Fig.~\ref{fig:test2ann}, where the analytical prediction is compared to the simulation outcome at $t = 10$ Gyr for different values of $h_A$. A graphical comparison between the analytical curve and the colored lines suggests that eq.~\eqref{eq:rhoann_time} does not provide a good fit to the simulation data at low scales, independently of the choice of $h_A$, although it accurately captures the physics at large radii. We confirmed this observation by fitting the simulation data with eq.~\eqref{eq:rhoann_time}, in which $\rho_{\text{core}} (t)$ was left as a free parameter, leading to agreement within $1\sigma$ between the best fit value and that computed by eq.~\eqref{eq:rhocore_time} at different times $t$, independently of the value of $h_A$.

In analogy to the case of scattering, as shown in the top right panel of Fig.~\ref{fig:test2scatt}, we investigated the cored-Hernquist profile function given by eq.~\eqref{eq:coredHernquist} for fitting the annihilation data displayed in the bottom panel of Fig.~\ref{fig:test2ann}, leaving $r_c$ and $\beta$ as free parameters. The results of this fit are shown with colored dotted lines in the same panel, with best-fit values $r_c /a \in (0.06, 0.48)$ and $\beta \in (1.2, 2.0)$.  They do not give significant improvement with respect to eq.~\eqref{eq:rhoann_time}.

\end{subappendices}

\section{Upper bounds for the DM-number violating mass}\label{app:dm}

In this appendix, we will investigate the upper bound on the Majorana mass $\delta m$. This parameter determines the timescale on which annihilations recouple after the initial asymmetric dark matter freezeout epoch. For convenience, we define $\alpha = Y_{11}-Y_{22}, \beta=Y_{12}-Y_{21}, \theta=Y_{12}+Y_{21},$ $\gamma=Y_{11}+Y_{22}$, $s=\bar{s} m_\chi^3/x^3$, $H=\kappa m_\chi^2/x^2$, where $\bar{s}=\frac{2\pi^2}{45} g_{\ast s}$, $\kappa = \frac{1.66}{M_p} \sqrt{g_\ast}$.

\begin{subappendices}
\subsection{Flavor-blind interactions}

From the Boltzmann equations after freeze-out we get
\begin{equation}
    x^2 \beta' - \left( \frac{ \bar{s} \left< \sigma v \right>_a   m_\chi}{\kappa} \eta_{DM} \right) \beta - \left( \frac{2i\delta m}{\kappa \, m_{\chi}^2} \eta_{DM} \right) x^3 = 0,
\end{equation}
with $\beta (\bar{x})=0$ as initial condition. Here we used $\alpha \approx Y_{11} \approx \eta_{DM}$ (this does not imply $\alpha'=0$), as we are working before the moment of residual annihilations. The solution to this equation can be approximated to $\beta (x) \approx i Bx(A+x)/2$, where
\begin{equation}
    A \equiv \frac{  \bar{s} \left< \sigma v \right>_a \, m_{\chi} }{\kappa}\eta_{DM},  \hspace{0.25in}  B \equiv \frac{2 \, \delta m }{\kappa \, m_{\chi}^2}\eta_{DM}.
\end{equation}
Plugging this result into the Boltzmann equations for $Y_{11}$ and $Y_{22}$, we get
\begin{eqnarray}
    16 \, \eta_{DM} \, Y'_{11} &=& B^2 (A+x) \left( A(A+x) - 4 \right) , \nonumber \\
    16 \, \eta_{DM} \, Y'_{22} &=& B^2 (A+x) \left( A(A+x) + 4 \right), \nonumber \\
\end{eqnarray}
with initial conditions $Y_{11}(\bar{x}) = \eta_{DM}$ and $Y_{22}(\bar{x}) = 0$.

Taking the solutions for $Y_{11}$ and $Y_{22}$ and solving for $x$ when $Y_{11}(\bar{\bar{x}})=Y_{22}(\bar{\bar{x}})$ gives
\begin{equation}
    \bar{\bar{x}} = 1.53 \frac{m_\chi}{\sqrt{\delta m \, M_{p}}} g_{\ast}^{1/4}.
    \label{eqD4}
\end{equation}
Now that we have found $\gamma=Y_{11}+Y_{22}$ near the epoch of residual annihilations, let us calculate how much it can deviate from $\eta_{DM}$. Defining the fractional change  in the dark matter comoving density from $\gamma=\eta_{DM} \, (1-\delta_{\eta})$, we get
\begin{equation}
 \delta m \lesssim  \frac{342}{\sqrt{g_{\ast}}} \frac{\delta_{\eta}^{1/2}}{\left< \sigma v \right>_a^2 \eta_{DM}^2 M_p^3},
 \end{equation}
for $x > \bar{\bar{x}}$. As a numerical example, for our set of parameters and taking $\delta_{\eta} \simeq 3\%$ (as limited by the change in the dark matter density after the formation of the CMB \cite{Poulin:2016nat}), we get $\delta m \lesssim 3 \times 10^{-30}$ eV. This bound will be relaxed if the second epoch of annihilation freezes out before the formation of the CMB \cite{Bringmann:2018jpr}.

\subsection{Flavor-sensitive interactions}

In this case, the equation for $\beta$ reads
\begin{equation}
    x^{5/2}\beta'+\left( \frac{3 I_s g'^4 m_\chi^3 \bar{s} \, \eta_{DM} }{8 \pi \kappa m_V^4} \right) \beta - \left( \frac{2i \delta m \, \eta_{DM}}{\kappa m_\chi^2}  \right) x^{7/2}=0
    \label{eqD6}
\end{equation}
where we considered $\alpha \approx \eta_{DM}$ and
\begin{equation}
    \left< \sigma v \right>_s = \frac{I_s g'^4 m_\chi^2}{4\pi m_V^4} \frac{1}{\sqrt{x}} = \overline{\left< \sigma v \right>}_s \frac{1}{\sqrt{x}}.
\end{equation}
This time, $A$ is redefined to
\begin{equation}
    A \equiv \frac{3 I_s g'^4 m_\chi^3 \bar{s} \, \eta_{DM} }{8 \pi \kappa m_V^4}. 
\end{equation}
Working with our set of parameters, it is possible to approximate the solution of eq.~(\ref{eqD6}) to
\begin{equation}
    \beta(x) \approx i \left(\frac{2}{3}\right)^{7/3} A^{4/3}B \, \Gamma \bigg(-\frac{4}{3},\frac{2A}{3x^{3/2}} \bigg),
\end{equation}
where the incomplete gamma function is defined by $\Gamma(a,z)=\int_{z}^{\infty} t^{a-1} e^{-t}dt.$ Now we can use this result and solve for $\alpha$. Taking the limit $\Gamma(s,r)/r^s = -1/s$ when $r \rightarrow 0$ for $Re(s)<0$, we get
\begin{equation}
    \alpha (x) = \eta_{DM}  \left( 1- \frac{\delta m^2}{2 \kappa^2 m_{\chi}^4} x^4 \right).
    \label{eqD11}
\end{equation}
Solving $\alpha=0$ for $x$ gives us the previous result of eq.~(\ref{eqD4}).

Now, let us rearrange the Boltzmann equations as an equation for the total DM comoving density $\gamma$ 
\begin{equation}
xH\gamma'=-\frac{1}{2} \left< \sigma v\right>_a \, s \left( \gamma^2 - \Upsilon^2  \right)
\end{equation}
and an equation for its ``late-time equilibrium'' function $\Upsilon=\frac{\sqrt{f(x)}}{2\delta m} $,
\begin{equation}
    f' = - 3 \overline{\left< \sigma v \right>}_s \frac{s}{\sqrt{x}} \eta_{DM} x H (\alpha')^2,
    \label{eqD13}
\end{equation}
where 
\begin{equation}
    f = (xH \alpha')^2 + 4 \delta m^2 \alpha^2.
    \label{eqD14}
\end{equation}
We will not attempt to solve the full set of equations from before freeze-out to today. Instead, let's try to evolve our functions from their states in the flat land to new states in the region of residual annihilations.

As we are working with smaller and smaller values of $\delta m$, let us explore what happens when $\delta m \rightarrow 0$. In this limit, there should be no residual annihilations, i.e. the total DM density must follow a constant equilibrium function, $ \lim_{\delta m \rightarrow 0} \Upsilon = \eta_{DM}$. Consequently,
\begin{equation}
  \lim_{\delta m \rightarrow 0} f(x) = 4 \delta m^2 \eta_{DM}^2.
\end{equation}
From this result, and equations (\ref{eqD13}) and (\ref{eqD14}) we get, $\alpha' \rightarrow 0$ and $\alpha \rightarrow \eta_{DM}$, in this limit. Also, $\beta \rightarrow 0$. Thus, $\Upsilon^2 \rightarrow \alpha^2$. Now, eq.~(\ref{eqD11}) was obtained using $\alpha \approx \eta_{DM}$. Performing the inverse substitution we get
\begin{equation}
    \alpha(z) = \frac{\eta_{DM}}{ \left( 1+ \frac{z^2}{2} \right)},
\end{equation}
where we defined $z \equiv \frac{\delta m}{\kappa m_\chi^2} x^2$. From now on, we will use $z$ instead of $x$. For example, from eq.~(\ref{eqD11}), the moment when $Y_{11}=Y_{22}$, i.e. $\alpha=0$ , is given by $z=\sqrt{2}$. Now, the equation we need to solve is,
\begin{equation}
    z^{3/2} \delta_{\eta}'(z)=W \, \eta_{DM} \, \left[ 1-2 \delta_{\eta}(z) - \frac{1}{\left( 1+ \frac{z^2}{2} \right)^2} \right],
\end{equation}
where we have parametrized the total DM density as $\gamma = \eta_{DM} (1-\delta_{\eta})$, where $\delta_{\eta} \ll 1$ and we have used $(1-\delta_{\eta})^2 \approx 1-2\delta_{\eta}$. Also, we defined
\begin{equation}
    W \equiv \frac{\left< \sigma v \right>_a \bar{s}}{4 \kappa^{3/2}} \sqrt{\delta m}= \overline{W}\sqrt{\delta m}.
\end{equation}
To a good approximation, we obtain
\begin{equation}
    \delta_{\eta}(z) \approx \frac{\overline{W} \, \eta_{DM}}{2} \sqrt{\delta m} \frac{z^{3/2}}{2+z^2}.
\end{equation}
In this way,
\begin{equation}
    \delta m \approx \frac{1521}{\sqrt{g_{\ast}}}  \frac{\delta_{\eta}^2}{\left< \sigma v \right>^2_a \eta_{DM}^2 M_{p}^3} \frac{(2+z^2)^2}{z^3}.
    \label{eqD20}
\end{equation}

Before getting an upper bound for $\delta m$, let us go back to eq.~(\ref{eqD13}) and eq.~(\ref{eqD14}). These can be merged into
\begin{equation}
    x^{5/2}\alpha'' + \left( 2A-x^{3/2}\right) \alpha'+\left( \frac{B}{\eta_{DM}} \right)^2 x^{9/2}\alpha=0.
\end{equation}
We notice there is a dramatic change in this equation when the damping term changes sign. For this reason, we will take 
\begin{equation}
    \bar{\bar{x}} = \left[  \frac{3}{1.66 \sqrt{g_\ast}} \overline{\left<\sigma v \right>}_s \bar{s} \, m_\chi M_{p} \, \eta_{DM} \right]^{2/3}
\end{equation}
as a better approximation for the starting point for residual annihilations. Numerically, this gives us $\bar{\bar{z}} \ll 1$, so we can make the following approximation
\begin{equation}
    \frac{(2+\bar{\bar{z}}^2)^2}{\bar{\bar{z}}^3} \rightarrow \frac{4}{\bar{\bar{z}}^3}.
\end{equation}
Since this is our starting point, for any $z>\bar{\bar{z}}$ we have from eq.~(\ref{eqD20})
\begin{equation}
    \delta m < 16.3 \frac{m_\chi^{1/2}}{g_{\ast}^{1/4}}  \frac{\delta^{1/2}_{\eta}}{ \overline{\left< \sigma v \right>}_s \left< \sigma v \right>_a^{1/2} \eta_{DM}^{3/2} M_{p}^{5/2}  }. 
\end{equation}
As we can see, for a fixed $\delta m$, the change in the DM comoving density $\delta_{\eta}$ goes to zero when we turn off scatterings. For our parameters, we obtain $\delta m < 5 \times 10^{-28}$ eV.

\end{subappendices}
\end{appendix}

\bibliography{references}

\providecommand{\href}[2]{#2}\begingroup\raggedright\begin{thebibliography}{100}

\bibitem{Salucci:2018hqu}
P.~Salucci, ``{The distribution of dark matter in galaxies},''
  \href{http://dx.doi.org/10.1007/s00159-018-0113-1}{{\em Astron. Astrophys.
  Rev.} {\bfseries 27} no.~1, (2019) 2},
  \href{http://arxiv.org/abs/1811.08843}{{\ttfamily arXiv:1811.08843
  [astro-ph.GA]}}.

\bibitem{Weinberg:2013aya}
D.~H. Weinberg, J.~S. Bullock, F.~Governato, R.~Kuzio~de Naray, and A.~H.~G.
  Peter, ``{Cold dark matter: controversies on small scales},''
  \href{http://dx.doi.org/10.1073/pnas.1308716112}{{\em Proc. Nat. Acad. Sci.}
  {\bfseries 112} (2015) 12249--12255},
  \href{http://arxiv.org/abs/1306.0913}{{\ttfamily arXiv:1306.0913
  [astro-ph.CO]}}.

\bibitem{Spergel:1999mh}
D.~N. Spergel and P.~J. Steinhardt, ``{Observational evidence for
  selfinteracting cold dark matter},''
  \href{http://dx.doi.org/10.1103/PhysRevLett.84.3760}{{\em Phys. Rev. Lett.}
  {\bfseries 84} (2000) 3760--3763},
  \href{http://arxiv.org/abs/astro-ph/9909386}{{\ttfamily
  arXiv:astro-ph/9909386}}.

\bibitem{Tulin:2017ara}
S.~Tulin and H.-B. Yu, ``{Dark Matter Self-interactions and Small Scale
  Structure},'' \href{http://dx.doi.org/10.1016/j.physrep.2017.11.004}{{\em
  Phys. Rept.} {\bfseries 730} (2018) 1--57},
  \href{http://arxiv.org/abs/1705.02358}{{\ttfamily arXiv:1705.02358
  [hep-ph]}}.

\bibitem{Markevitch:2003at}
M.~Markevitch, A.~Gonzalez, D.~Clowe, A.~Vikhlinin, L.~David, W.~Forman,
  C.~Jones, S.~Murray, and W.~Tucker, ``{Direct constraints on the dark matter
  self-interaction cross-section from the merging galaxy cluster 1E0657-56},''
  \href{http://dx.doi.org/10.1086/383178}{{\em Astrophys. J.} {\bfseries 606}
  (2004) 819--824}, \href{http://arxiv.org/abs/astro-ph/0309303}{{\ttfamily
  arXiv:astro-ph/0309303}}.

\bibitem{Kim:2016ujt}
S.~Y. Kim, A.~H. Peter, and D.~Wittman, ``{In the Wake of Dark Giants: New
  Signatures of Dark Matter Self Interactions in Equal Mass Mergers of Galaxy
  Clusters},'' \href{http://dx.doi.org/10.1093/mnras/stx896}{{\em Mon. Not.
  Roy. Astron. Soc.} {\bfseries 469} no.~2, (2017) 1414--1444},
  \href{http://arxiv.org/abs/1608.08630}{{\ttfamily arXiv:1608.08630
  [astro-ph.CO]}}.

\bibitem{Rocha:2012jg}
M.~Rocha, A.~H. Peter, J.~S. Bullock, M.~Kaplinghat, S.~Garrison-Kimmel,
  J.~Onorbe, and L.~A. Moustakas, ``{Cosmological Simulations with
  Self-Interacting Dark Matter I: Constant Density Cores and Substructure},''
  \href{http://dx.doi.org/10.1093/mnras/sts514}{{\em Mon. Not. Roy. Astron.
  Soc.} {\bfseries 430} (2013) 81--104},
  \href{http://arxiv.org/abs/1208.3025}{{\ttfamily arXiv:1208.3025
  [astro-ph.CO]}}.

\bibitem{Vogelsberger:2012ku}
M.~Vogelsberger, J.~Zavala, and A.~Loeb, ``{Subhaloes in Self-Interacting
  Galactic Dark Matter Haloes},''
  \href{http://dx.doi.org/10.1111/j.1365-2966.2012.21182.x}{{\em Mon. Not. Roy.
  Astron. Soc.} {\bfseries 423} (2012) 3740},
  \href{http://arxiv.org/abs/1201.5892}{{\ttfamily arXiv:1201.5892
  [astro-ph.CO]}}.

\bibitem{Kaplinghat:2015aga}
M.~Kaplinghat, S.~Tulin, and H.-B. Yu, ``{Dark Matter Halos as Particle
  Colliders: Unified Solution to Small-Scale Structure Puzzles from Dwarfs to
  Clusters},'' \href{http://dx.doi.org/10.1103/PhysRevLett.116.041302}{{\em
  Phys. Rev. Lett.} {\bfseries 116} no.~4, (2016) 041302},
  \href{http://arxiv.org/abs/1508.03339}{{\ttfamily arXiv:1508.03339
  [astro-ph.CO]}}.

\bibitem{McDermott:2017vyk}
S.~D. McDermott, ``{Is Self-Interacting Dark Matter Undergoing Dark Fusion?},''
  \href{http://dx.doi.org/10.1103/PhysRevLett.120.221806}{{\em Phys. Rev.
  Lett.} {\bfseries 120} no.~22, (2018) 221806},
  \href{http://arxiv.org/abs/1711.00857}{{\ttfamily arXiv:1711.00857
  [hep-ph]}}.

\bibitem{Chu:2018fzy}
X.~Chu, C.~Garcia-Cely, and H.~Murayama, ``{Velocity Dependence from Resonant
  Self-Interacting Dark Matter},''
  \href{http://dx.doi.org/10.1103/PhysRevLett.122.071103}{{\em Phys. Rev.
  Lett.} {\bfseries 122} no.~7, (2019) 071103},
  \href{http://arxiv.org/abs/1810.04709}{{\ttfamily arXiv:1810.04709
  [hep-ph]}}.

\bibitem{Chu:2019awd}
X.~Chu, C.~Garcia-Cely, and H.~Murayama, ``{A Practical and Consistent
  Parametrization of Dark Matter Self-Interactions},''
  \href{http://dx.doi.org/10.1088/1475-7516/2020/06/043}{{\em JCAP} {\bfseries
  06} (2020) 043}, \href{http://arxiv.org/abs/1908.06067}{{\ttfamily
  arXiv:1908.06067 [hep-ph]}}.

\bibitem{Tsai:2020vpi}
Y.-D. Tsai, R.~McGehee, and H.~Murayama, ``{Resonant Self-Interacting Dark
  Matter from Dark QCD},'' \href{http://arxiv.org/abs/2008.08608}{{\ttfamily
  arXiv:2008.08608 [hep-ph]}}.

\bibitem{Chu:2018faw}
X.~Chu, C.~Garcia-Cely, and H.~Murayama, ``{Finite-size dark matter and its
  effect on small-scale structure},''
  \href{http://dx.doi.org/10.1103/PhysRevLett.124.041101}{{\em Phys. Rev.
  Lett.} {\bfseries 124} no.~4, (2020) 041101},
  \href{http://arxiv.org/abs/1901.00075}{{\ttfamily arXiv:1901.00075
  [hep-ph]}}.

\bibitem{Chu:2018nki}
X.~Chu and C.~Garcia-Cely, ``{Core formation from self-heating dark matter},''
  \href{http://dx.doi.org/10.1088/1475-7516/2018/07/013}{{\em JCAP} {\bfseries
  07} (2018) 013}, \href{http://arxiv.org/abs/1803.09762}{{\ttfamily
  arXiv:1803.09762 [hep-ph]}}.

\bibitem{Kamada:2019wjo}
A.~Kamada and H.~J. Kim, ``{Escalating core formation with dark matter
  self-heating},'' \href{http://dx.doi.org/10.1103/PhysRevD.102.043009}{{\em
  Phys. Rev. D} {\bfseries 102} no.~4, (2020) 043009},
  \href{http://arxiv.org/abs/1911.09717}{{\ttfamily arXiv:1911.09717
  [hep-ph]}}.

\bibitem{Kamada:2017gfc}
A.~Kamada, H.~J. Kim, H.~Kim, and T.~Sekiguchi, ``{Self-Heating Dark Matter via
  Semiannihilation},''
  \href{http://dx.doi.org/10.1103/PhysRevLett.120.131802}{{\em Phys. Rev.
  Lett.} {\bfseries 120} no.~13, (2018) 131802},
  \href{http://arxiv.org/abs/1707.09238}{{\ttfamily arXiv:1707.09238
  [hep-ph]}}.

\bibitem{Kamada:2018hte}
A.~Kamada, H.~J. Kim, and H.~Kim, ``{Self-heating of Strongly Interacting
  Massive Particles},''
  \href{http://dx.doi.org/10.1103/PhysRevD.98.023509}{{\em Phys. Rev. D}
  {\bfseries 98} no.~2, (2018) 023509},
  \href{http://arxiv.org/abs/1805.05648}{{\ttfamily arXiv:1805.05648
  [hep-ph]}}.

\bibitem{Kamada:2020buc}
A.~Kamada, H.~J. Kim, and T.~Kuwahara, ``{Maximally self-interacting dark
  matter: models and predictions},''
  \href{http://arxiv.org/abs/2007.15522}{{\ttfamily arXiv:2007.15522
  [hep-ph]}}.

\bibitem{Braaten:2018xuw}
E.~Braaten, D.~Kang, and R.~Laha, ``{Production of dark-matter bound states in
  the early universe by three-body recombination},''
  \href{http://dx.doi.org/10.1007/JHEP11(2018)084}{{\em JHEP} {\bfseries 11}
  (2018) 084}, \href{http://arxiv.org/abs/1806.00609}{{\ttfamily
  arXiv:1806.00609 [hep-ph]}}.

\bibitem{Braaten:2019ohj}
E.~Braaten, D.~Kang, and R.~Laha, ``{Dark Matter Bound States from Three-Body
  Recombination},'' \href{http://dx.doi.org/10.1007/978-3-030-32357-8_156}{{\em
  Springer Proc. Phys.} {\bfseries 238} (2020) 1001--1006},
  \href{http://arxiv.org/abs/1905.04558}{{\ttfamily arXiv:1905.04558
  [hep-ph]}}.

\bibitem{Kaplinghat:2000vt}
M.~Kaplinghat, L.~Knox, and M.~S. Turner, ``{Annihilating the cold dark matter
  cusp crisis},'' \href{http://dx.doi.org/10.1103/PhysRevLett.85.3335}{{\em
  Phys. Rev. Lett.} {\bfseries 85} (2000) 3335},
  \href{http://arxiv.org/abs/astro-ph/0005210}{{\ttfamily
  arXiv:astro-ph/0005210}}.

\bibitem{Cohen:2009fz}
T.~Cohen and K.~M. Zurek, ``{Leptophilic Dark Matter from the Lepton
  Asymmetry},'' \href{http://dx.doi.org/10.1103/PhysRevLett.104.101301}{{\em
  Phys. Rev. Lett.} {\bfseries 104} (2010) 101301},
  \href{http://arxiv.org/abs/0909.2035}{{\ttfamily arXiv:0909.2035 [hep-ph]}}.

\bibitem{buckley2012}
M.~R. Buckley and S.~Profumo, ``Regenerating a symmetry in asymmetric dark
  matter,'' \href{http://dx.doi.org/10.1103/PhysRevLett.108.011301}{{\em Phys.
  Rev. Lett.} {\bfseries 108} (Jan, 2012) 011301}.
  \url{https://link.aps.org/doi/10.1103/PhysRevLett.108.011301}.

\bibitem{Cirelli:2011ac}
M.~Cirelli, P.~Panci, G.~Servant, and G.~Zaharijas, ``{Consequences of
  DM/antiDM Oscillations for Asymmetric WIMP Dark Matter},''
  \href{http://dx.doi.org/10.1088/1475-7516/2012/03/015}{{\em JCAP} {\bfseries
  03} (2012) 015}, \href{http://arxiv.org/abs/1110.3809}{{\ttfamily
  arXiv:1110.3809 [hep-ph]}}.

\bibitem{Tulin:2012re}
S.~Tulin, H.-B. Yu, and K.~M. Zurek, ``{Oscillating Asymmetric Dark Matter},''
  \href{http://dx.doi.org/10.1088/1475-7516/2012/05/013}{{\em JCAP} {\bfseries
  05} (2012) 013}, \href{http://arxiv.org/abs/1202.0283}{{\ttfamily
  arXiv:1202.0283 [hep-ph]}}.

\bibitem{Poulin:2016nat}
V.~Poulin, P.~D. Serpico, and J.~Lesgourgues, ``{A fresh look at linear
  cosmological constraints on a decaying dark matter component},''
  \href{http://dx.doi.org/10.1088/1475-7516/2016/08/036}{{\em JCAP} {\bfseries
  08} (2016) 036}, \href{http://arxiv.org/abs/1606.02073}{{\ttfamily
  arXiv:1606.02073 [astro-ph.CO]}}.

\bibitem{Bringmann:2018jpr}
T.~Bringmann, F.~Kahlhoefer, K.~Schmidt-Hoberg, and P.~Walia, ``{Converting
  nonrelativistic dark matter to radiation},''
  \href{http://dx.doi.org/10.1103/PhysRevD.98.023543}{{\em Phys. Rev. D}
  {\bfseries 98} no.~2, (2018) 023543},
  \href{http://arxiv.org/abs/1803.03644}{{\ttfamily arXiv:1803.03644
  [astro-ph.CO]}}.

\bibitem{Newman:2012nv}
A.~B. Newman, T.~Treu, R.~S. Ellis, D.~J. Sand, C.~Nipoti, J.~Richard, and
  E.~Jullo, ``{The Density Profiles of Massive, Relaxed Galaxy Clusters: I. The
  Total Density Over 3 Decades in Radius},''
  \href{http://dx.doi.org/10.1088/0004-637X/765/1/24}{{\em Astrophys. J.}
  {\bfseries 765} (2013) 24}, \href{http://arxiv.org/abs/1209.1391}{{\ttfamily
  arXiv:1209.1391 [astro-ph.CO]}}.

\bibitem{Bellazzini:2013foa}
B.~Bellazzini, M.~Cliche, and P.~Tanedo, ``{Effective theory of
  self-interacting dark matter},''
  \href{http://dx.doi.org/10.1103/PhysRevD.88.083506}{{\em Phys. Rev. D}
  {\bfseries 88} no.~8, (2013) 083506},
  \href{http://arxiv.org/abs/1307.1129}{{\ttfamily arXiv:1307.1129 [hep-ph]}}.

\bibitem{Kahlhoefer:2017umn}
F.~Kahlhoefer, K.~Schmidt-Hoberg, and S.~Wild, ``{Dark matter self-interactions
  from a general spin-0 mediator},''
  \href{http://dx.doi.org/10.1088/1475-7516/2017/08/003}{{\em JCAP} {\bfseries
  08} (2017) 003}, \href{http://arxiv.org/abs/1704.02149}{{\ttfamily
  arXiv:1704.02149 [hep-ph]}}.

\bibitem{Agrawal:2020lea}
P.~Agrawal, A.~Parikh, and M.~Reece, ``{Systematizing the Effective Theory of
  Self-Interacting Dark Matter},''
  \href{http://dx.doi.org/10.1007/JHEP10(2020)191}{{\em JHEP} {\bfseries 10}
  (2020) 191}, \href{http://arxiv.org/abs/2003.00021}{{\ttfamily
  arXiv:2003.00021 [hep-ph]}}.

\bibitem{Blinov:2020uvz}
N.~Blinov, C.~Keith, and D.~Hooper, ``{Warm Decaying Dark Matter and the Hubble
  Tension},'' \href{http://dx.doi.org/10.1088/1475-7516/2020/06/005}{{\em JCAP}
  {\bfseries 06} (2020) 005}, \href{http://arxiv.org/abs/2004.06114}{{\ttfamily
  arXiv:2004.06114 [astro-ph.CO]}}.

\bibitem{Abellan:2020pmw}
G.~F. Abellan, R.~Murgia, V.~Poulin, and J.~Lavalle, ``{Hints for decaying dark
  matter from $S_8$ measurements},''
  \href{http://arxiv.org/abs/2008.09615}{{\ttfamily arXiv:2008.09615
  [astro-ph.CO]}}.

\bibitem{Cyburt:2015mya}
R.~H. Cyburt, B.~D. Fields, K.~A. Olive, and T.-H. Yeh, ``{Big Bang
  Nucleosynthesis: 2015},''
  \href{http://dx.doi.org/10.1103/RevModPhys.88.015004}{{\em Rev. Mod. Phys.}
  {\bfseries 88} (2016) 015004},
  \href{http://arxiv.org/abs/1505.01076}{{\ttfamily arXiv:1505.01076
  [astro-ph.CO]}}.

\bibitem{Fields:2019pfx}
B.~D. Fields, K.~A. Olive, T.-H. Yeh, and C.~Young, ``{Big-Bang Nucleosynthesis
  After Planck},'' \href{http://dx.doi.org/10.1088/1475-7516/2020/03/010}{{\em
  JCAP} {\bfseries 03} (2020) 010},
  \href{http://arxiv.org/abs/1912.01132}{{\ttfamily arXiv:1912.01132
  [astro-ph.CO]}}.

\bibitem{Aghanim:2018eyx}
{\bfseries Planck} Collaboration, N.~Aghanim {\em et~al.}, ``{Planck 2018
  results. VI. Cosmological parameters},''
  \href{http://arxiv.org/abs/1807.06209}{{\ttfamily arXiv:1807.06209
  [astro-ph.CO]}}.

\bibitem{Adshead:2016xxj}
P.~Adshead, Y.~Cui, and J.~Shelton, ``{Chilly Dark Sectors and Asymmetric
  Reheating},'' \href{http://dx.doi.org/10.1007/JHEP06(2016)016}{{\em JHEP}
  {\bfseries 06} (2016) 016}, \href{http://arxiv.org/abs/1604.02458}{{\ttfamily
  arXiv:1604.02458 [hep-ph]}}.

\bibitem{Adshead:2019uwj}
P.~Adshead, P.~Ralegankar, and J.~Shelton, ``{Reheating in two-sector
  cosmology},'' \href{http://dx.doi.org/10.1007/JHEP08(2019)151}{{\em JHEP}
  {\bfseries 08} (2019) 151}, \href{http://arxiv.org/abs/1906.02755}{{\ttfamily
  arXiv:1906.02755 [hep-ph]}}.

\bibitem{Green:2019glg}
D.~Green {\em et~al.}, ``{Messengers from the Early Universe: Cosmic Neutrinos
  and Other Light Relics},'' {\em Bull. Am. Astron. Soc.} {\bfseries 51} no.~7,
  (2019) 159, \href{http://arxiv.org/abs/1903.04763}{{\ttfamily
  arXiv:1903.04763 [astro-ph.CO]}}.

\bibitem{Kamada:2016euw}
A.~Kamada, M.~Kaplinghat, A.~B. Pace, and H.-B. Yu, ``{How the Self-Interacting
  Dark Matter Model Explains the Diverse Galactic Rotation Curves},''
  \href{http://dx.doi.org/10.1103/PhysRevLett.119.111102}{{\em Phys. Rev.
  Lett.} {\bfseries 119} no.~11, (2017) 111102},
  \href{http://arxiv.org/abs/1611.02716}{{\ttfamily arXiv:1611.02716
  [astro-ph.GA]}}.

\bibitem{Lokas:2000mu}
E.~L. Lokas and G.~A. Mamon, ``{Properties of spherical galaxies and clusters
  with an NFW density profile},''
  \href{http://dx.doi.org/10.1046/j.1365-8711.2001.04007.x}{{\em Mon. Not. Roy.
  Astron. Soc.} {\bfseries 321} (2001) 155},
  \href{http://arxiv.org/abs/astro-ph/0002395}{{\ttfamily
  arXiv:astro-ph/0002395}}.

\bibitem{1988ApJ...332L..33C}
C.~{Carignan} and K.~C. {Freeman}, ``{DDO 154: A ``Dark'' Galaxy?},''
  \href{http://dx.doi.org/10.1086/185260}{{\em \apjl} {\bfseries 332} (Sept.,
  1988) L33}.

\bibitem{Oh:2015xoa}
S.-H. Oh {\em et~al.}, ``{High-resolution mass models of dwarf galaxies from
  LITTLE THINGS},'' \href{http://dx.doi.org/10.1088/0004-6256/149/6/180}{{\em
  Astron. J.} {\bfseries 149} (2015) 180},
  \href{http://arxiv.org/abs/1502.01281}{{\ttfamily arXiv:1502.01281
  [astro-ph.GA]}}.

\bibitem{Dutton:2014xda}
A.~A. Dutton and A.~V. Macciò, ``{Cold dark matter haloes in the Planck era:
  evolution of structural parameters for Einasto and NFW profiles},''
  \href{http://dx.doi.org/10.1093/mnras/stu742}{{\em Mon. Not. Roy. Astron.
  Soc.} {\bfseries 441} no.~4, (2014) 3359--3374},
  \href{http://arxiv.org/abs/1402.7073}{{\ttfamily arXiv:1402.7073
  [astro-ph.CO]}}.

\bibitem{Springel:2005mi}
V.~Springel, ``{The Cosmological simulation code GADGET-2},''
  \href{http://dx.doi.org/10.1111/j.1365-2966.2005.09655.x}{{\em Mon. Not. Roy.
  Astron. Soc.} {\bfseries 364} (2005) 1105--1134},
  \href{http://arxiv.org/abs/astro-ph/0505010}{{\ttfamily
  arXiv:astro-ph/0505010}}.

\bibitem{Springel:2000yr}
V.~Springel, N.~Yoshida, and S.~D. White, ``{GADGET: A Code for collisionless
  and gasdynamical cosmological simulations},''
  \href{http://dx.doi.org/10.1016/S1384-1076(01)00042-2}{{\em New Astron.}
  {\bfseries 6} (2001) 79},
  \href{http://arxiv.org/abs/astro-ph/0003162}{{\ttfamily
  arXiv:astro-ph/0003162}}.

\bibitem{Hernquist:1990be}
L.~Hernquist, ``{An Analytical Model for Spherical Galaxies and Bulges},''
  \href{http://dx.doi.org/10.1086/168845}{{\em Astrophys. J.} {\bfseries 356}
  (1990) 359}.

\bibitem{Springel:2004kf}
V.~Springel, T.~Di~Matteo, and L.~Hernquist, ``{Modeling feedback from stars
  and black holes in galaxy mergers},''
  \href{http://dx.doi.org/10.1111/j.1365-2966.2005.09238.x}{{\em Mon. Not. Roy.
  Astron. Soc.} {\bfseries 361} (2005) 776--794},
  \href{http://arxiv.org/abs/astro-ph/0411108}{{\ttfamily
  arXiv:astro-ph/0411108}}.

\bibitem{Robertson:2016xjh}
A.~Robertson, R.~Massey, and V.~Eke, ``{What does the Bullet Cluster tell us
  about self-interacting dark matter?},''
  \href{http://dx.doi.org/10.1093/mnras/stw2670}{{\em Mon. Not. Roy. Astron.
  Soc.} {\bfseries 465} no.~1, (2017) 569--587},
  \href{http://arxiv.org/abs/1605.04307}{{\ttfamily arXiv:1605.04307
  [astro-ph.CO]}}.

\bibitem{Newman:2013}
A.~B. Newman, T.~Treu, R.~S. Ellis, and D.~J. Sand, ``The density profiles of
  massive, relaxed galaxy clusters. ii. separating luminous and dark matter in
  cluster cores,'' \href{http://dx.doi.org/10.1088/0004-637x/765/1/25}{{\em
  Astrophys. J.} {\bfseries 765} (2013) 25},
  \href{http://arxiv.org/abs/1209.1392}{{\ttfamily arXiv:1209.1392
  [astro-ph.CO]}}.

\bibitem{Sagunski:2020spe}
L.~Sagunski, S.~Gad-Nasr, B.~Colquhoun, A.~Robertson, and S.~Tulin,
  ``{Velocity-dependent Self-interacting Dark Matter from Groups and Clusters
  of Galaxies},'' \href{http://arxiv.org/abs/2006.12515}{{\ttfamily
  arXiv:2006.12515 [astro-ph.CO]}}.

\bibitem{Eliasdottir:2007md}
A.~Eliasdottir, M.~Limousin, J.~Richard, J.~Hjorth, J.-P. Kneib, P.~Natarajan,
  K.~Pedersen, E.~Jullo, and D.~Paraficz, ``{Where is the matter in the Merging
  Cluster Abell 2218?},'' \href{http://arxiv.org/abs/0710.5636}{{\ttfamily
  arXiv:0710.5636 [astro-ph]}}.

\bibitem{Chabrier:2003ki}
G.~Chabrier, ``{Galactic stellar and substellar initial mass function},''
  \href{http://dx.doi.org/10.1086/376392}{{\em Publ. Astron. Soc. Pac.}
  {\bfseries 115} (2003) 763--796},
  \href{http://arxiv.org/abs/astro-ph/0304382}{{\ttfamily
  arXiv:astro-ph/0304382}}.

\bibitem{Read:2018pft}
J.~Read, M.~Walker, and P.~Steger, ``{The case for a cold dark matter cusp in
  Draco},'' \href{http://dx.doi.org/10.1093/mnras/sty2286}{{\em Mon. Not. Roy.
  Astron. Soc.} {\bfseries 481} no.~1, (2018) 860--877},
  \href{http://arxiv.org/abs/1805.06934}{{\ttfamily arXiv:1805.06934
  [astro-ph.GA]}}.

\bibitem{Hayashi:2020jze}
K.~Hayashi, M.~Chiba, and T.~Ishiyama, ``{Diversity of dark matter density
  profiles in the Galactic dwarf spheroidal satellites},''
  \href{http://arxiv.org/abs/2007.13780}{{\ttfamily arXiv:2007.13780
  [astro-ph.GA]}}.

\bibitem{Hayashi:2020syu}
K.~Hayashi, M.~Ibe, S.~Kobayashi, Y.~Nakayama, and S.~Shirai, ``{Probing Dark
  Matter Self-interaction with Ultra-faint Dwarf Galaxies},''
  \href{http://arxiv.org/abs/2008.02529}{{\ttfamily arXiv:2008.02529
  [astro-ph.CO]}}.

\bibitem{Kummer:2019yrb}
J.~Kummer, M.~Br\"uggen, K.~Dolag, F.~Kahlhoefer, and K.~Schmidt-Hoberg,
  ``{Simulations of core formation for frequent dark matter
  self-interactions},'' \href{http://dx.doi.org/10.1093/mnras/stz1261}{{\em
  Mon. Not. Roy. Astron. Soc.} {\bfseries 487} no.~1, (2019) 354--363},
  \href{http://arxiv.org/abs/1902.02330}{{\ttfamily arXiv:1902.02330
  [astro-ph.CO]}}.

\bibitem{Robles:2019mfq}
V.~H. Robles, T.~Kelley, J.~S. Bullock, and M.~Kaplinghat, ``{The Milky
  Way\textquoteright{}s halo and subhaloes in self-interacting dark matter},''
  \href{http://dx.doi.org/10.1093/mnras/stz2345}{{\em Mon. Not. Roy. Astron.
  Soc.} {\bfseries 490} no.~2, (2019) 2117--2123},
  \href{http://arxiv.org/abs/1903.01469}{{\ttfamily arXiv:1903.01469
  [astro-ph.GA]}}.

\bibitem{Sameie:2019zfo}
O.~Sameie, H.-B. Yu, L.~V. Sales, M.~Vogelsberger, and J.~Zavala,
  ``{Self-Interacting Dark Matter Subhalos in the Milky Way\textquoteright{}s
  Tides},'' \href{http://dx.doi.org/10.1103/PhysRevLett.124.141102}{{\em Phys.
  Rev. Lett.} {\bfseries 124} no.~14, (2020) 141102},
  \href{http://arxiv.org/abs/1904.07872}{{\ttfamily arXiv:1904.07872
  [astro-ph.GA]}}.

\bibitem{Kahlhoefer:2019oyt}
F.~Kahlhoefer, M.~Kaplinghat, T.~R. Slatyer, and C.-L. Wu, ``{Diversity in
  density profiles of self-interacting dark matter satellite halos},''
  \href{http://dx.doi.org/10.1088/1475-7516/2019/12/010}{{\em JCAP} {\bfseries
  12} (2019) 010}, \href{http://arxiv.org/abs/1904.10539}{{\ttfamily
  arXiv:1904.10539 [astro-ph.GA]}}.

\bibitem{2020arXiv200702958C}
C.~A. {Correa}, ``{Constraining Velocity-dependent Self-Interacting Dark Matter
  with the Milky Way's dwarf spheroidal galaxies},'' {\em arXiv e-prints}
  (July, 2020) arXiv:2007.02958,
  \href{http://arxiv.org/abs/2007.02958}{{\ttfamily arXiv:2007.02958
  [astro-ph.GA]}}.

\bibitem{Gondolo:1990dk}
P.~Gondolo and G.~Gelmini, ``{Cosmic abundances of stable particles: Improved
  analysis},'' \href{http://dx.doi.org/10.1016/0550-3213(91)90438-4}{{\em Nucl.
  Phys. B} {\bfseries 360} (1991) 145--179}.

\bibitem{Vogelsberger:2018bok}
M.~Vogelsberger, J.~Zavala, K.~Schutz, and T.~R. Slatyer, ``{Evaporating the
  Milky Way halo and its satellites with inelastic self-interacting dark
  matter},'' \href{http://arxiv.org/abs/1805.03203}{{\ttfamily arXiv:1805.03203
  [astro-ph.GA]}}.

\bibitem{Monaghan1985}
J.~J. {Monaghan} and J.~C. {Lattanzio}, ``{A refined particle method for
  astrophysical problems},'' {\em Astron. and Astrophys.} {\bfseries 149}
  no.~1, (Aug., 1985) 135--143.

\bibitem{Kochanek:2000pi}
C.~Kochanek and M.~J. White, ``{A Quantitative study of interacting dark matter
  in halos},'' \href{http://dx.doi.org/10.1086/317149}{{\em Astrophys. J.}
  {\bfseries 543} (2000) 514},
  \href{http://arxiv.org/abs/astro-ph/0003483}{{\ttfamily
  arXiv:astro-ph/0003483}}.

\bibitem{Yoshida:2000uw}
N.~Yoshida, V.~Springel, S.~D. White, and G.~Tormen, ``{Weakly self-interacting
  dark matter and the structure of dark halos},''
  \href{http://dx.doi.org/10.1086/317306}{{\em Astrophys. J.} {\bfseries 544}
  (2000) L87--L90}, \href{http://arxiv.org/abs/astro-ph/0006134}{{\ttfamily
  arXiv:astro-ph/0006134}}.

\bibitem{Dave:2000ar}
R.~Dave, D.~N. Spergel, P.~J. Steinhardt, and B.~D. Wandelt, ``{Halo properties
  in cosmological simulations of selfinteracting cold dark matter},''
  \href{http://dx.doi.org/10.1086/318417}{{\em Astrophys. J.} {\bfseries 547}
  (2001) 574--589}, \href{http://arxiv.org/abs/astro-ph/0006218}{{\ttfamily
  arXiv:astro-ph/0006218}}.

\bibitem{Koda:2011yb}
J.~Koda and P.~R. Shapiro, ``{Gravothermal collapse of isolated
  self-interacting dark matter haloes: N-body simulation versus the fluid
  model},'' \href{http://dx.doi.org/10.1111/j.1365-2966.2011.18684.x}{{\em Mon.
  Not. Roy. Astron. Soc.} {\bfseries 415} (2011) 1125},
  \href{http://arxiv.org/abs/1101.3097}{{\ttfamily arXiv:1101.3097
  [astro-ph.CO]}}.

\bibitem{Fry:2015rta}
A.~B. Fry, F.~Governato, A.~Pontzen, T.~Quinn, M.~Tremmel, L.~Anderson,
  H.~Menon, A.~Brooks, and J.~Wadsley, ``{All about baryons: revisiting SIDM
  predictions at small halo masses},''
  \href{http://dx.doi.org/10.1093/mnras/stv1330}{{\em Mon. Not. Roy. Astron.
  Soc.} {\bfseries 452} no.~2, (2015) 1468--1479},
  \href{http://arxiv.org/abs/1501.00497}{{\ttfamily arXiv:1501.00497
  [astro-ph.CO]}}.

\bibitem{Elbert:2016dbb}
O.~D. Elbert, J.~S. Bullock, M.~Kaplinghat, S.~Garrison-Kimmel, A.~S. Graus,
  and M.~Rocha, ``{A Testable Conspiracy: Simulating Baryonic Effects on
  Self-Interacting Dark Matter Halos},''
  \href{http://dx.doi.org/10.3847/1538-4357/aa9710}{{\em Astrophys. J.}
  {\bfseries 853} no.~2, (2018) 109},
  \href{http://arxiv.org/abs/1609.08626}{{\ttfamily arXiv:1609.08626
  [astro-ph.GA]}}.

\bibitem{Blinnikov:1983gh}
S.~Blinnikov and M.~Khlopov, ``{Possible astronomical effects of mirror
  particles},'' {\em Sov. Astron.} {\bfseries 27} (1983) 371--375.

\bibitem{Berezhiani:1995am}
Z.~Berezhiani, A.~Dolgov, and R.~Mohapatra, ``{Asymmetric inflationary
  reheating and the nature of mirror universe},''
  \href{http://dx.doi.org/10.1016/0370-2693(96)00219-5}{{\em Phys. Lett. B}
  {\bfseries 375} (1996) 26--36},
  \href{http://arxiv.org/abs/hep-ph/9511221}{{\ttfamily arXiv:hep-ph/9511221}}.

\bibitem{Foot:2004pa}
R.~Foot, ``{Mirror matter-type dark matter},''
  \href{http://dx.doi.org/10.1142/S0218271804006449}{{\em Int. J. Mod. Phys. D}
  {\bfseries 13} (2004) 2161--2192},
  \href{http://arxiv.org/abs/astro-ph/0407623}{{\ttfamily
  arXiv:astro-ph/0407623}}.

\bibitem{Cline:2012is}
J.~M. Cline, Z.~Liu, and W.~Xue, ``{Millicharged Atomic Dark Matter},''
  \href{http://dx.doi.org/10.1103/PhysRevD.85.101302}{{\em Phys. Rev. D}
  {\bfseries 85} (2012) 101302},
  \href{http://arxiv.org/abs/1201.4858}{{\ttfamily arXiv:1201.4858 [hep-ph]}}.

\bibitem{Cline:2013pca}
J.~M. Cline, Z.~Liu, G.~Moore, and W.~Xue, ``{Scattering properties of dark
  atoms and molecules},''
  \href{http://dx.doi.org/10.1103/PhysRevD.89.043514}{{\em Phys. Rev. D}
  {\bfseries 89} no.~4, (2014) 043514},
  \href{http://arxiv.org/abs/1311.6468}{{\ttfamily arXiv:1311.6468 [hep-ph]}}.

\bibitem{Cline:2013zca}
J.~M. Cline, Z.~Liu, G.~Moore, and W.~Xue, ``{Composite strongly interacting
  dark matter},'' \href{http://dx.doi.org/10.1103/PhysRevD.90.015023}{{\em
  Phys. Rev. D} {\bfseries 90} no.~1, (2014) 015023},
  \href{http://arxiv.org/abs/1312.3325}{{\ttfamily arXiv:1312.3325 [hep-ph]}}.

\bibitem{CyrRacine:2012fz}
F.-Y. Cyr-Racine and K.~Sigurdson, ``{Cosmology of atomic dark matter},''
  \href{http://dx.doi.org/10.1103/PhysRevD.87.103515}{{\em Phys. Rev. D}
  {\bfseries 87} no.~10, (2013) 103515},
  \href{http://arxiv.org/abs/1209.5752}{{\ttfamily arXiv:1209.5752
  [astro-ph.CO]}}.

\bibitem{Feng:2009mn}
J.~L. Feng, M.~Kaplinghat, H.~Tu, and H.-B. Yu, ``{Hidden Charged Dark
  Matter},'' \href{http://dx.doi.org/10.1088/1475-7516/2009/07/004}{{\em JCAP}
  {\bfseries 07} (2009) 004}, \href{http://arxiv.org/abs/0905.3039}{{\ttfamily
  arXiv:0905.3039 [hep-ph]}}.

\bibitem{Foot:2014uba}
R.~Foot and S.~Vagnozzi, ``{Dissipative hidden sector dark matter},''
  \href{http://dx.doi.org/10.1103/PhysRevD.91.023512}{{\em Phys. Rev. D}
  {\bfseries 91} (2015) 023512},
  \href{http://arxiv.org/abs/1409.7174}{{\ttfamily arXiv:1409.7174 [hep-ph]}}.

\bibitem{Foot:2016wvj}
R.~Foot and S.~Vagnozzi, ``{Solving the small-scale structure puzzles with
  dissipative dark matter},''
  \href{http://dx.doi.org/10.1088/1475-7516/2016/07/013}{{\em JCAP} {\bfseries
  07} (2016) 013}, \href{http://arxiv.org/abs/1602.02467}{{\ttfamily
  arXiv:1602.02467 [astro-ph.CO]}}.

\bibitem{Boddy:2014yra}
K.~K. Boddy, J.~L. Feng, M.~Kaplinghat, and T.~M.~P. Tait, ``{Self-Interacting
  Dark Matter from a Non-Abelian Hidden Sector},''
  \href{http://dx.doi.org/10.1103/PhysRevD.89.115017}{{\em Phys. Rev. D}
  {\bfseries 89} no.~11, (2014) 115017},
  \href{http://arxiv.org/abs/1402.3629}{{\ttfamily arXiv:1402.3629 [hep-ph]}}.

\bibitem{Boddy:2016bbu}
K.~K. Boddy, M.~Kaplinghat, A.~Kwa, and A.~H.~G. Peter, ``{Hidden Sector
  Hydrogen as Dark Matter: Small-scale Structure Formation Predictions and the
  Importance of Hyperfine Interactions},''
  \href{http://dx.doi.org/10.1103/PhysRevD.94.123017}{{\em Phys. Rev. D}
  {\bfseries 94} no.~12, (2016) 123017},
  \href{http://arxiv.org/abs/1609.03592}{{\ttfamily arXiv:1609.03592
  [hep-ph]}}.

\bibitem{Feng:2009hw}
J.~L. Feng, M.~Kaplinghat, and H.-B. Yu, ``{Halo Shape and Relic Density
  Exclusions of Sommerfeld-Enhanced Dark Matter Explanations of Cosmic Ray
  Excesses},'' \href{http://dx.doi.org/10.1103/PhysRevLett.104.151301}{{\em
  Phys. Rev. Lett.} {\bfseries 104} (2010) 151301},
  \href{http://arxiv.org/abs/0911.0422}{{\ttfamily arXiv:0911.0422 [hep-ph]}}.

\bibitem{Buckley:2009in}
M.~R. Buckley and P.~J. Fox, ``{Dark Matter Self-Interactions and Light Force
  Carriers},'' \href{http://dx.doi.org/10.1103/PhysRevD.81.083522}{{\em Phys.
  Rev. D} {\bfseries 81} (2010) 083522},
  \href{http://arxiv.org/abs/0911.3898}{{\ttfamily arXiv:0911.3898 [hep-ph]}}.

\bibitem{Tulin:2013teo}
S.~Tulin, H.-B. Yu, and K.~M. Zurek, ``{Beyond Collisionless Dark Matter:
  Particle Physics Dynamics for Dark Matter Halo Structure},''
  \href{http://dx.doi.org/10.1103/PhysRevD.87.115007}{{\em Phys. Rev. D}
  {\bfseries 87} no.~11, (2013) 115007},
  \href{http://arxiv.org/abs/1302.3898}{{\ttfamily arXiv:1302.3898 [hep-ph]}}.

\bibitem{Born:1926yhp}
M.~Born, ``{Quantenmechanik der Stoßvorgänge},''
  \href{http://dx.doi.org/10.1007/BF01397184}{{\em Z. Phys.} {\bfseries 38}
  no.~11-12, (1926) 803--827}.

\bibitem{Ibe:2009mk}
M.~Ibe and H.-b. Yu, ``{Distinguishing Dark Matter Annihilation Enhancement
  Scenarios via Halo Shapes},''
  \href{http://dx.doi.org/10.1016/j.physletb.2010.07.026}{{\em Phys. Lett. B}
  {\bfseries 692} (2010) 70--73},
  \href{http://arxiv.org/abs/0912.5425}{{\ttfamily arXiv:0912.5425 [hep-ph]}}.

\bibitem{Kahlhoefer:2013dca}
F.~Kahlhoefer, K.~Schmidt-Hoberg, M.~T. Frandsen, and S.~Sarkar, ``{Colliding
  clusters and dark matter self-interactions},''
  \href{http://dx.doi.org/10.1093/mnras/stt2097}{{\em Mon. Not. Roy. Astron.
  Soc.} {\bfseries 437} no.~3, (2014) 2865--2881},
  \href{http://arxiv.org/abs/1308.3419}{{\ttfamily arXiv:1308.3419
  [astro-ph.CO]}}.

\bibitem{Robertson:2016qef}
A.~Robertson, R.~Massey, and V.~Eke, ``{Cosmic particle colliders: simulations
  of self-interacting dark matter with anisotropic scattering},''
  \href{http://dx.doi.org/10.1093/mnras/stx463}{{\em Mon. Not. Roy. Astron.
  Soc.} {\bfseries 467} no.~4, (2017) 4719--4730},
  \href{http://arxiv.org/abs/1612.03906}{{\ttfamily arXiv:1612.03906
  [astro-ph.CO]}}.

\bibitem{Zavala:2012us}
J.~Zavala, M.~Vogelsberger, and M.~G. Walker, ``{Constraining Self-Interacting
  Dark Matter with the Milky Way's dwarf spheroidals},''
  \href{http://dx.doi.org/10.1093/mnrasl/sls053}{{\em Mon. Not. Roy. Astron.
  Soc.} {\bfseries 431} (2013) L20--L24},
  \href{http://arxiv.org/abs/1211.6426}{{\ttfamily arXiv:1211.6426
  [astro-ph.CO]}}.

\bibitem{Vogelsberger:2013}
M.~Vogelsberger and J.~Zavala, ``{Direct detection of self-interacting dark
  matter},'' \href{http://dx.doi.org/10.1093/mnras/sts712}{{\em Mon. Not. Roy.
  Astron. Soc.} {\bfseries 430} no.~3, (2013) 1722--1735},
  \href{http://arxiv.org/abs/1211.1377}{{\ttfamily 1211.1377}}.

\bibitem{Vogelsberger:2014pda}
M.~Vogelsberger, J.~Zavala, C.~Simpson, and A.~Jenkins, ``{Dwarf galaxies in
  CDM and SIDM with baryons: observational probes of the nature of dark
  matter},'' \href{http://dx.doi.org/10.1093/mnras/stu1713}{{\em Mon. Not. Roy.
  Astron. Soc.} {\bfseries 444} no.~4, (2014) 3684--3698},
  \href{http://arxiv.org/abs/1405.5216}{{\ttfamily arXiv:1405.5216
  [astro-ph.CO]}}.

\bibitem{Vogelsberger:2015gpr}
M.~Vogelsberger, J.~Zavala, F.-Y. Cyr-Racine, C.~Pfrommer, T.~Bringmann, and
  K.~Sigurdson, ``{ETHOS -- an effective theory of structure formation: dark
  matter physics as a possible explanation of the small-scale CDM problems},''
  \href{http://dx.doi.org/10.1093/mnras/stw1076}{{\em Mon. Not. Roy. Astron.
  Soc.} {\bfseries 460} no.~2, (2016) 1399--1416},
  \href{http://arxiv.org/abs/1512.05349}{{\ttfamily arXiv:1512.05349
  [astro-ph.CO]}}.

\bibitem{Robertson:2020pxj}
A.~Robertson, R.~Massey, V.~Eke, J.~Schaye, and T.~Theuns, ``{The surprising
  accuracy of isothermal Jeans modelling of self-interacting dark matter
  density profiles},'' \href{http://arxiv.org/abs/2009.07844}{{\ttfamily
  arXiv:2009.07844 [astro-ph.CO]}}.

\bibitem{Feng:2010zp}
J.~L. Feng, M.~Kaplinghat, and H.-B. Yu, ``{Sommerfeld Enhancements for Thermal
  Relic Dark Matter},''
  \href{http://dx.doi.org/10.1103/PhysRevD.82.083525}{{\em Phys. Rev. D}
  {\bfseries 82} (2010) 083525},
  \href{http://arxiv.org/abs/1005.4678}{{\ttfamily arXiv:1005.4678 [hep-ph]}}.

\bibitem{Khrapak:2004}
S.~A. e.~a. Khrapak, ``{Scattering in the Attractive Yukawa Potential:
  Application to the Ion-Drag Force in Complex Plasmas},''
  \href{http://dx.doi.org/10.1109/TPS.2004.826073}{{\em IEEE Transactions on
  Plasma Science} {\bfseries 32} no.~2, (2004) 555--560}.

\bibitem{Iwanus:2017mue}
N.~Iwanus, P.~J. Elahi, and G.~F. Lewis, ``{Dark matter annihilation feedback
  in cosmological simulations -- I: Code convergence and idealized haloes},''
  \href{http://dx.doi.org/10.1093/mnras/stx1974}{{\em Mon. Not. Roy. Astron.
  Soc.} {\bfseries 472} no.~1, (2017) 1214--1225},
  \href{http://arxiv.org/abs/1707.06770}{{\ttfamily arXiv:1707.06770
  [astro-ph.CO]}}.

\bibitem{List:2019jrl}
F.~List, N.~Iwanus, P.~J. Elahi, and G.~F. Lewis, ``{A Novel Scheme for Dark
  Matter Annihilation Feedback in Cosmological Simulations},''
  \href{http://dx.doi.org/10.1093/mnras/stz2287}{{\em Mon. Not. Roy. Astron.
  Soc.} {\bfseries 489} no.~3, (2019) 4217--4232},
  \href{http://arxiv.org/abs/1908.05812}{{\ttfamily arXiv:1908.05812
  [astro-ph.CO]}}.

\bibitem{Neumann:1951}
J.~von Neumann, ``Various techniques used in connection with random digits,''
  in {\em Monte Carlo Method}, A.~S. Householder, G.~E. Forsythe, and H.~H.
  Germond, eds., vol.~12 of {\em National Bureau of Standards Applied
  Mathematics Series}, ch.~13, pp.~36--38.
\newblock US Government Printing Office, Washington, DC, 1951.

\bibitem{Aarseth:1974}
S.~J. Aarseth, M.~Henon, and R.~Wielen, ``{A Comparison of Numerical Methods
  for the Study of Star Cluster Dynamics},'' {\em Astron. \& Astrophys.}
  {\bfseries 37} no.~1, (1974) 183--187.

\end{thebibliography}\endgroup
\bibliographystyle{utphys}

\end{document}